\documentclass[trackchanges]{aastex701}
\usepackage{booktabs}
\usepackage{tabularx}
\usepackage{rotating}
\usepackage{longtable}
\usepackage{afterpage}
\usepackage{amsmath}

\newcommand{\degree}{^\circ}
\newcommand{\kms}{$\mathrm{km~s^{-1}}$}
\defcitealias{Hunt2023}{[H\&R23,24]}
\defcitealias{Cantat-Gaudin2018}{[C-G18]}
\defcitealias{Cantat-Gaudin2019}{[C-G19]}
\defcitealias{Cantat-Gaudin2020}{[C-G20]}
\defcitealias{Dias2002}{[D02]}
\defcitealias{Dias2014}{[D14]}
\defcitealias{Dias2018}{[D18]}
\defcitealias{Dias2021}{[D21]}
\defcitealias{Sampedro2017}{[S17]}
\defcitealias{Lynga1964}{[L64]}
\defcitealias{Lynga1987}{[L87]}
\defcitealias{Carrera2019}{[C19]}
\defcitealias{Kounkel2020}{[K20]}
\defcitealias{Cavallo2024}{[C24]}
\defcitealias{Carraro2006}{[C06]}
\defcitealias{Kharchenko2013}{[K13]}
\defcitealias{Just2023}{[J23]}
\defcitealias{Perren2023}{[P23]}
\defcitealias{Plevne2026}{[P26]}
\defcitealias{Zhong2020}{[Z20]}
\defcitealias{Fu2022}{[F22]}
\defcitealias{Randich2022}{[R22]}
\newcolumntype{C}{>{$}c<{$}}  % centered math mode 
\accepted{July 20, 2026}
\begin{document}

\title{One cluster, two clusters, no cluster? -- The curious case of HSC 2686 and Lynga 3}

% Author Orchid ID: enter ID or remove command
%\newcommand{\orcidauthorA}{0000-0003-0869-4847} % Add \orcidA{} behind the author's name
%\newcommand{\orcidauthorB}{0000-0000-0000-000X} % Add \orcidB{} behind the author's name

% Authors, for the paper (add full first names)
\author[orcid=0000-0003-0869-4847,gname='Andreas',sname='Ritter']{Andreas Ritter}
%\altaffiliation{Kitt Peak National Observatory}
\affiliation{Laboratory for Space Research, The University of Hong Kong, Hong Kong (S.A.R.)}
\email[show]{azuri.ritter@gmail.com}  

\author[orcid=0000-0002-2062-0173,gname='Quentin A.',sname='Parker']{Quentin A. Parker}
\affiliation{Laboratory for Space Research, The University of Hong Kong, Hong Kong (S.A.R.)}
\email[]{quentinp@hku.hk}  

\author[orcid=0009-0008-7342-5217,gname='Xizhe',sname='Xie']{Xizhe Xie}
\affiliation{Department of Physics, Imperial College London, London SW7 2AZ, United Kingdom}
\email[]{julia.xie22@imperial.ac.uk}

\author[orcid=0000-0003-4979-5273,gname='Xingyi',sname='Guo']{Xingyi Guo}
\affiliation{Department of Astronomy, The University of Tokyo, Hongo, Bunkyo-ku, Tokyo 113-0033, Japan}
\affiliation{Division of Science, National Astronomical Observatory of Japan, Osawa, Mitaka, Tokyo 181-8588, Japan}
\email[]{xingyi.guo@grad.nao.ac.jp}
\begin{abstract}
This study was motivated by the search for Planetary Nebulae (PNe) and Open Cluster (OCs) associations.
Only 5 PNe are currently in Galactic OCs among the previous $\sim4,800$ OCs known. They are valuable because as cluster members their properties can be linked to their progenitor stars, something not possible for general field PNe. Since the GAIA astrometric satellite first data release, thousands of new OC candidates have been identified. This offers fresh motivation to search them as hosts for known PNe. 
Statistically we expect 30 OC-PN pairs to be present. We searched these GAIA candidate OCs, found by Hunt \& Reffert (2023), for known central stars of PNe (CSPNe) as a practical proxy for the PNe themselves and one suited to automated searches. 
We identified 1 potential CSPN (IRAS 15127-5811) of "Likely" PN [GKF2010] MN18 in the HASH database, as a claimed member of the putative cluster HSC~2686. A second cluster (Lynga~3) with similar parameters and in very close angular proximity could also be a potential host. Here we reclassify  CSPN IRAS 15127-5811 as a blue supergiant with a surrounding bi-polar nebula. Nevertheless, we collected all available information on proposed member stars for both these clusters. We conclude that neither claimed cluster is real. Our results shed considerable doubt on the veracity of many newly identified OCs that have modest numbers of stellar members. 
\end{abstract}



\section{Introduction} 
Planetary Nebulae (PNe) are the glowing, ionised shrouds of the expelled outer atmospheres of low- to 
intermediate-mass stars ($0.8-8\ M_{\odot}$) during their transformation from AGB stars to White Dwarfs  
\citep[e.g. see reviews by][]{Kwitter2022, 2022FrASS...9.5287P}. With a life span typically of only a few 
tens of thousands of years  \citep[e.g. see][]{Badenes2015,2016A&A...588A..25M}
they represent a short, colourful and important late stage of low mass stellar evolution and play a key role in enriching the Interstellar Medium (ISM). According to the HASH PNe database\footnote{\href{http://hashpn.space}{http://hashpn.space}} \citep{Parker2016} 
there are currently about 4,000 True, Likely, and Possible PNe known in our Galaxy. Only 5 of those are 
confirmed members of Galactic Open Clusters (OCs hereafter) -- PN PHR~J1724-3859 \citep[a member of Trumpler~25,][]{2026ApJ...996...90F}; PN NGC2818  \citep[a member of NGC2818A,][]{Fragkou2025};  PN G177.5+03.1 \citep[a member of M37,][]{Fragkou2022}; 
BMP J1613-5406 \citep[a member of NGC 6067,][]{Fragkou2022a}; and PHR J1315-6555 \citep[a member of ESO 96-SC04,][]{Parker2011}. 
Associations between PNe and OCs, although exceedingly rare, are very important as they can provide accurate ages 
and therefore decent progenitor mass estimates of the PN central stars (CSPNe) as well as other key parameters. 
Finding other PNe inside OCs is important because we can directly link their 
properties to their progenitor stars, something not possible for general field PNe.
Spectrophotometric observations provide key PN properties, while progenitor star properties such as main sequence
turn-off mass come from cluster colour-magnitude diagrams (CMD) and theoretical 
isochrones \citep{2012MNRAS.427..127B} because cluster stars share some common physical characteristics.

With space-based high-precision astrometry with \textit{GAIA} the number of known and 
potential OCs in the Galaxy has increased substantially, e.g. \cite{Cantat-Gaudin2019} (hereafter \citetalias{Cantat-Gaudin2019}), \cite{Chi2023}, \cite{Hunt2023,Hunt2024} (hereafter \citetalias{Hunt2023}). 
Here we look for potential CSPNe within the 7,167 OC candidates identified in \citetalias{Hunt2023}. 
The paper is organised as follows: 
In section \ref{sec:candidate_selection} we describe our sample selection. Section \ref{sec:OC_catalogue} 
describes the OC catalogue by Hunt and Reffert. In \ref{sec:methods} we describe our investigation, while in \ref{sec:results} we present our results.

%\endnote{This is an endnote.} % To use endnotes, please un-comment \printendnotes below (before References). Only journal Laws uses \footnote.

% The order of the section titles is different for some journals. Please refer to the "Instructions for Authors” on the journal homepage.

\section{Selection of CSPNe}
\label{sec:candidate_selection}
Two major attempts have been made to identify CSPNe using GAIA data \citep[see][]{Gonzalez-Santamaria2021,Chornay2021}. These catalogues are valuable but they cannot be taken as a "black box" 
because \cite{Parker2022} have shown that many true CSPNe are beyond GAIA limits (g$\sim$21). 
The approach adopted by both \citet{Gonzalez-Santamaria2021} and  \citet{Chornay2021} is to select the bluest Gaia star closest to any PNe's geometric centre. However, this leads to a $\sim20\%$ CSPNe misidentification rate \citep{Parker2022}. Hence, we limit our search to the 1,297 CSPNe of True, Likely, and Possible PNe currently identified in the HASH database from deeper, multi-colour imaging where the true, usually faint, blue CSPNe can be properly identified. 

\section{The New putative Open Cluster Catalogue}
\label{sec:OC_catalogue}
\citepalias{Hunt2023} used an unsupervised "Hierarchical Density-Based Spatial Clustering of Applications with Noise" (HDBSCAN) algorithm to recover new OCs. They validated their OC candidates using a statistical density test \citep[Mann-Whitney U test,][]{Mann1947} and a Bayesian convolutional neural network for colour-magnitude diagram classification. They then inferred basic astrometric parameters, ages, extinctions, and distances for the OCs in the resulting catalogue. 
They identified 7,167 OCs in the Milky Way with these techniques - ostensibly the largest OC catalogue to date including 2,387 new candidate OCs. Of those new OC candidates, 2213 have less than 100, 1,833 less than 50 and 1,265 less than 30 identified members. The reliability of these OC identifications is a major aspect of this current work but investigating all new candidate OCs is beyond the scope of this paper.

\section{Methods}
\label{sec:methods}
\begin{figure}[h]
\centerline{\includegraphics[width=10.5 cm]{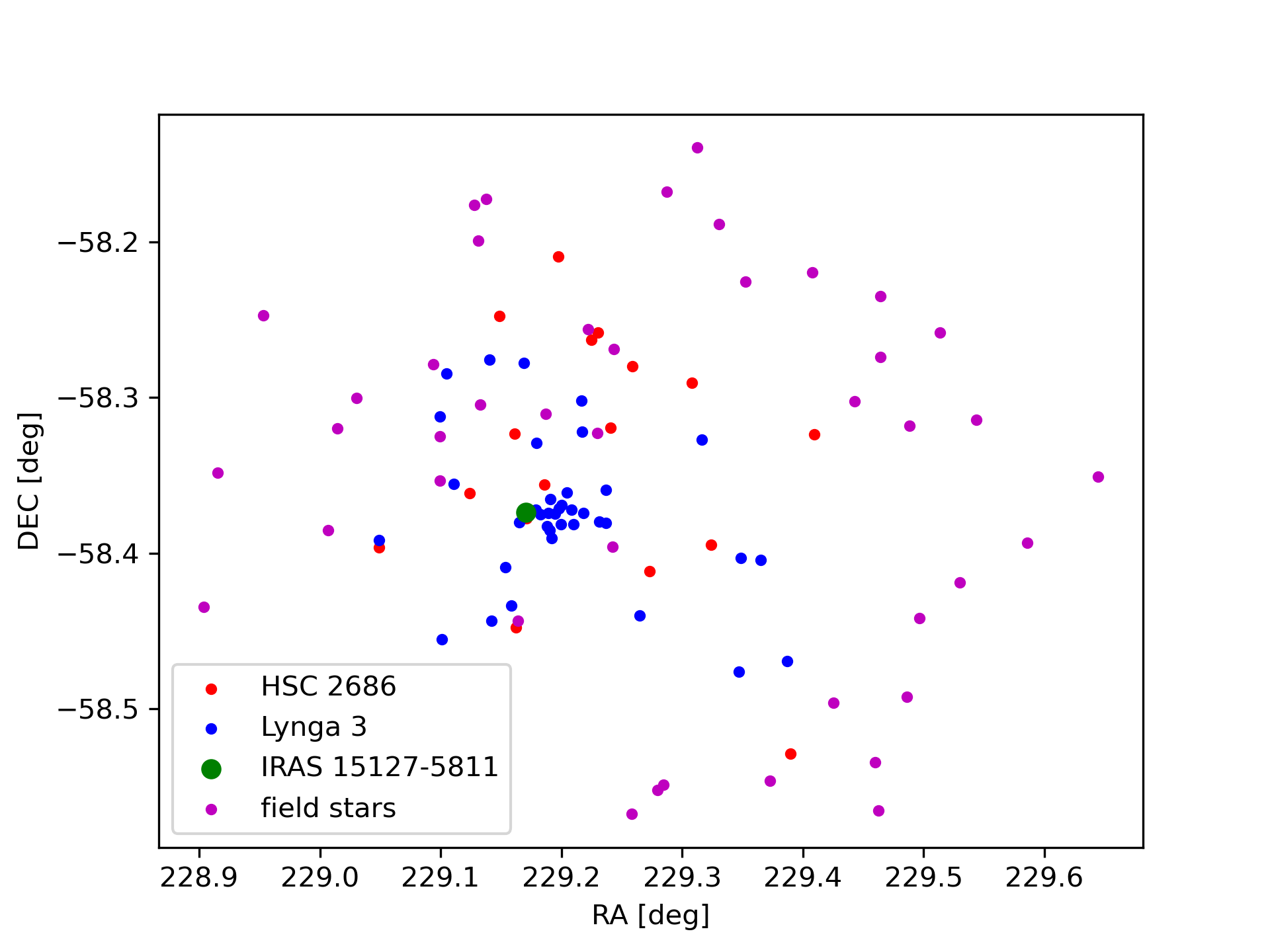}}
\vspace*{-2mm}
\centerline{\includegraphics[width=10.5 cm]{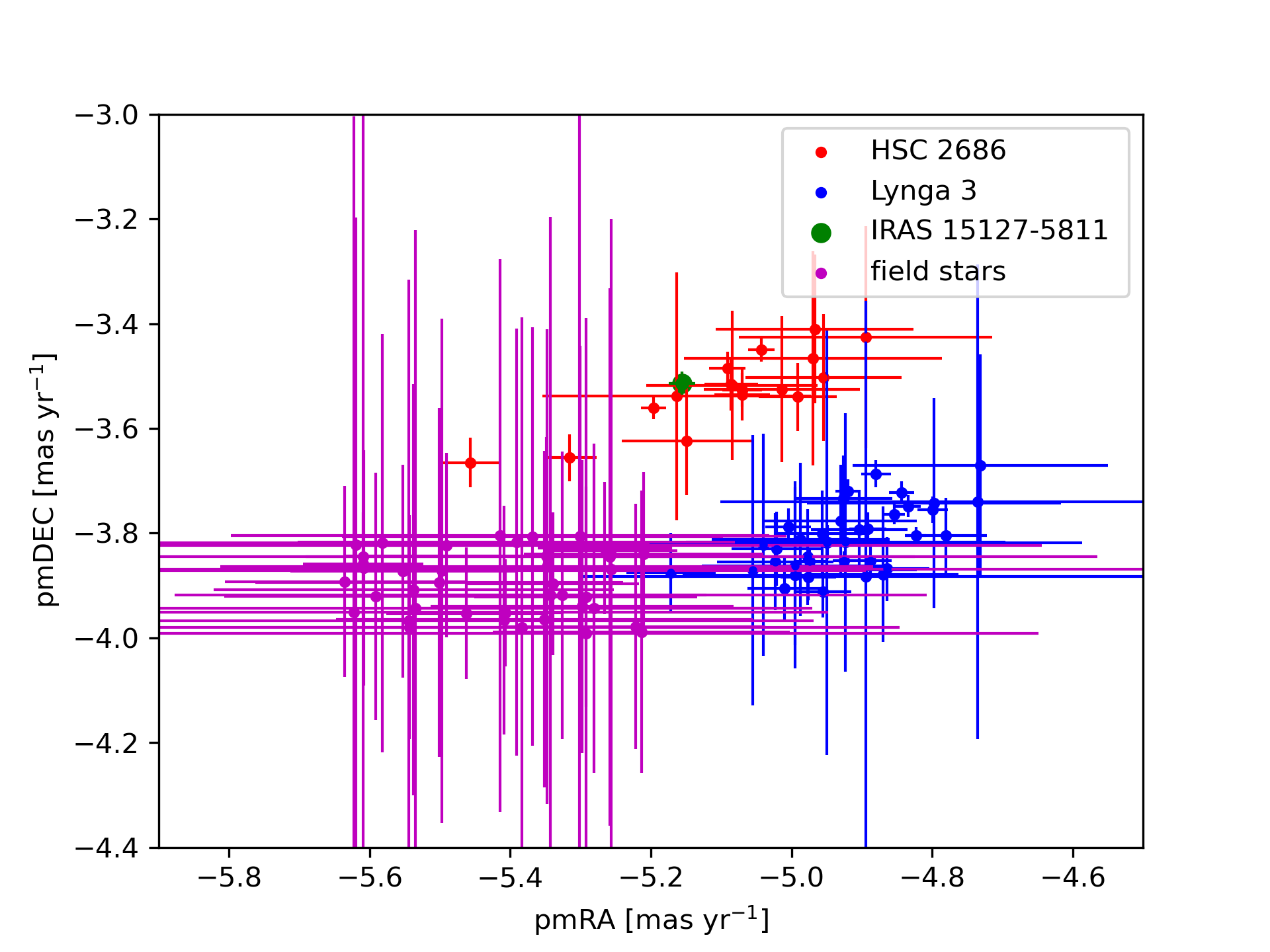}}
\vspace*{-2mm}
\centerline{\includegraphics[width=10.5 cm]{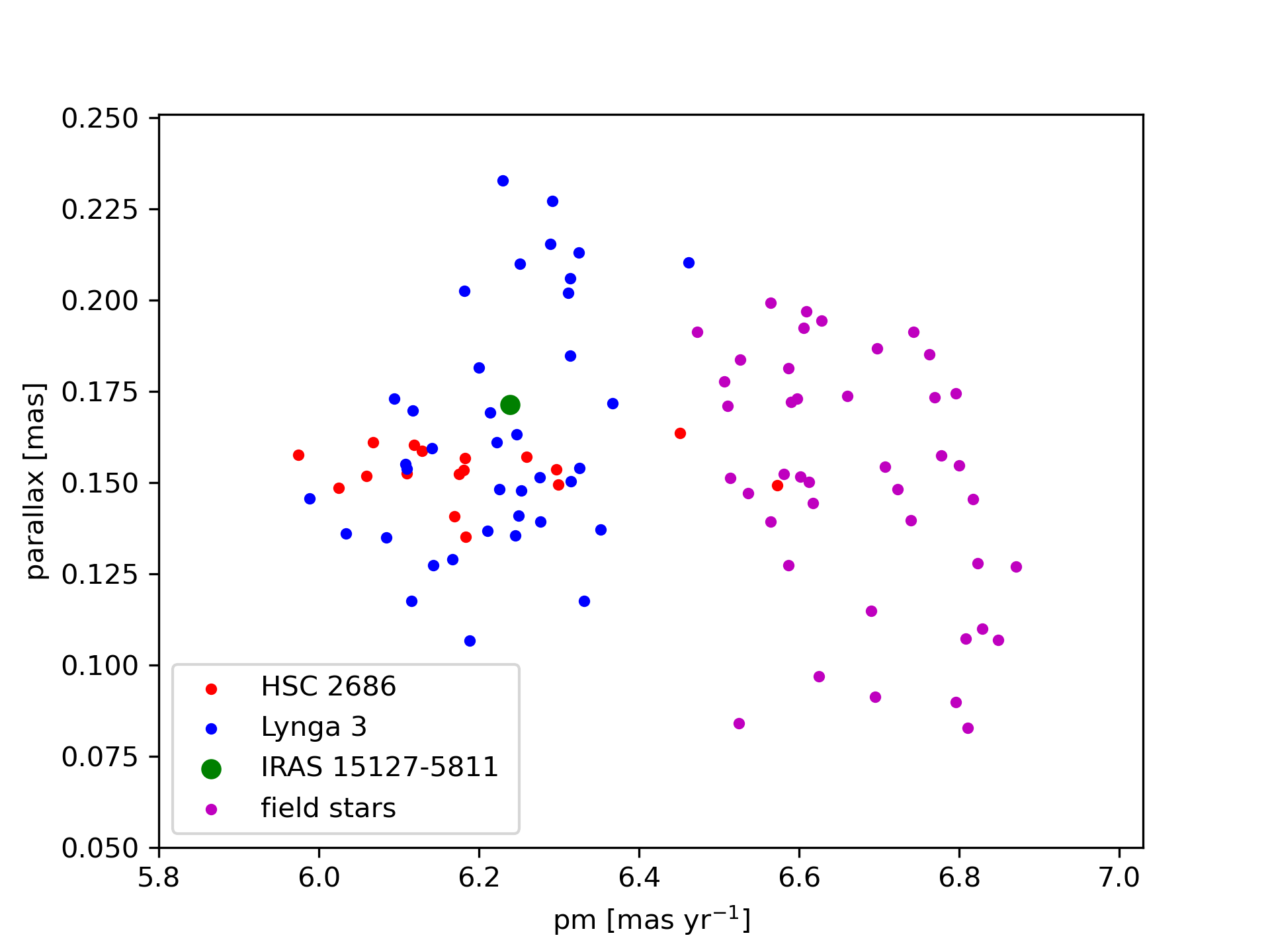}}
\caption{(\textbf{top}) Positions of the stars in HSC~2686, Lynga~3, and the field stars sample. (\textbf{middle}) 
Proper motions for the same clusters and the field stars sample. (\textbf{bottom}) Parallaxes versus proper motions for the same clusters and the field stars sample. Note that the uncertainties are not plotted in the top and bottom panels to preserve visibility of the individual stars.}
\label{fig:RA_DEC_pm_par}
\end{figure}

We cross-matched the HASH CSPNe with the latest H\&R catalogue of member stars of their 7,167 clusters. 
One CSPN (IRAS~15127-5811) of Likely PN [GKF2010] MN18 is noted as a member of the OC HSC~2686 with a membership probability of 0.978. The cluster itself is in close angular proximity to another cluster, Lynga~3, which has similar parameters. 
Interestingly, \cite{Lynga1964} (\citetalias{Lynga1964} hereafter), as well as \citet{Sampedro2017} (\citetalias{Sampedro2017} hereafter), \cite{Cantat-Gaudin2018,Cantat-Gaudin2020} (\citetalias{Cantat-Gaudin2018} and \citetalias{Cantat-Gaudin2020} hereafter), \citet{Dias2014, Dias2018, Dias2021} (\citetalias{Dias2014}, \citetalias{Dias2018}, and \citetalias{Dias2021} hereafter), listed IRAS~15127-5811 as a member of Lynga 3. The H\&R OC 
parameters for HSC~2686 and Lynga~3 compared to IRAS~15127-5811 are listed in Tab.~\ref{tab1}. Both clusters are distant and affected by high extinction, with parallaxes close to the usable \textit{Gaia} DR3 limit. As can be seen in the table, the positions, proper motions, parallaxes, distances, extinctions, and even the listed ages of both clusters pretty much agree within the uncertainties, leaving room for the interpretation that both clusters are actually the same OC. We collected all member stars of both clusters, created diagnostic plots (Fig.~\ref{fig:RA_DEC_pm_par}) and fitted the isochrones with an adapted version of \textit{OCFit}\footnote{https://github.com/hektor-monteiro/OCFit/} (Monteiro \& Dias 2016) using the PARSEC v1.2 isochrones by \cite{Bressan2012} and the CMD 3.8 
website\footnote{http://stev.oapd.inaf.it/cgi-bin/cmd\_3.8}. For each cluster we collected all available parameters from the literature and ran OCFit 10 times to determine the mean parameter values and their uncertainties. For a comparison we also fitted the isochrones 
of the well-known and well studied  OCs NGC~2244 and NGC~6530, as well as a sample of field stars consisting of 39 stars close to 
HSC~2686 and Lynga~3 with similar positions, proper motions, and parallaxes (see Fig.~\ref{fig:RA_DEC_pm_par}). As can be seen in Fig.~\ref{fig:RA_DEC_pm_par}, HSC~2686 and Lynga~3 are overlapping in RA, DEC, and parallax, and are only separated as 2 distinct groups in the proper motions, although still close together.
%\unskip

\begin{table}[h] 
\caption{OC parameters for HSC~2686 and Lynga~3 from \citetalias{Hunt2023} compared to IRAS~15127-5811.\label{tab1}}
%\begin{adjustwidth}{-\extralength}{0cm}
%\begin{tabular}{\textwidth}{|C|CC|CC|C|}
\begin{tabular}{|C|C|C|C|}
\toprule
\textbf{Parameter}	& \textbf{HSC 2686}	&  \textbf{Lynga 3} &  \textbf{IRAS 15127-5811}\\
\midrule
\mathrm{\# member~stars}		& 18	& 38 &\\
\mathrm{RA~[deg]}		& 229.1715		&  229.1970  & 229.1708 \\
\mathrm{DEC~[deg]}		& -58.3735		&  -58.3750 & -58.3739\\
\mathrm{Glon~[deg]}		& 321.0244	&  321.0350  & 321.0239\\
\mathrm{Glat~[deg]}		& -0.6991	&  -0.7074 & -0.6992\\
\mathrm{astrometric~SNR} & 4.4 & 7.4 & \\
\mathrm{r50~[pc]}  & 6.75 & 1.38 & \\
\mathrm{r_{core}~[pc]} & 15.0 & 0.94 & \\
\mathrm{r_{tidal}~[pc]} & 15.0 & 7.0 & \\
\mathrm{r_{total}~[pc]} & 17.5 & 12.1 & \\
\mathrm{pmRA~[mas~yr^{-1}]} & -5.09 & -4.91 & -5.16\\
\mathrm{\sigma_{pmRA}~[mas~yr^{-1}]} & 0.14 & 0.08 & \\
\mathrm{pmDEC~[mas~yr^{-1}]} & -3.53 & -3.81 & -3.51\\
\mathrm{\sigma_{pmDEC} ~[mas~yr^{-1}]} & 0.07 & 0.06 & \\
\mathrm{parallax~[mas]} & 0.155 & 0.161 & 0.171\pm0.019\\
\mathrm{\sigma_{parallax}} & 0.007 & 0.029 & \\
\mathrm{dist16~[pc]} & 4996 & 4929 & \\
\mathrm{dist50~[pc]} & 5209  & 5056 & 5257 \\
\mathrm{dist84~[pc]} & 5442 & 5191 & \\
\mathrm{N_{dist}} & 18 & 34 & \\
\mathrm{v_{rad}~[km~s^{-1}]} & -56.4 & -4.8  &\\
\mathrm{\sigma_{v_{rad}}~[km~s^{-1}]} & 7.2 & 23.4  &\\
\mathrm{N_{vrad}} & 1 & 2 & \\
\mathrm{logAge16~[yr]} & 6.46 & 6.40 & \\
\mathrm{logAge50~[yr]} & 7.02 & 6.87 & \approx 7~\text{\citep{Gvaramadze2015}} \\
\mathrm{logAge84~[yr]} & 7.31 & 7.09 & \\
\mathrm{AV16} & 5.8 & 5.9 & \\
\mathrm{AV50} & 6.2 & 6.3 & \\
\mathrm{AV84} & 6.6 & 6.7 & \\
\mathrm{CMDCl2.5} & 0.03 & 0.55 & \\
\mathrm{CMDCl16} & 0.29 & 0.82 & \\
\mathrm{CMDCl50} & 0.49 & 0.88 & \\
\mathrm{CMDCl84} & 0.95 & 0.98 & \\
\mathrm{CMDCl97.5} & 0.95 & 0.98 & \\

%\textsuperscript{1}\\
\bottomrule
\end{tabular}
%\end{adjustwidth}
%\noindent{\footnotesize{\textsuperscript{1} Tables may have a footer.}}
\end{table}

For validation and comparison of our isochrone-fitting procedure we chose 2 well studied clusters -- NGC~2244 \citep[e.g.][]{Bonatto2009} and NGC~6530 \citep[e.g.][]{Prisinzano2019}. NGC~2244 has 360 member stars listed in H\&R, while NGC~6530 has 680. Individual OC parameters from the literature are given in Tab.~\ref{tab:literature}. 
%We first provide brief details of the two OC candidates HSC~2686 and Lynga~3 that might host [GKF2010] MN18. Two known OCs NGC~2244 and NGC~6530 and some field stars from the general area of HSC~2686 and Lynga~3 are also examined for comparison. Together these comprise this brief study.

%\afterpage{
%\iffalse % LaTeX ignores everything after this
\begin{sidewaystable}[h]
\begin{longtable}{|C|C|C|C|C|C|C|C|C|C|}
    \caption{OC parameters for HSC 2686, Lynga 3, NGC 2244, and NGC 6530 from the literature. Parameters from H\&R are printed in red, fitted parameters in blue.\label{tab:literature}}\\
        \hline
        \text{\textbf{Cluster name}} & \text{\textbf{N}} & \text{\textbf{distance}} & \text{\textbf{log age/[yr]}} & \text{\textbf{metallicity}} & \text{\textbf{A}}_{\mathbf{V}} & \text{\textbf{parallax}} & \text{\textbf{pm}}_{\mathbf{x}} & \text{\textbf{pm}}_{\mathbf{y}} & \text{\textbf{reference}}\\
        & \text{\textbf{(P}}\mathbf{\geq0.5)} & \text{\textbf{[pc]}} & & \text{\textbf{[dex]}} & \text{\textbf{[mag]}} & \text{\textbf{[mas\,yr}}^{\mathbf{-1}}\text{\textbf{]}} & \text{\textbf{[mas\,yr}}^{\mathbf{-1}}\text{\textbf{]}} & \text{\textbf{[mas\,yr}}^{\mathbf{-1}}\text{\textbf{]}} &\\
        \hline
        \endfirsthead
        \endhead
        \hline
        \endfoot
        \mathrm{HSC~2686} & \textcolor{red}{18} &\textcolor{red}{5209\pm230} & \textcolor{red}{7.02^{+0.29}_{-0.48}} & \textcolor{blue}{-0.17\pm0.01} & \textcolor{red}{6.18^{+0.44}_{-0.34}} & \textcolor{red}{0.1552\pm0.0017} & \textcolor{red}{-5.093\pm0.034} & \textcolor{red}{-3.526\pm0.017} & \textcolor{red}{\text{\citetalias{Hunt2023}}} \\
        & & \textcolor{blue}{6362\pm 566} & \textcolor{blue}{7.26\pm0.97} & \textcolor{blue}{0.12\pm0.21} & \textcolor{blue}{4.4\pm0.9} &  & & & \textcolor{blue}{\textit{OCFit}}\\
        \hline
        \mathrm{HSC~2686} & & 6452\pm27 & 6.79\pm0.22 & 0.48\pm0.3 & 5.47\pm0.25 & 0.155\pm0.037 & & & \text{\citetalias{Cavallo2024}}\\
        \hline
        \hline
        \mathrm{Lynga~3} & \textcolor{red}{38} & \textcolor{red}{5056^{+134}_{-127}} & \textcolor{red}{6.87^{+0.22}_{-0.46}} & \textcolor{blue}{-0.41\pm0.24} & \textcolor{red}{6.26^{+0.39}_{-0.33}} & \textcolor{red}{0.161\pm0.005} & \textcolor{red}{-4.914\pm0.015} & \textcolor{red}{-3.809\pm0.011} & \text{\textcolor{red}{\citetalias{Hunt2023}}}\\
        & & \textcolor{blue}{6269\pm 789} & \textcolor{blue}{6.74\pm0.15} & \textcolor{blue}{0.22\pm0.12} & \textcolor{blue}{4.64\pm0.04} & & & & \textcolor{blue}{\textit{OCFit}}\\
        \hline
        \mathrm{Lynga~3} & 25 & \textcolor{blue}{1579\pm1334} & \textcolor{blue}{8.549\pm0.31} & \textcolor{blue}{0.04\pm0.49} & \textcolor{blue}{2.65\pm0.49} & \textcolor{blue}{0.600\pm0.553} & \textcolor{blue}{-5.453\pm3.458} & \textcolor{blue}{-4.142\pm3.424} & \text{\citetalias{Lynga1964}~\textcolor{blue}{\textit{OCFit}}}\\
        \hline
        \mathrm{Lynga~3} & 75^5 & 1369 & 8.55 &  & 4.033 &  & -10.04\pm0.53 & -3.89\pm0.46 & \text{\citetalias{Dias2002} v3.4b} \\
        \hline
        \mathrm{Lynga~3} & 216^3 & 4164 & 8.915 & \textcolor{blue}{0.12\pm0.11} & 4.842^1 & & -2.39\pm0.73 & 0.01\pm0.73 & \text{\citetalias{Kharchenko2013}}\\
        & & \textcolor{blue}{7836\pm687} & \textcolor{blue}{8.866\pm0.072} & \textcolor{blue}{0.173\pm0.019} & \textcolor{blue}{1.550\pm0.557} & & & & \textcolor{blue}{\textit{OCFit}}\\
        \hline
        \mathrm{Lynga~3} & 83 &  &  &  &  &  & -10.04\pm4.55 & -3.89\pm4.01 & \text{\citetalias{Dias2014}} \\
        & & \textcolor{blue}{959\pm307} & \textcolor{blue}{8.705\pm0.338} & \textcolor{blue}{-0.228\pm0.376} & \textcolor{blue}{1.832\pm0.631} & & & & \textcolor{blue}{\textit{OCFit}}\\
        \hline
        \mathrm{Lynga~3} & 39^3 & 1369 & 8.55 & \textcolor{blue}{-0.16\pm0.34} & 4.033^1 & & & & \text{\citetalias{Sampedro2017}}\\
        & & \textcolor{blue}{1179\pm2519} & \textcolor{blue}{8.766\pm0.410} & \textcolor{blue}{0.202\pm0.241} & \textcolor{blue}{1.802\pm0.284} & & & & \textcolor{blue}{\textit{OCFit}}\\
        \hline
        \mathrm{Lynga~3} & 38 & 7292\pm500 &  & & & 0.108\pm0.012 & -4.853\pm0.018 & -3.727\pm0.028 & \text{\citetalias{Cantat-Gaudin2018}}\\
        & & \textcolor{blue}{4183\pm1309} & \textcolor{blue}{6.764\pm0.153} & \textcolor{blue}{0.304\pm0.137} & \textcolor{blue}{4.580\pm0.036} & & & & \textcolor{blue}{\textit{OCFit}}\\
        \hline
        \mathrm{Lynga~3} & 101 &  &  &  &  &  & -4.17\pm0.11 & -3.44\pm0.1 & \text{\citetalias{Dias2018}} \\
        & & \textcolor{blue}{982\pm1239} & \textcolor{blue}{8.772\pm0.421} & \textcolor{blue}{0.307\pm0.427} & \textcolor{blue}{1.674\pm0.731} & & & & \textcolor{blue}{\textit{OCFit}}\\
        \hline
        \mathrm{Lynga~3} & 24^2 &  &  &  & &0.108\pm0.076 & -4.853\pm0.115 & -3.727\pm0.148 & \text{\citetalias{Cantat-Gaudin2020}}\\
        & & \textcolor{blue}{4856\pm823} & \textcolor{blue}{6.763\pm0.055} & \textcolor{blue}{0.280\pm0.137} & \textcolor{blue}{4.581\pm0.062} & & & & \textcolor{blue}{\textit{OCFit}}\\
        \hline
        \mathrm{Lynga~3} & & 3310 & 8.68 & & 4.39 & & & & \text{\citetalias{Kounkel2020}}\\
        \hline
        \mathrm{Lynga~3} & 55 & 6922\pm2613 & 6.708\pm1.045 & 0.067\pm0.124 & 4.687\pm0.6 & 0.117\pm0.077 & -4.858\pm0.111 & -3.756\pm0.174 & \text{\citetalias{Dias2021}} \\
        & & \textcolor{blue}{4086\pm963} & \textcolor{blue}{6.806\pm0.679} & \textcolor{blue}{0.218\pm0.196} & \textcolor{blue}{4.618\pm0.337} & & & & \textcolor{blue}{\textit{OCFit}}\\
        \hline
        \mathrm{Lynga~3} &  & 4164 & 8.915 &  &  & & & & \text{\citetalias{Just2023}}\\
        \hline
        \mathrm{Lynga~3} & 47 & 5750 & 8.386 & -0.051 & 4.93 & 0.148 & -4.887 & -3.794 & \text{\citetalias{Perren2023}}\\
        \hline
        \mathrm{Lynga~3} &  & 6135\pm30 & 6.26\pm0.25 & -0.17\pm0.3 & 5.79\pm0.2 & 0.163\pm0.033 & & & \text{\citetalias{Cavallo2024}}\\
        \hline
        \mathrm{Lynga~3} & 11^4 & 6452\pm35 & 6.75^{+0.48}_{-0.42} & -0.41_{-0.06}^{+0.14} & 8.10_{-0.91}^{+0.78}\ ^5 & & & & \text{\citetalias{Plevne2026}}\\
        \hline
        \textcolor{blue}{\mathrm{Lynga~3~combined}} & \textcolor{blue}{313} & \textcolor{blue}{1298\pm674} & \textcolor{blue}{8.460\pm0.258} & \textcolor{blue}{-0.296\pm0.221} & \textcolor{blue}{2.302\pm0.683} & & & & \textcolor{blue}{\textit{OCFit}}\\
        \hline
        \textcolor{blue}{\mathrm{Field ~stars}} & \textcolor{blue}{39} & \textcolor{blue}{5536\pm1465} & \textcolor{blue}{8.78\pm0.490} & \textcolor{blue}{0.18\pm0.36} & \textcolor{blue}{4.11\pm0.36} & & & & \textcolor{blue}{\textit{OCFit}}\\
        \hline
        \hline
        \mathrm{NGC~2244} & \textcolor{red}{360}& \textcolor{red}{1415\pm4} & \textcolor{red}{6.61^{+0.11}_{-0.13}} & \textcolor{blue}{-0.21\pm0.09} & \textcolor{red}{1.63^{+0.25}_{-0.22}} & \textcolor{red}{0.6567\pm0.0036} & \textcolor{red}{-1.737\pm0.008} & \textcolor{red}{0.253\pm0.010} & \textcolor{red}{\text{\citetalias{Hunt2023}}}\\
        & & \textcolor{blue}{1384\pm36} & \textcolor{blue}{6.710\pm0.043} & \textcolor{blue}{-0.052\pm0.077} & \textcolor{blue}{1.476\pm0.046} & & & & \textcolor{blue}{\textit{OCFit}}\\
        \hline
        \mathrm{NGC~2244} & & 1700 & 6.48 & & 1.395 & & & & \text{\citetalias{Lynga1987}}\\
        \mathrm{NGC~2244} & 360^3 & 1532 & 6.7 &  & 1.677^1 & & -0.84\pm0.25 & -0.94\pm0.25 & \text{\citetalias{Kharchenko2013}}\\
        \mathrm{NGC~2244} & 639 &  &  &  &  &  & -0.23\pm1.65 & -1.13\pm2.79 & \text{\citetalias{Dias2014}} \\
        \mathrm{NGC~2244} & & 1660 & 6.28 & & 1.457^1 & & & & \text{\citetalias{Sampedro2017}}\\
        \bottomrule
    \end{longtable}
\end{sidewaystable}

\begin{sidewaystable}[h]
\begin{longtable}{|C|C|C|C|C|C|C|C|C|C|}
    %\caption{OC parameters for HSC 2686, Lynga 3, NGC 2244, and NGC 6530 from the literature.\label{tab:literature}}\\
        \hline
        \multicolumn{10}{c}{Continuation of Table \ref{tab:literature}} \\ 
        \hline
        \text{\textbf{Cluster name}} & \text{\textbf{N}} & \text{\textbf{distance}} & \text{\textbf{log age/[yr]}} & \text{\textbf{metallicity}} & \text{\textbf{A}}_{\mathbf{V}} & \text{\textbf{parallax}} & \text{\textbf{pm}}_{\mathbf{x}} & \text{\textbf{pm}}_{\mathbf{y}} & \text{\textbf{reference}}\\
         & \text{\textbf{(P}}\geq\text{\textbf{0.5)}} & \text{\textbf{[pc]}} &  & \text{\textbf{[dex]}} & \text{\textbf{[mag]}} & \text{\textbf{[mas\,yr}}^{\mathbf{-1}}\text{\textbf{]}} & \text{\textbf{[mas\,yr}}^{\mathbf{-1}}\text{\textbf{]}} & \text{\textbf{[mas\,yr}}^{\mathbf{-1}}\text{\textbf{]}} &\\
        \hline
        \endfirsthead
        \hline
        \endhead
        \hline
        \endfoot
    % MOVE YOUR NOTES HERE
    \hline
    \multicolumn{10}{l}{\footnotesize \textsuperscript{1} calculated from E(B-V) assuming $\mathrm{R_V}=3.1$.} \\
    \multicolumn{10}{l}{\footnotesize \textsuperscript{2} P$\geq$0.7} \\
    \multicolumn{10}{l}{\footnotesize \textsuperscript{3} all membership probabilities stated \begin{math}\geq\end{math}0.5} \\
    \multicolumn{10}{l}{\footnotesize \textsuperscript{4} P$\geq$0.75} \\
    \multicolumn{10}{l}{\footnotesize \textsuperscript{5} no membership probabilities given} \\
    \multicolumn{10}{l}{\footnotesize \textsuperscript{6} A$_{\mathrm{G}}$} \\
%    \multicolumn{10}{l}{\footnotesize [C24] = \cite{Cavallo2024}}\\
%    \multicolumn{10}{l}{\footnotesize [K13] = \cite{Kharchenko2013}}\\
%    \multicolumn{10}{l}{\footnotesize [S17] = \cite{Sampedro2017}}\\
%    \multicolumn{10}{l}{\footnotesize [C-G18] = \cite{Cantat-Gaudin2018}}\\
%    \multicolumn{10}{l}{\footnotesize [C-G20] = \cite{Cantat-Gaudin2020}}\\
%    \multicolumn{10}{l}{\footnotesize [D02] = \cite{Dias2002}}\\
%    \multicolumn{10}{l}{\footnotesize [D14] = \cite{Dias2014}}\\
%    \multicolumn{10}{l}{\footnotesize [D18] = \cite{Dias2018}}\\
%    \multicolumn{10}{l}{\footnotesize [D21] = \cite{Dias2021}}\\
%    \multicolumn{10}{l}{\footnotesize [J23] = \cite{Just2023}}\\
%    \multicolumn{10}{l}{\footnotesize [P26] = \cite{Plevne2026}}\\
%    \multicolumn{10}{l}{\footnotesize [L87] = \cite{Lynga1987}}\\
    \multicolumn{10}{l}{\footnotesize \citetalias{Carrera2019} = \cite{Carrera2019}}\\
    \multicolumn{10}{l}{\footnotesize \citetalias{Zhong2020} = \cite{Zhong2020}}\\
    \multicolumn{10}{l}{\footnotesize \citetalias{Fu2022} = \cite{Fu2022}}\\
    \multicolumn{10}{l}{\footnotesize \citetalias{Randich2022} = \cite{Randich2022}}\\
%    \multicolumn{10}{l}{\footnotesize \textsuperscript{11} = }\\
    \endlastfoot
        \mathrm{NGC~2244} & 623 & 1549\pm10 & & & & 0.617\pm0.006 & -1.598\pm0.019 & 0.179\pm0.017 &\text{\citetalias{Cantat-Gaudin2018}}\\
        \mathrm{NGC~2244} & 580 &  &  &  &  &  & -2.03\pm1.46 & 0.62\pm2.28 & \text{\citetalias{Dias2018}} \\
        \mathrm{NGC~2244} & & & & -0.23\pm0.09 & & & & & \text{\citetalias{Carrera2019}}\\
        \mathrm{NGC~2244} & 460^2 & 1478 & 7.1 &  & 1.46 & 0.617\pm0.127 & -1.598\pm0.3 & 0.179\pm0.352 &\text{\citetalias{Cantat-Gaudin2020}}\\
        \mathrm{NGC~2244} & & 1702\pm105 & 6.71\pm0.19 & & 2.5\pm0.3 & & & & \text{\citetalias{Kounkel2020}}\\
        \mathrm{NGC~2244} & & & & -0.177\pm0.059 & & & & & \text{\citetalias{Zhong2020}}\\
        \mathrm{NGC~2244} & 623 & 1254\pm89 & 7.111\pm0.090 & -0.214\pm0.084 & 1.571\pm0.063 & 0.605\pm0.136 & -1.563\pm0.475 & 0.186\pm0.433 & \text{\citetalias{Dias2021}} \\
        \mathrm{NGC~2244}& & & & -0.081\pm0.038 & & & & & \text{\citetalias{Fu2022}} \\
        \mathrm{NGC~2244} & & 1478 & 6.6 & -0.04\pm0.05 & & & & & \text{\citetalias{Randich2022}} \\
        \mathrm{NGC~2244} &  & 1531 & 6.7 &  &  &  & & & \text{\citetalias{Just2023}}\\
        \mathrm{NGC~2244} & 491 & 1450 & 6.70 & -0.196 & 1.46 & 0.653 & -1.708 & 0.234 & \text{\citetalias{Perren2023}}\\
        \mathrm{NGC~2244} &  & 1517\pm22 & 6.62\pm0.16 & 0.26\pm0.22 & 1.28\pm0.12 & 0.659\pm0.045 & & & \text{\citetalias{Cavallo2024}}\\
        \mathrm{NGC~2244} & 246 & 1491\pm15 & 6.73_{-0.44}^{+0.54} & -0.01_{-0.34}^{+0.35} & 1.18_{-0.73}^{+0.89}\ ^6 &  & & & \text{\citepalias{Plevne2026}}\\
        \hline
        \hline
        \mathrm{NGC~6530} & \textcolor{red}{476} & \textcolor{red}{1230\pm3} & \textcolor{red}{6.59^{+0.09}_{-0.14}} & \textcolor{blue}{0.14\pm0.09} & \textcolor{red}{1.48^{+0.25}_{-0.19}} & \textcolor{red}{0.7641\pm0.0028} & \textcolor{red}{1.333\pm0.015} & \textcolor{red}{-2.058\pm0.011} & \textcolor{red}{\text{\citetalias{Hunt2023}}}\\
        & & \textcolor{blue}{1204\pm26} & \textcolor{blue}{6.687\pm0.026} & \textcolor{blue}{0.242\pm0.094} & \textcolor{blue}{1.125\pm0.057} & & & & \textcolor{blue}{\textit{OCFit}}\\
        \hline
        \mathrm{NGC~6530} & & 1600 & 6.3 & & 0.93^1 & & & & \text{\citetalias{Lynga1987}}\\
        \mathrm{NGC~6530} & 62 & 1365 & 6.67 & & 6.677 & & 1.06\pm0.39 & -2.85\pm0.39 & \text{\citetalias{Kharchenko2013}}\\
        \mathrm{NGC~6530} & 473 &  &  &  &  &  & 0.35\pm0.1 & -2.11\pm0.92 & \text{\citetalias{Dias2014}}\\
        \mathrm{NGC~6530} & & 1330 & 6.87 & & 1.023^1 & & & & \text{\citetalias{Sampedro2017}}\\
        \mathrm{NGC~6530} & 446 &  &  &  &  &  & 0.5\pm0.1 & -2.99\pm0.1 & \text{\citetalias{Dias2018}}\\
        \mathrm{NGC~6530} & 80 & 1206\pm39 & 6.728\pm0.045 & 0.373\pm0.137 & 1.163\pm0.037 & 0.762\pm0.111 & 1.375\pm0.352 & -1.992\pm0.307 & \text{\citetalias{Dias2021}}\\
        \mathrm{NGC~6530} & & 1325 & 6.3 & -0.02\pm0.08 & & & & & \text{\citetalias{Randich2022}} \\
        \mathrm{NGC~6530} & & 1365 & 6.67 &  &  &  &  & & \text{\citetalias{Just2023}}\\
        \mathrm{NGC~6530} & 616 & 1325 & 6.70 & -0.025 & 1.05 & 0.762 & 1.286 & -2.052 & \text{\citetalias{Perren2023}}\\
        \mathrm{NGC~6530} & & 1309\pm26 & 6.87\pm0.24 & -0.05\pm0.26 & 1.5\pm0.2 & 0.764\pm0.038 &  & & \text{\citetalias{Cavallo2024}}\\
        \bottomrule
    \end{longtable}
\end{sidewaystable}
\clearpage

\section{Results}
\label{sec:results}

\subsection{The putative PN [GKF2010] MN18 (IRAS 15127-5811)}
\label{subsec:MN18}
\begin{figure}[h]
%\vspace*{-15mm}%\hspace*{-1cm}
\centerline{\includegraphics[width=50mm,height=45mm]{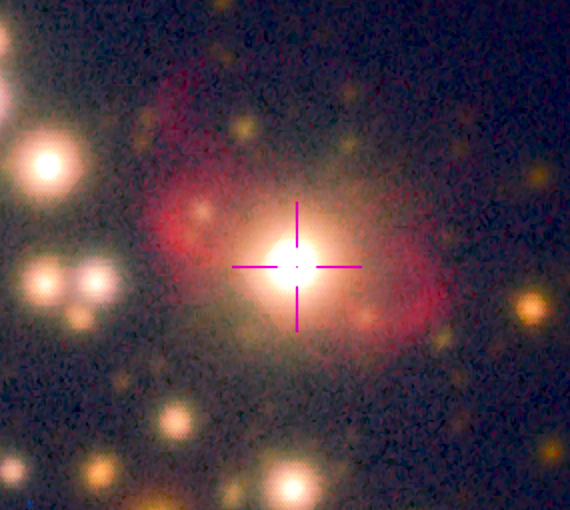}\hspace*{2mm}\includegraphics[width=50mm,height=45mm]{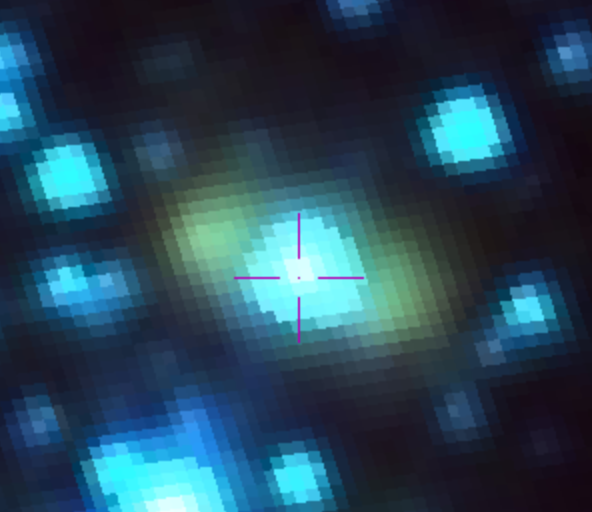}}
\caption{The blue supergiant IRAS 15127-5811 with its bi-polar nebula as seen by VPHAS+ \citep[left,][]{Drew2014} and IRAC/SPITZER \citep[right,][]{Werner2004,Fazio2004}.}
\label{fig:MN18}
\end{figure}

For this work we find that the likely PN [GKF2010] MN18 (IRAS 15127-5811) is in fact a blue supergiant surrounded by a bi-polar nebula evident in the optical and Mid-Infrared imagery (Fig.~\ref{fig:MN18}). The nebula spectrum has no obvious emission lines in the blue due to extinction but exhibits narrow, strong [NII] and H-alpha optical emission lines in the ratio of $\sim$0.7 (hence the original confusion with a PNe). A convincing and detailed study on this source is given by \cite{Gvaramadze2015}. HASH has been updated to reflect the new classification. 

\subsection{Open Cluster parameters}
The main OC parameters (distance, age, metallicity, $A_V$, parallax, and proper motion) from H\&R (first row for each cluster) and from \textit{OCFit} (second row for each cluster) are presented in Tab.~\ref{tab:literature}, together with the literature values from the individual OC catalogues. \citetalias{Hunt2023} (as well as most other catalogues) did not publish metallicities. We therefore used \textit{OCFit} to get the best fitting metallicities for the fixed OC parameters from each catalogue and listed them in Tab.~\ref{tab:literature} as blue values.\\

\subsubsection{HSC 2686}
\label{subsubsec:HSC2686}
\begin{figure}[h]
%\vspace*{-15mm}%\hspace*{-1cm}
\centerline{\includegraphics[width=75mm]{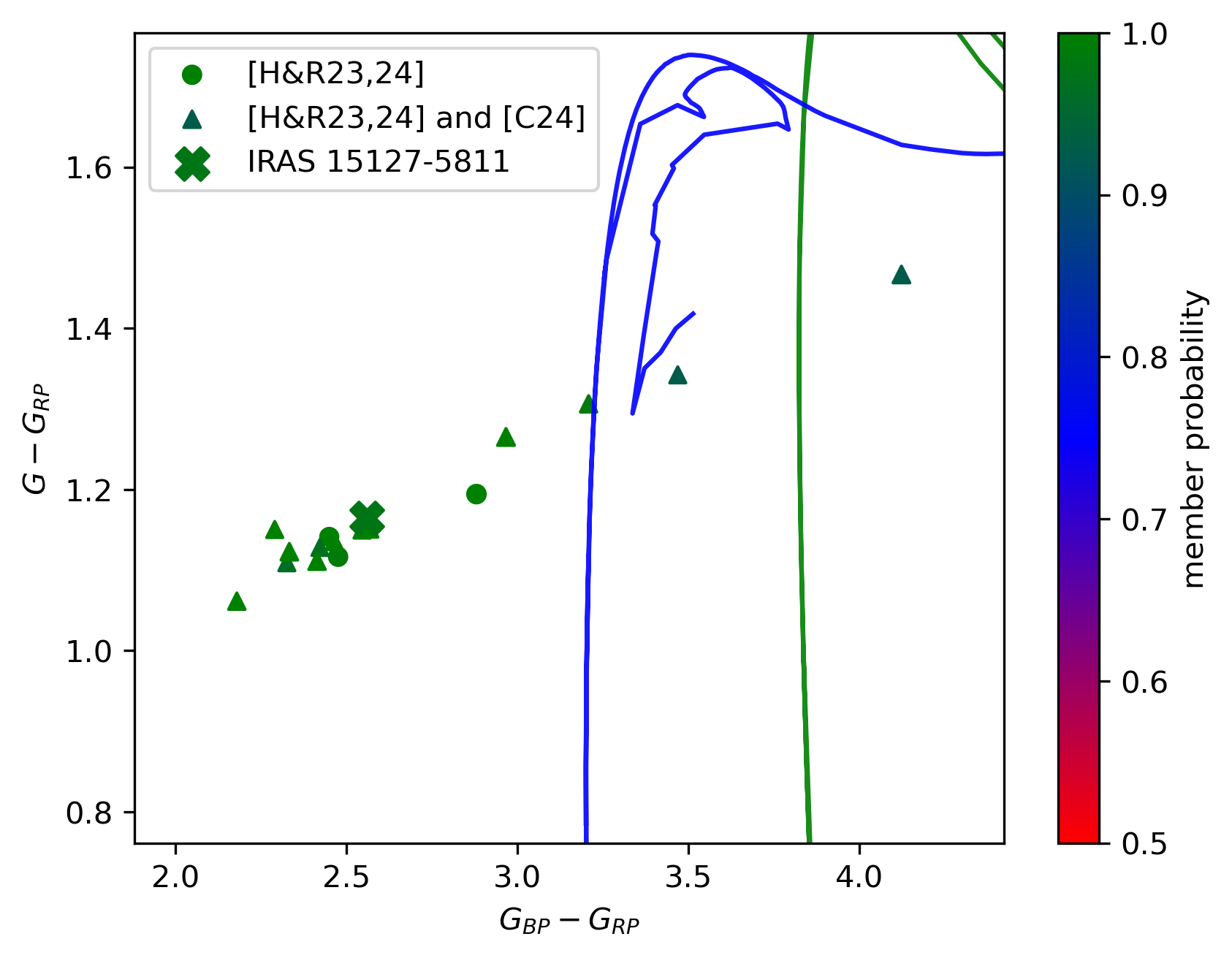} \includegraphics[width=75mm]{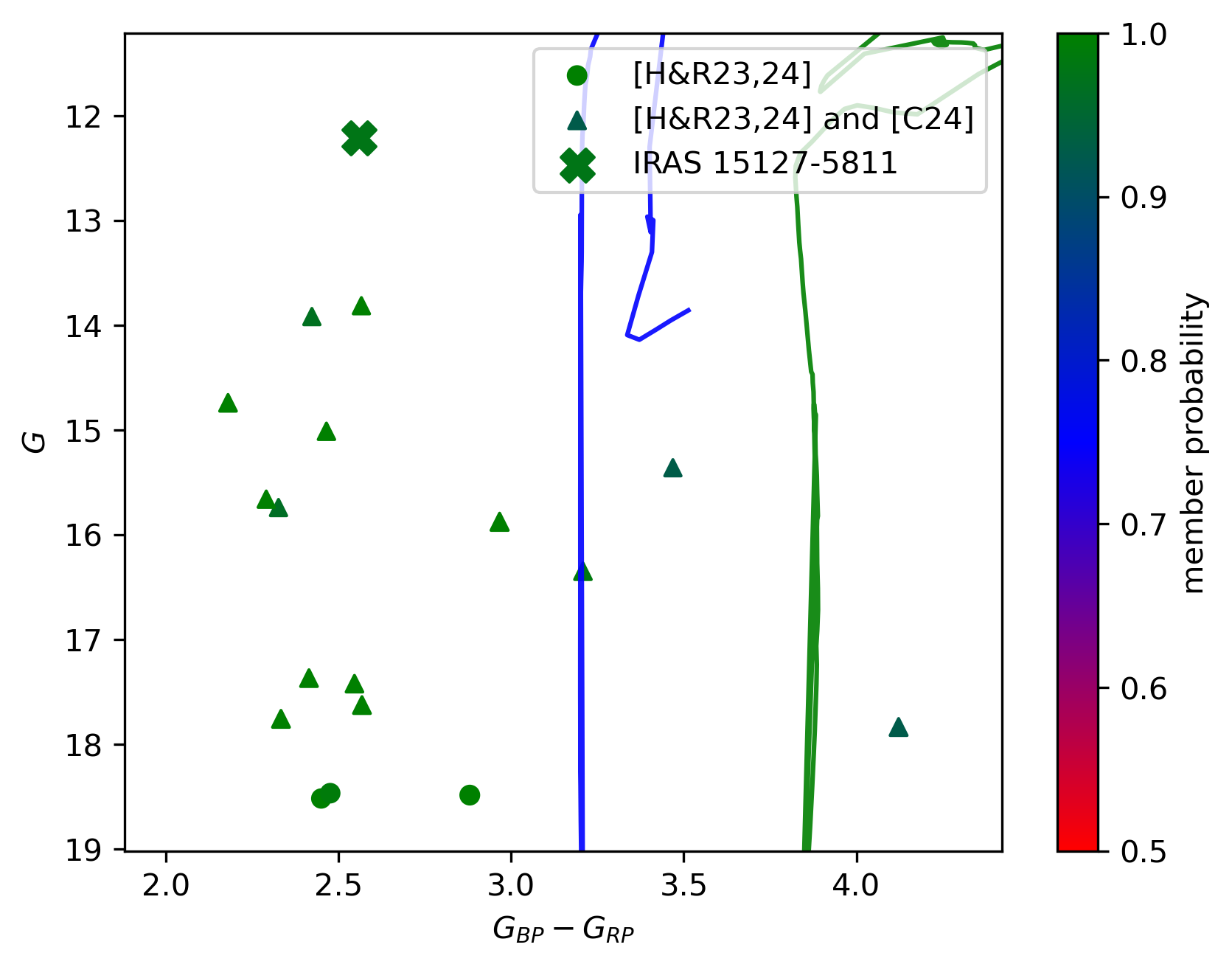}}
%\vspace*{-5mm}
%\hspace*{-1cm}
\centerline{\includegraphics[width=75mm]{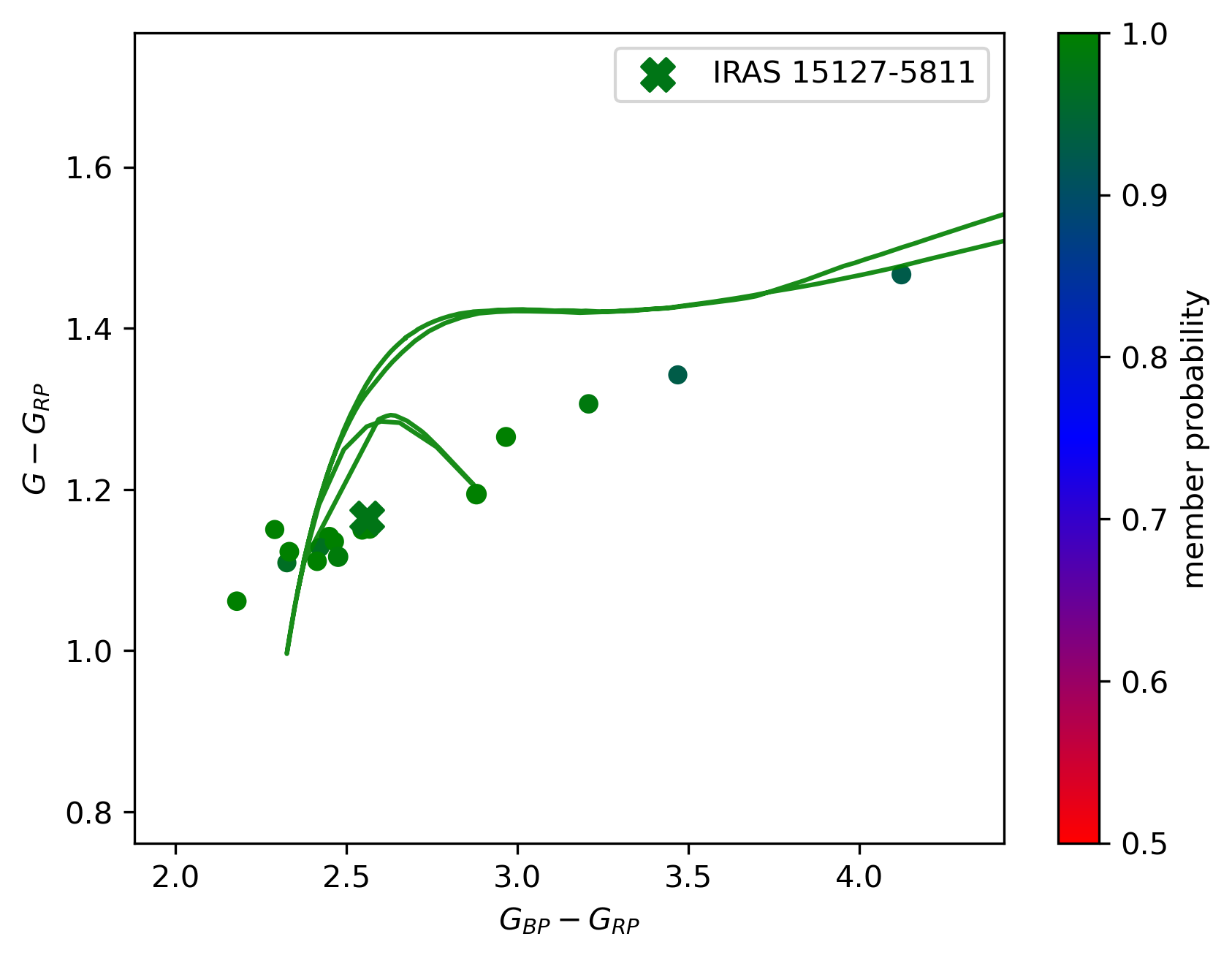} \includegraphics[width=75mm]{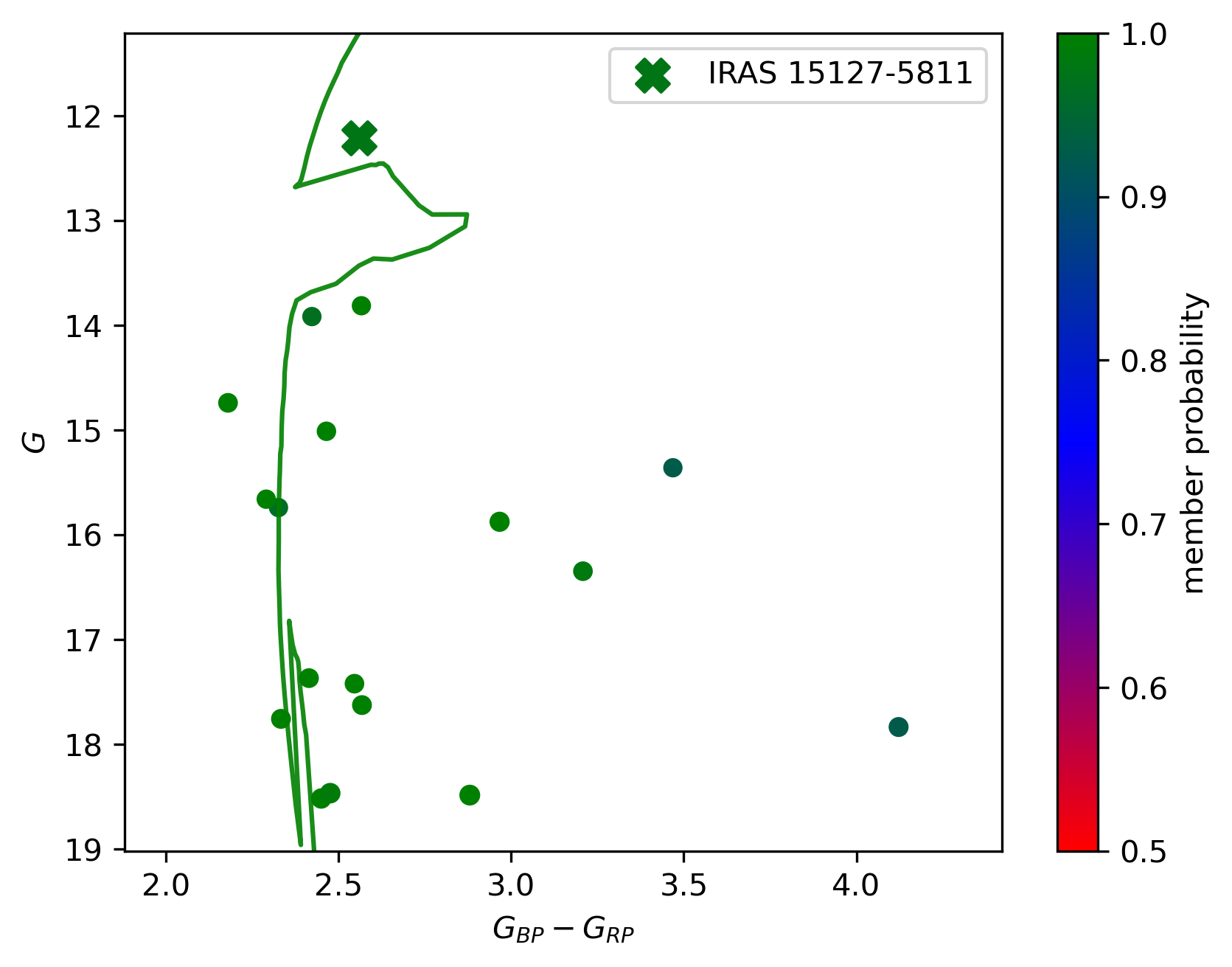}}
\caption{CCDs (left) and CMDs (right) for HSC 2686 with best-fitting isochrones; (\textbf{1st row}) green line: using the parameters from H\&R and only fitting the metallicity. blue line: using the OC parameters provided by \citetalias{Cavallo2024}.  (\textbf{2nd row}) fitting all OC parameters with \textit{OCFit}. The size of the symbols in the CCDs is proportional to the uncertainty in the \textit{Gaia~G} magnitude. Note that the size of IRAS 15127-5811 was scaled up by a factor of 4 for better visibility. The colour corresponds to the cluster membership probability.}\label{fig:HSC2686_isochrones}
\end{figure}

\citetalias{Hunt2023} list only 18 member stars for HSC 2686. The cluster parameters are shown in Tab.~\ref{tab1} and Tab.~\ref{tab:literature}. Only 1 star has a published radial velocity of -56.4~\kms, so one major diagnostic tool for cluster membership, their tight velocity agreement to 1-3~\kms  \citep{Mermilliod2009}, is not currently available. In Fig.~\ref{fig:RA_DEC_pm_par} the stars identified as comprising OC HSC~2686 appear to have a tight spread in parallax of $\sim 0.125-0.170$ mas and an approximate angular size of $\sim0.35\degree$.\\
Comparing the OC parameter values from \citetalias{Hunt2023} and \cite{Cavallo2024} (\citetalias{Cavallo2024} hereafter), the H\&R distance is $\approx$1.2 kpc less than the distance given in \citetalias{Cavallo2024}; ages agree within the stated uncertainties, while the extinction is slightly larger in H\&R. While \citetalias{Cavallo2024} did not publish a members list, they used the members from H\&R with \textit{Gaia} $G<18$~mag. The fitted metallicity for the given H\&R OC parameters is $-0.17\pm0.01$ dex, and $0.12\pm0.21$ dex when all parameters are fitted, compared to $0.48\pm0.3$ dex stated by \citetalias{Cavallo2024}.
OC parameter values fitted with \textit{OCFit} to the H\&R member stars yield a better agreement in the distance with \cite{Cavallo2024}. The fitted age is even older (though with large uncertainties), while the extinction is much lower compared to both H\&R and \citetalias{Cavallo2024}. The best-fitting isochrones in the CCDs and CMDs are shown in Fig.~\ref{fig:HSC2686_isochrones}. The top row shows the isochrones for the H\&R OC parameters with fitted metallicity as well as for the parameters from \citetalias{Cavallo2024}. No reasonable fit is achieved for neither the parameter values given in \citetalias{Hunt2023} nor \citetalias{Cavallo2024}, with the distance being too low (for H\&R parameters) and $A_V$ too high (both parameter sets). The second row shows the resulting isochrone from \textit{OCFit} with the fitted parameter values given in the second row in Tab.~\ref{tab:literature}. While the fitted isochrones appear more reasonable, the parameter uncertainties are very large compared to the real OCs NGC~2244 and NGC~6530 (where uncertainties are given). Each run of \textit{OCFit} produces different results (leading to the large uncertainties), meaning that the fitting procedure does not converge. The large uncertainties of the fitted OC parameters as well as the CCD and CMD are more consistent with field stars, indicating that HSC~2686 is likely not a real OC. 

\subsubsection{Lynga~3}
\label{subsubsec:Lynga3}
\begin{figure}[h]
\centerline{\includegraphics[width=9 cm]{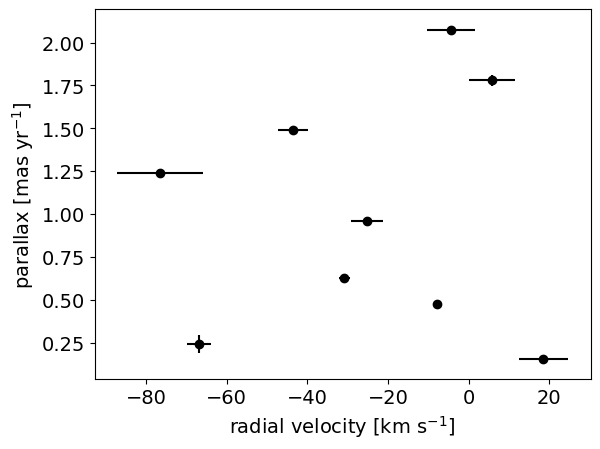}\hfill\includegraphics[width=9 cm]{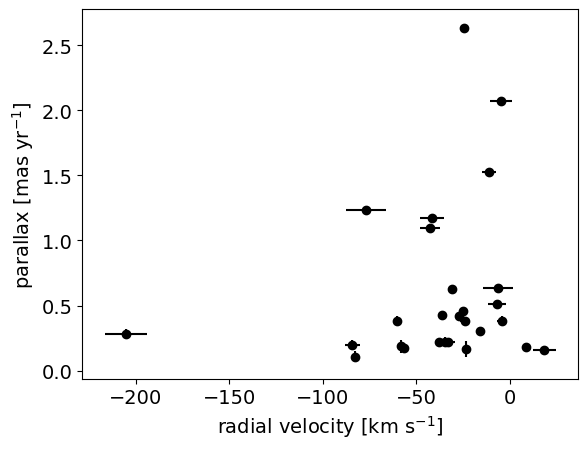}}
\centerline{\includegraphics[width=9 cm]{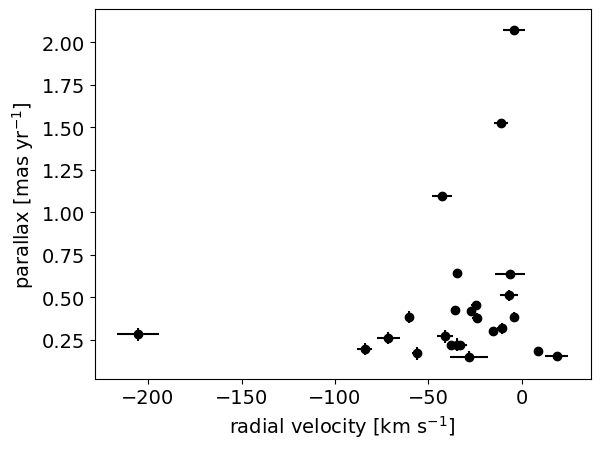}\hfill\includegraphics[width=9 cm]{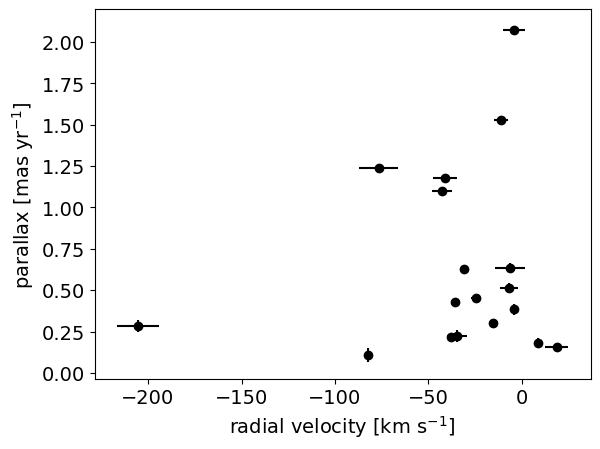}}
\centerline{\includegraphics[width=9 cm]{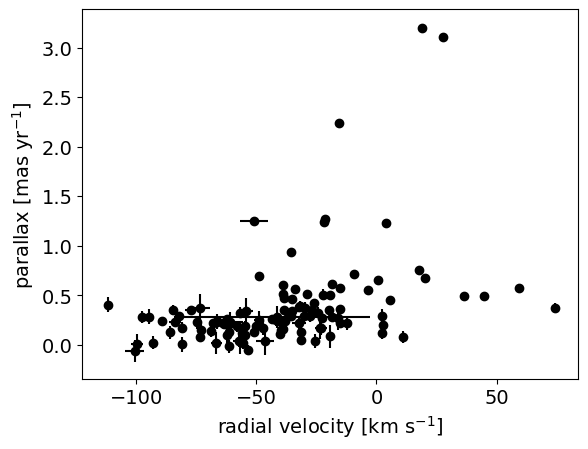}}
\caption{Parallax vs. radial velocity for Lynga~3 data sets with \textit{GAIA} radial velocities for more than 3 stars with membership probabilities $\geq0.5$. \textbf{top left}: \citetalias{Lynga1964}, \textbf{top right}: \citetalias{Dias2014}, \textbf{centre left}: \citetalias{Dias2018}, \textbf{center right}: \citetalias{Sampedro2017}, \textbf{bottom center}: \citetalias{Kharchenko2013}.}
\label{fig:parallax_vs_vrad}
\end{figure}

Contrary to HSC 2686, Lynga 3 shows a distinct, compact clump in RA and DEC (see Fig.~\ref{fig:RA_DEC_pm_par} upper panel). 
It was first identified as a possible OC by \citetalias{Lynga1964}. Even though Lynga did not publish any cluster parameters except for the position (for which he only gave a rough estimate), we identified all 25 member stars in \textit{Gaia} DR3 and produced the best-fitting isochrones and cluster parameters (see below).\\
Lynga 3 has since been further explored in a number of studies. \cite{Carraro2006} (\citetalias{Carraro2006} hereafter) did not find any significant difference of the Colour-Magnitude Diagram (CMD) with respect to a comparison field and concluded that Lynga~3 is not a real OC. However, as already noted by \cite{Gvaramadze2015}, they checked at the wrong position on the sky (more than $\approx0.6^\circ$ away from the actual position according to their stated coordinates). We re-visited this paper and found more inconsistencies. Firstly, their stated RA and DEC do not match their Galactic longitude and latitude (see Tab.~\ref{tab:Lynga3coords}). Secondly, we could not match their image of Lynga~3 (their Fig.~2) at either stated coordinate pairs. An astrometric solution achieved with the desktop version of \textit{astrometry.net} \citep{Lang2010} yielded that the field shown is actually centered at $\mathrm{RA}=233.331^\circ$, $\mathrm{DEC}=-55.234^\circ$, $3.9^\circ$(!) away from the actual position of Lynga~3, with North up and East to the right (not left). Consequently their findings regarding Lynga~3 are to be dismissed.\\
Other studies (\cite{Lynga1987} -- \citetalias{Lynga1987} hereafter, \cite{Dias2002} -- \citetalias{Dias2002} hereafter, \cite{Kharchenko2013} -- \citetalias{Kharchenko2013} hereafter, \citetalias{Sampedro2017}, \citetalias{Cantat-Gaudin2018} and \citetalias{Cantat-Gaudin2020}, \cite{Kounkel2020} -- \citetalias{Kounkel2020} hereafter, \citetalias{Dias2014}, \citetalias{Dias2018} and \citetalias{Dias2021}, \cite{Just2023} -- \citetalias{Just2023} hereafter, \cite{Perren2023} -- \citetalias{Perren2023} hereafter, \citetalias{Cavallo2024}, and most recently \cite{Plevne2026} -- \citetalias{Plevne2026} hereafter) did not list it as a dubious OC and produced coordinates consistent with the original \citetalias{Lynga1964} ones (See Tab.~\ref{tab:Lynga3coords}), with their basic parameters given in Tab.~\ref{tab:literature}. 
A comparison of those basic OC parameters shows large differences in the number of member stars (between 25 and 216 stars with membership probabilities $\geq0.5$), distances (between 959 pc and 7.8 kpc), ages (log age/[yrs] between 6.26 and 8.9), metallicities (between $-0.41$ and $0.067$ dex), and extinction ($\mathrm{A_V}$ between 2.65 and 8.1 mag), casting doubt on the classification of Lynga~3 as a real OC. Interestingly, \citetalias{Lynga1964} list 2 HSC~2686 stars as members of Lynga~3, \citetalias{Cantat-Gaudin2020} 1 star, \citetalias{Dias2014}, \citetalias{Dias2018}, and \citetalias{Sampedro2017} 3 stars, \citetalias{Cantat-Gaudin2018} and \citetalias{Dias2021} 6 stars.\\
As for HSC~2686, radial velocities are not available for the vast majority of the putative \citetalias{Hunt2023} member stars but a value of -4.8~km/s for Lynga~3 is noted from 2 identified OC stars with velocities of $-28.3\pm10.0$ and $18.5\pm6.1$ \kms. 
A cross-match of \citetalias{Lynga1964} member stars with \textit{GAIA} DR3 yields 9 radial velocities ($v_{rad}=-25.7\pm30$ \kms), \citetalias{Dias2014} 27 ($v_{rad}=-37.9\pm42$ \kms), \citetalias{Dias2018} 25 ($v_{rad}=-34.9\pm42$ \kms), \citetalias{Dias2021} 3 ($-28.3\pm10.0$ \kms, $18.5\pm6.1$ \kms, and $-23.8\pm2.7$ \kms), \citetalias{Sampedro2017} 18 ($v_{rad}=-35.2\pm49$ \kms), \citetalias{Cantat-Gaudin2018} 3 ($-23.8\pm 2.7$ \kms, $-28.3\pm10.0$ \kms, and $18.5\pm6.1$ \kms) and \citetalias{Kharchenko2013} 189 ($v_{rad}=-38.7\pm33$ \kms). The plots of the parallax vs. radial velocity for data sets with more than 3 radial velocities are shown in Fig.~\ref{fig:parallax_vs_vrad}.
All velocity distributions are highly incompatible with normal, very tight OC velocity dispersions which are typically 1-3km/s  \citep{Mermilliod2009}. 

\begin{figure}[h]
%\vspace*{-15mm}%\hspace*{-1cm}
\centerline{\includegraphics[width=75mm]{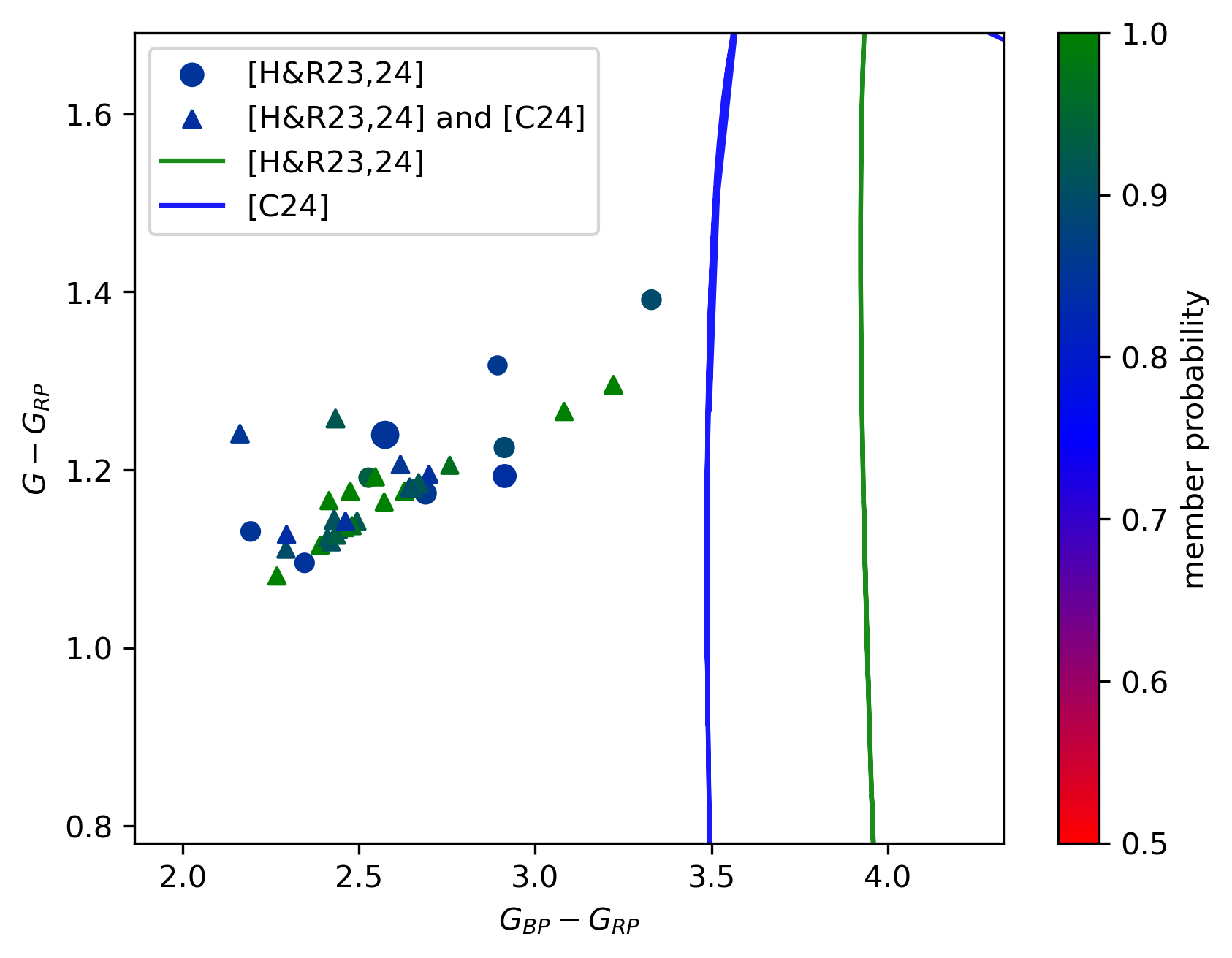} \includegraphics[width=75mm]{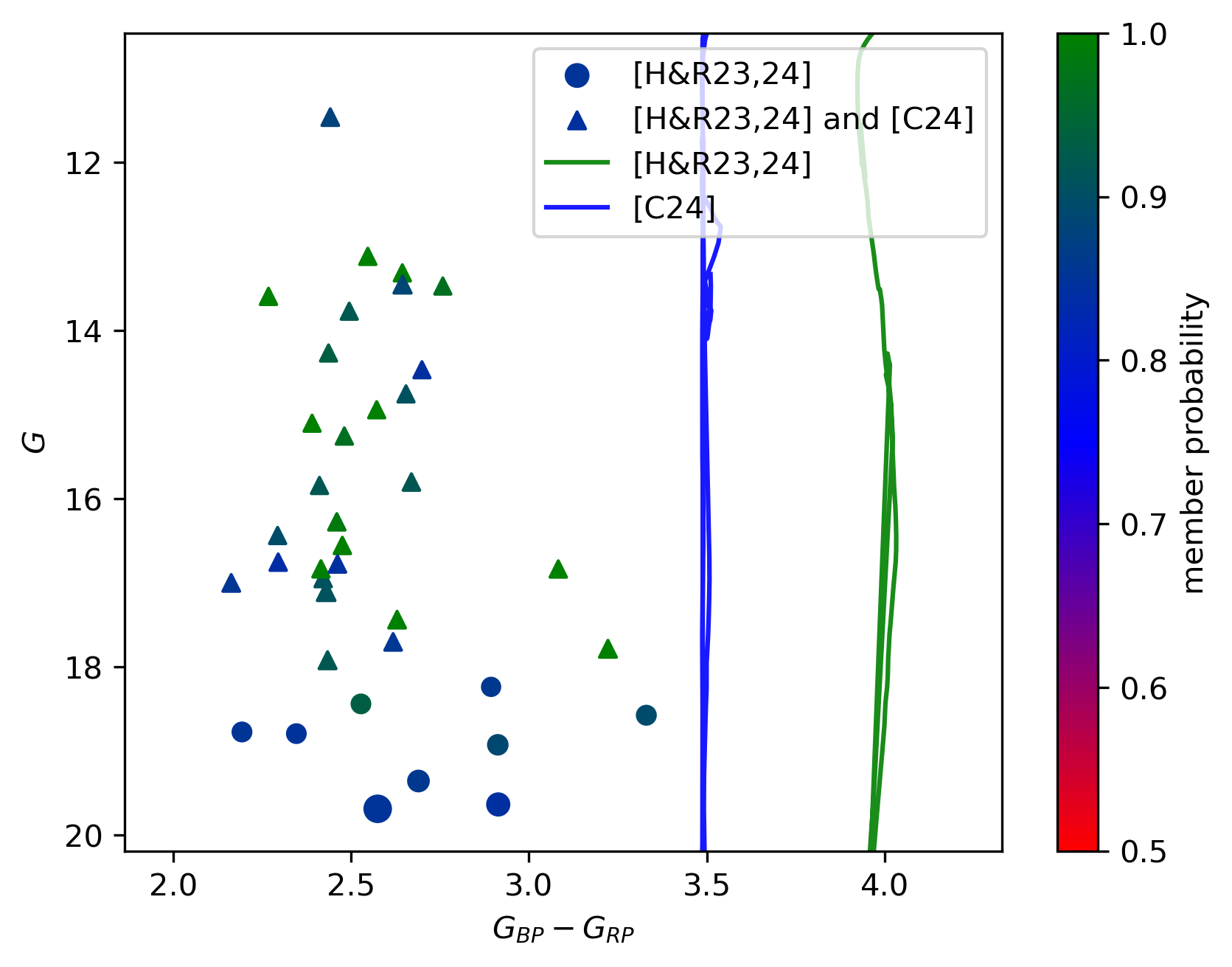}}
\centerline{\includegraphics[width=75mm]{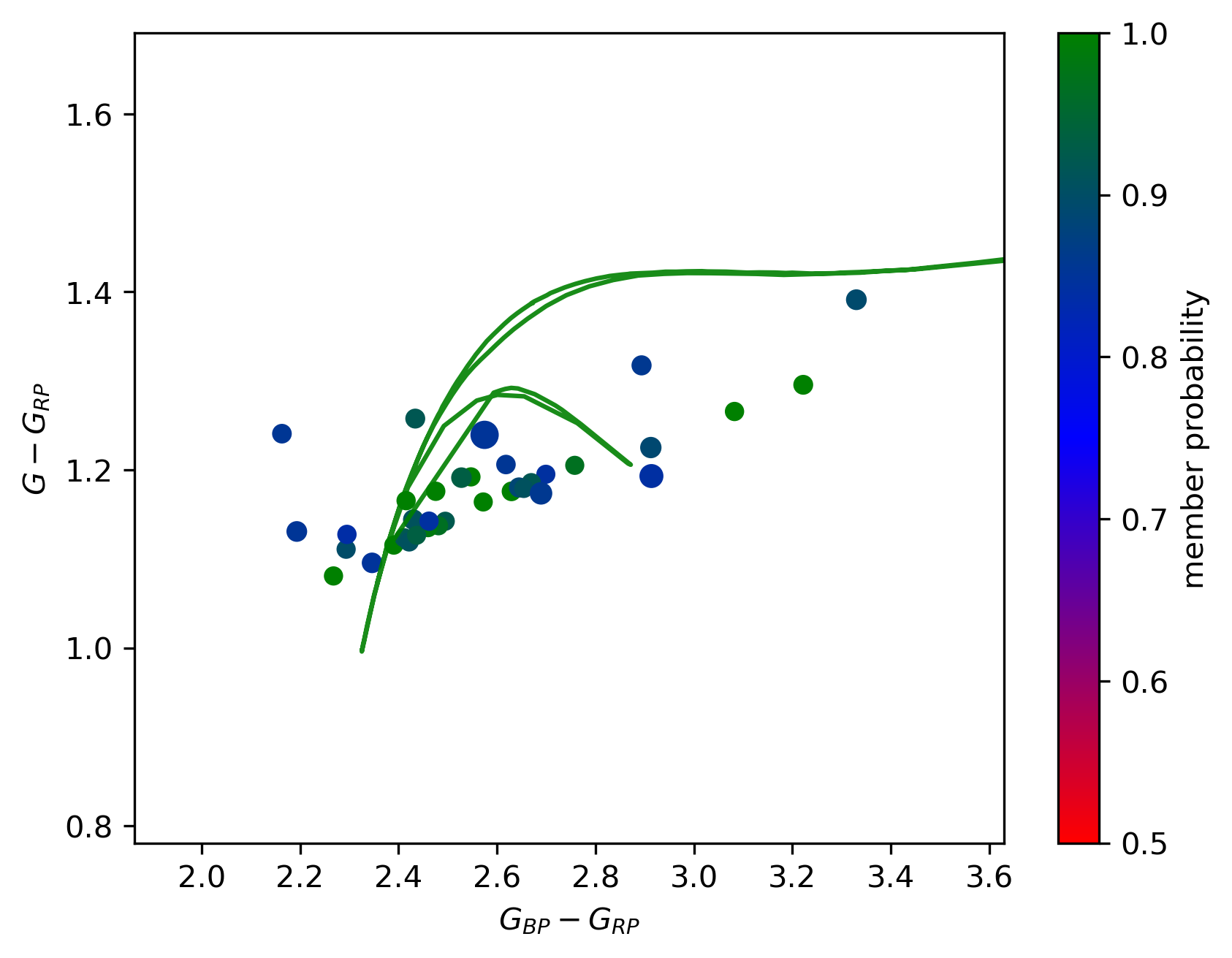} \includegraphics[width=75mm]{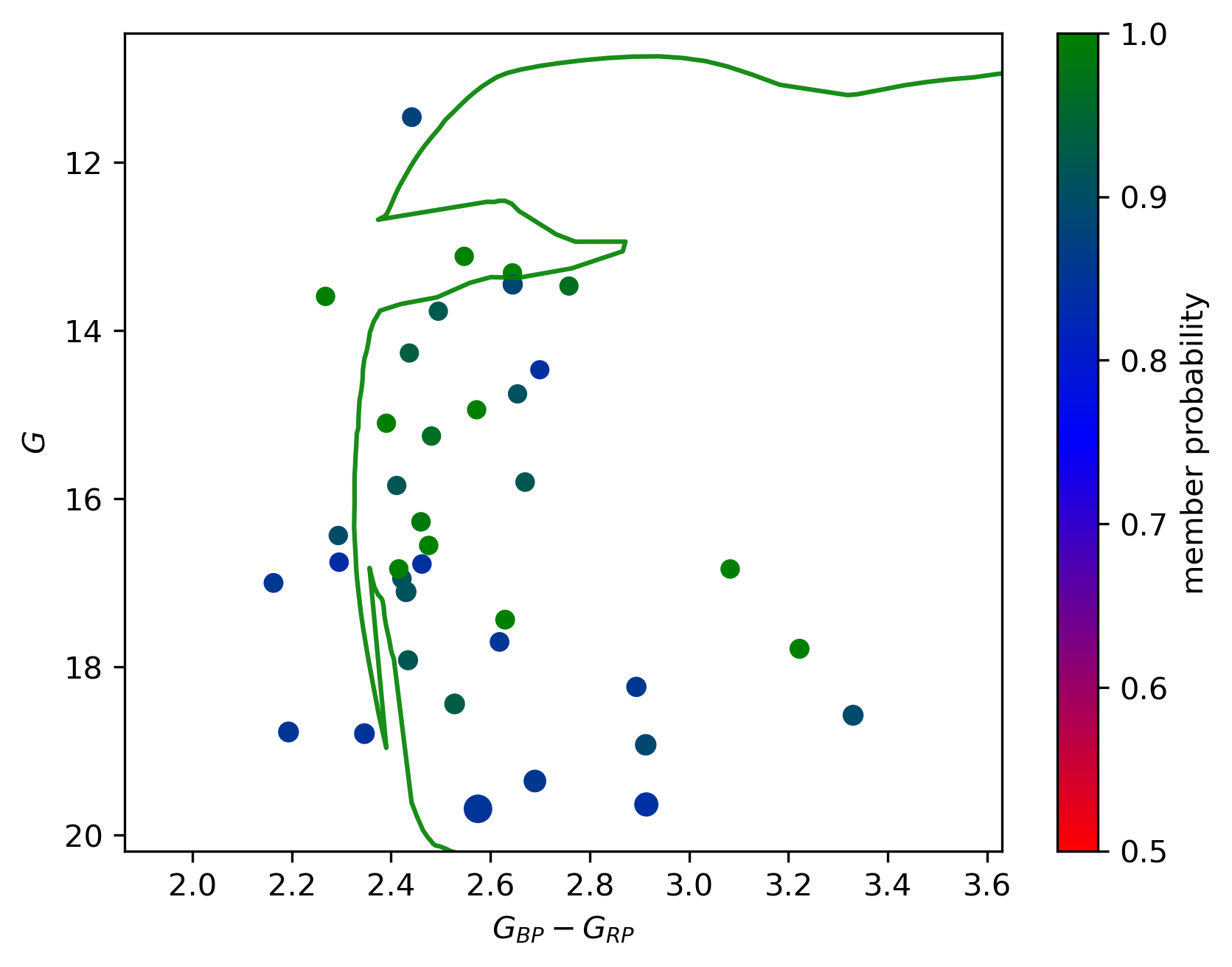}}
\caption{Same as Fig.~\ref{fig:HSC2686_isochrones} but for Lynga 3 member stars from \citetalias{Hunt2023}/\citetalias{Cavallo2024}.} \label{fig:Lynga3_isochrones}
\end{figure}

Similar to HSC~2686, no reasonable fit can be achieved for the OC parameters given in \citetalias{Hunt2023}, as can be seen in Fig.~\ref{fig:Lynga3_isochrones} (top). As with HSC~2686 the distance is underestimated while $A_V$ is overestimated. 
Fitting all OC parameters, the uncertainty in the cluster distance (Tab.~\ref{tab:literature}) is even larger compared to HSC~2686, while the uncertainties 
for the age, metallicity, and $A_V$ appear more realistic but still much larger compared to NGC~2244 and NGC~6530.\\
We produced CMDs and CCDs with the fitted isochrones for all catalogues with published member stars. These are shown in Fig.s~\ref{fig:Lynga3_isochrones_Lynga1964} for \citetalias{Lynga1964}, \ref{fig:Lynga3_isochrones_Kharchenko2013} for \citetalias{Kharchenko2013},~\ref{fig:Lynga3_isochrones_Sampedro2017} for \citetalias{Sampedro2017},~\ref{fig:Lynga3_isochrones_Cantat_Gaudin2018} for \citetalias{Cantat-Gaudin2018},~\ref{fig:Lynga3_isochrones_Cantat_Gaudin2020} for \citetalias{Cantat-Gaudin2020},~\ref{fig:Lynga3_isochrones_Dias2014} for \citetalias{Dias2014},~\ref{fig:Lynga3_isochrones_Dias2018} for \citetalias{Dias2018}, and~\ref{fig:Lynga3_isochrones_Dias2021} for \citetalias{Dias2021}. While \citetalias{Plevne2026} did state the number of member stars on their website, they did not publish the list of members. However, their CMD can be found at \href{https://ucc.ar/_clusters/lynga3/}{https://ucc.ar/\_clusters/lynga3/}.
The fitted isochrone for all member stars with membership probabilities $\ge0.5$ combined from all catalogues are shown in Fig.~\ref{fig:Lynga3_combined}, with the best-fitting OC parameters listed in Tab.~\ref{tab:literature}.

\subsubsection{Field star sample}
\label{subsubsec:field_stars}
\begin{figure}[h]
%\vspace*{-15mm}%\hspace*{-1cm}
\centerline{\includegraphics[width=75mm]{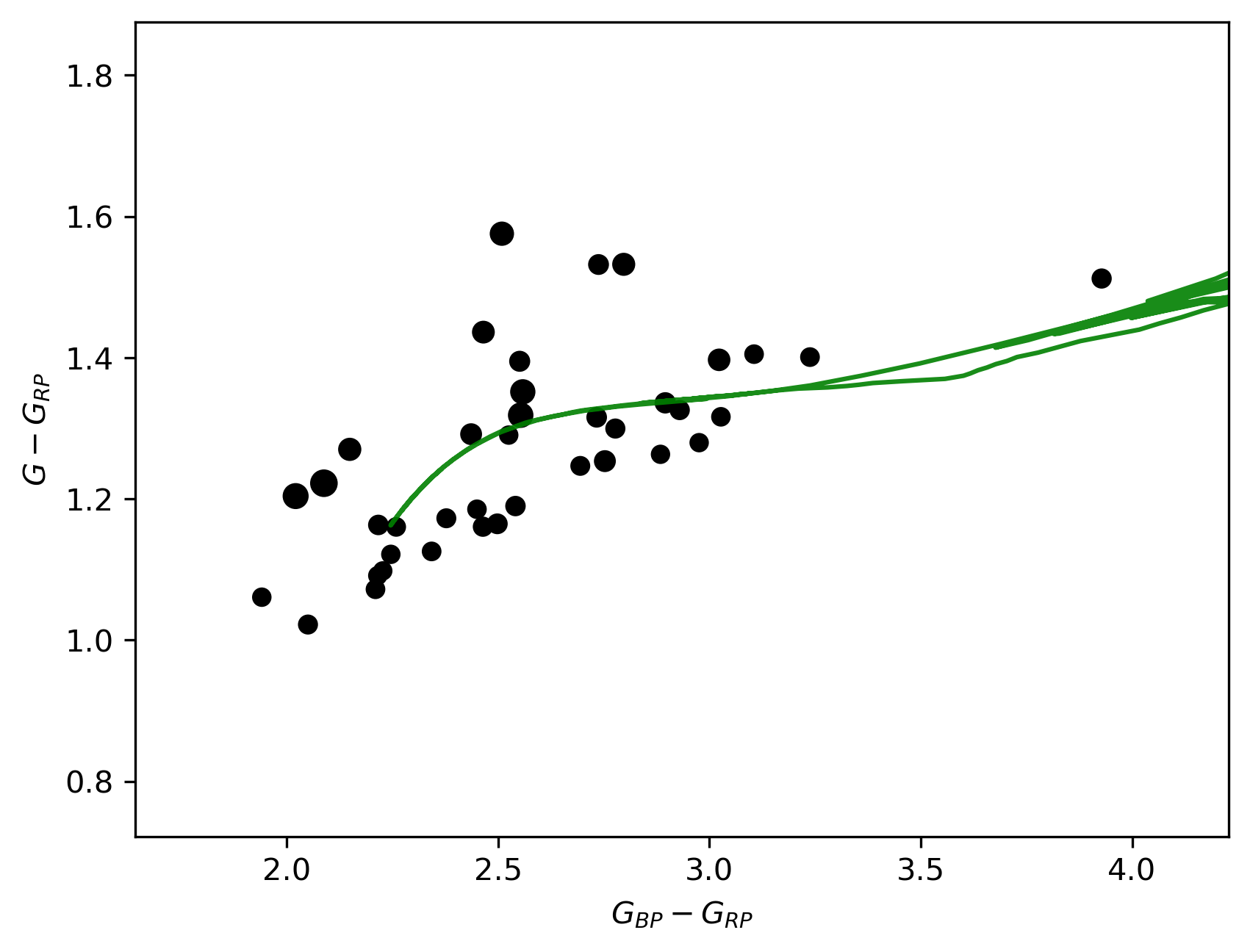} \includegraphics[width=75mm]{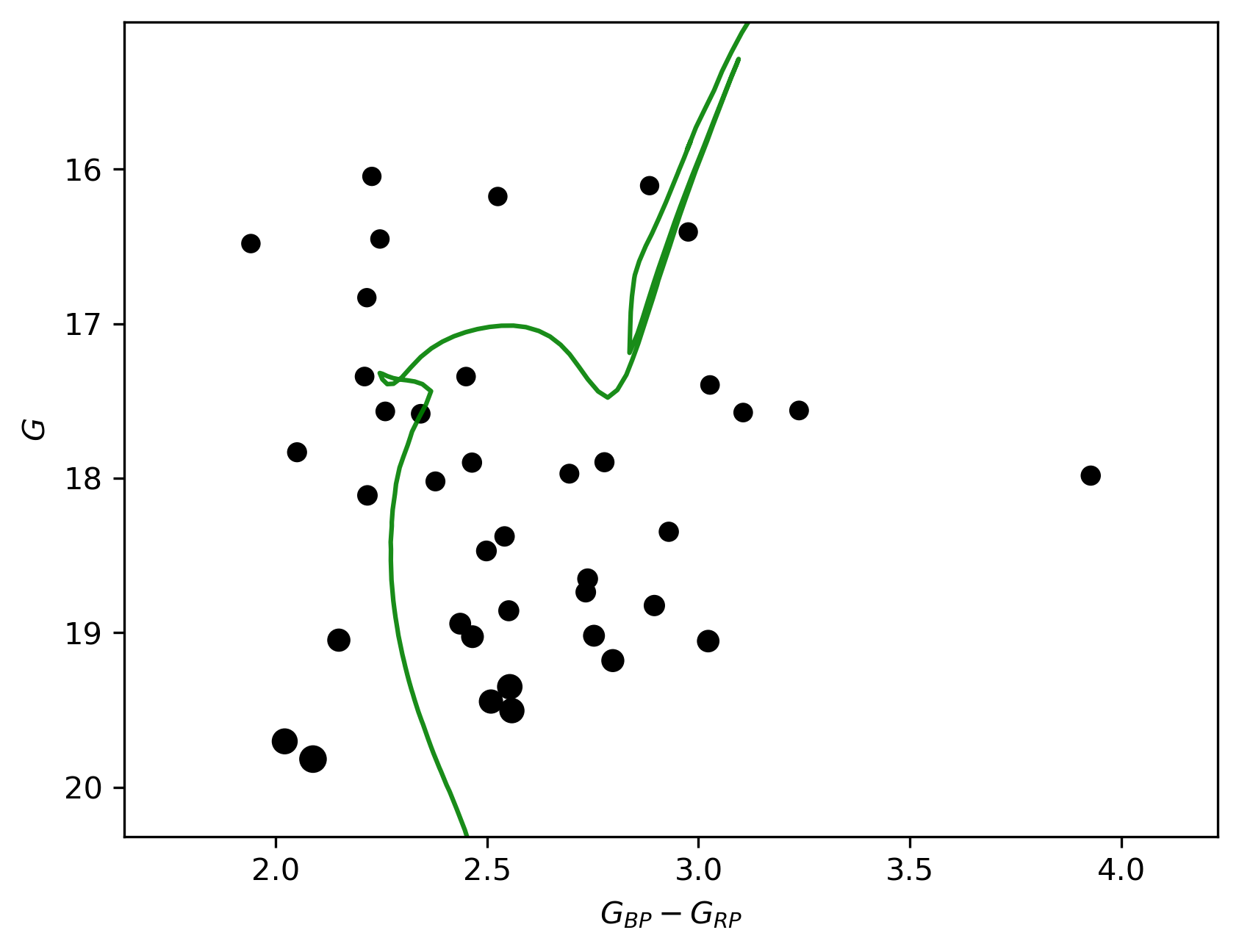}}
%\vspace*{-5mm}
%\hspace*{-1cm}
%\centerline{\includegraphics[width=75mm]{random_sample3_CCD_isocfit_Mar03.png} \includegraphics[width=75mm]{random_sample3_CMD_isocfit_Mar03.png}}
%\vspace*{-5mm}
%\hspace*{-1cm}
%\centerline{\includegraphics[width=75mm]{random_sample3_CCD_isocfit_Mar03-1.png} \includegraphics[width=75mm]{random_sample3_CMD_isocfit_Mar03-1.png}}
%\vspace*{-5mm}
%\vspace*{5mm}\hspace*{-4cm}
\caption{Same as Fig.~\ref{fig:HSC2686_isochrones} but for the field stars sample occupying a similar parameter space 
in position, parallax, and proper motion as HSC~2686 and Lynga~3. Note that the first row corresponding to the \citetalias{Hunt2023} OC parameters is 
missing here as this is not a cluster claimed by \citetalias{Hunt2023}. \label{fig:random_isochrones}}
\end{figure}

We selected a sample of field stars from the same $0.75 \times 0.5$ degree (RA and DEC) sky region that encompasses HSC~2686 
and Lynga~3 which also share similar proper motions ($-5.65 \leq \mathrm{pmRA/mas~yr^{-1}} \leq -5.2$, $-4 \leq 
\mathrm{pmDEC/mas~yr^{-1}} \leq -3.8$) and parallaxes ($0.08 \leq \mathrm{parallax / mas} \leq 0.2$, Fig.~\ref{fig:RA_DEC_pm_par}). This 
yielded a sample size of 39 field stars for the comparison to the putative clusters HSC 2686 and Lynga 3. Interestingly, this sample of field stars is larger than the number of stars in HSC~2686, and has a similar number of stars when compared to Lynga~3.\\
The isochrones fitted by \textit{OCFit} for our sample of field stars (see Fig.~\ref{fig:random_isochrones}) are basically 
indistinguishable from the fits to the putative HSC~2686 and Lynga~3 member stars. If anything the CCD actually looks better. The stated large parameter uncertainties are similar to both claimed OCs, adding confidence that both HSC~2686 and Lynga~3 are not real clusters.

\begin{table}[h]
    \centering
    \begin{tabular}{|l|c|c|c|c|}
    \hline
        Reference & RA [deg] J2000 & DEC [deg] J2000 & Glon [deg] & Glat [deg] \\
        \hline
        \citetalias{Lynga1964} & 228.485 & -59.086 & 320.348 & -1.120\\
        ~~~~observed & 229.211 & -58.393 & 321.032 & -0.727\\
        \citetalias{Lynga1987} & 229.128 & -58.301 & 321.043 & -0.625\\
        \citetalias{Carraro2006} & 228.150 (229.121) & -58.133 (-58.320) & 321.03 (320.691) & -0.64 (-0.214)\\
        ~~~~observed& 233.331 & -55.234 & 324.656 & 0.658\\
        \citetalias{Kharchenko2013} & 229.208 & -58.340 & 321.053 & -0.676\\
        \citetalias{Dias2002} v1.0 & 229.092 & -58.317 & 321.019 & -0.629\\
        \citetalias{Dias2002} v2.4 & 229.204 & -58.367 & 321.043 & -0.702\\
        \citetalias{Dias2014}/\citetalias{Sampedro2017} & 229.204 & -58.366 & 321.043 & -0.702\\
        %D2017 & 229.204 & -58.367 & 321.043 & -0.702\\
        \citetalias{Dias2021} & 229.193 & -58.365 & 321.038 & -0.698 \\
        \citetalias{Cantat-Gaudin2018}/\citetalias{Cantat-Gaudin2020} & 229.194 & -58.373 & 321.035 & -0.705 \\
        \citetalias{Hunt2023} & 229.197 & -58.375 & 321.024 & -0.699 \\
        \citetalias{Just2023} & 229.195 & -58.339 & 321.053 & -0.676 \\
        \citetalias{Perren2023} & 229.192 & -58.374 & 321.033 & -0.705 \\
        \citetalias{Cavallo2024} & 229.199 & -58.373 & 321.037 & -0.706\\
        %\textbf{\citetalias{Almeida2025}} & \textbf{229.193} & \textbf{-58.365} & \textbf{321.039} & \textbf{-0.698} \\
        \hline
%        \multicolumn{5}{l}{L64 = \cite{Lynga1964}}
    \end{tabular}
    \caption{Lynga 3 coordinates from the different OC catalogues. Note that the coordinates \citetalias{Lynga1964} stated in his paper were only rough approximations and the coordinates of the field he observed are given in the 2nd row. The values for Galactic longitude and latitude given in \citetalias{Carraro2006} differ from the values calculated from the stated values of RA and DEC, which are given in the parentheses. The RA and DEC calculated for the stated Galactic longitude and latitude are also given inside parentheses. The coordinates of the field that \citetalias{Carraro2006} actually observed are given in row 5.}
    \label{tab:Lynga3coords}
\end{table}

\subsection{NGC 2244}
\begin{figure}[h]
%\vspace*{-15mm}%\hspace*{-1cm}
\centerline{\includegraphics[width=75mm]{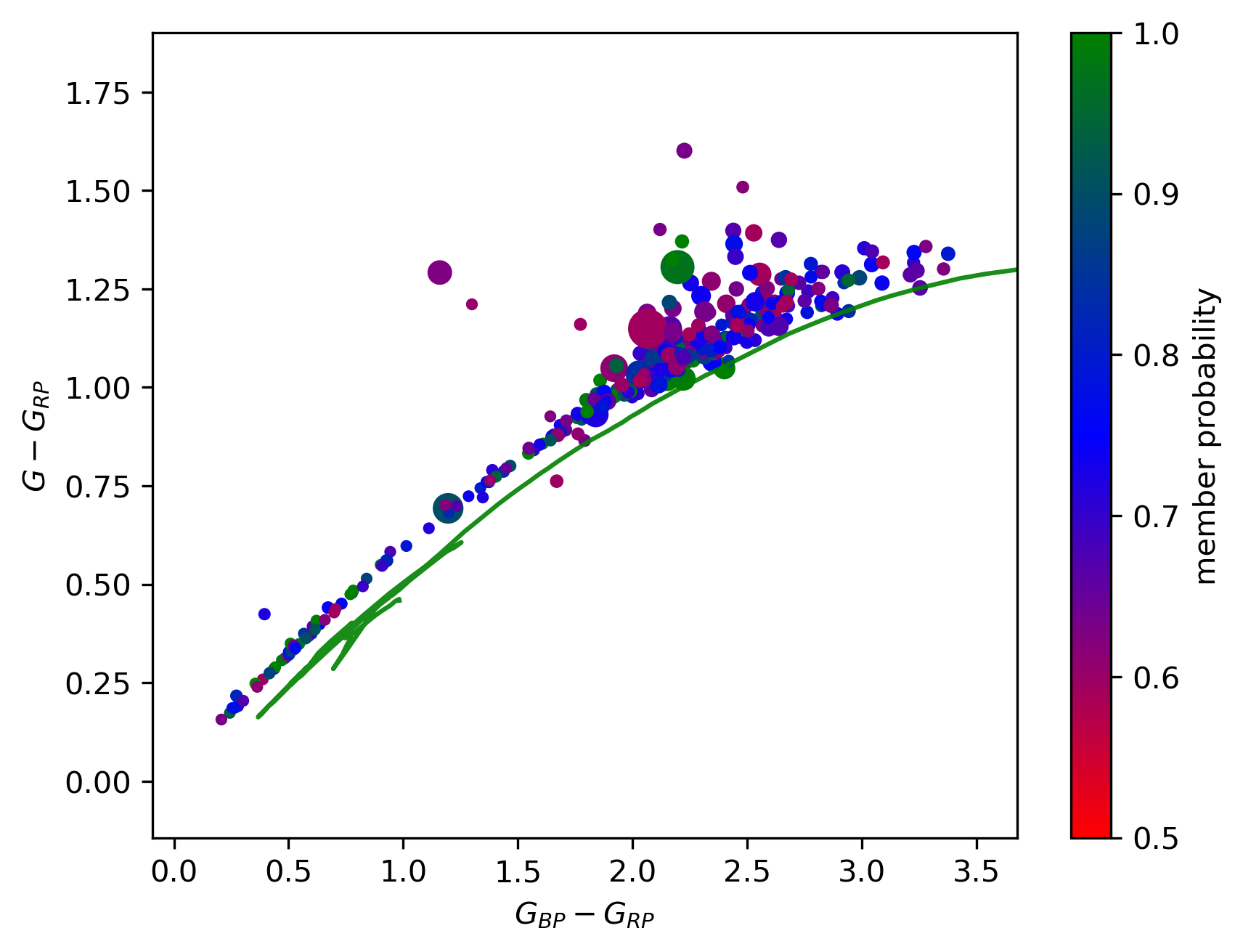} \includegraphics[width=75mm]{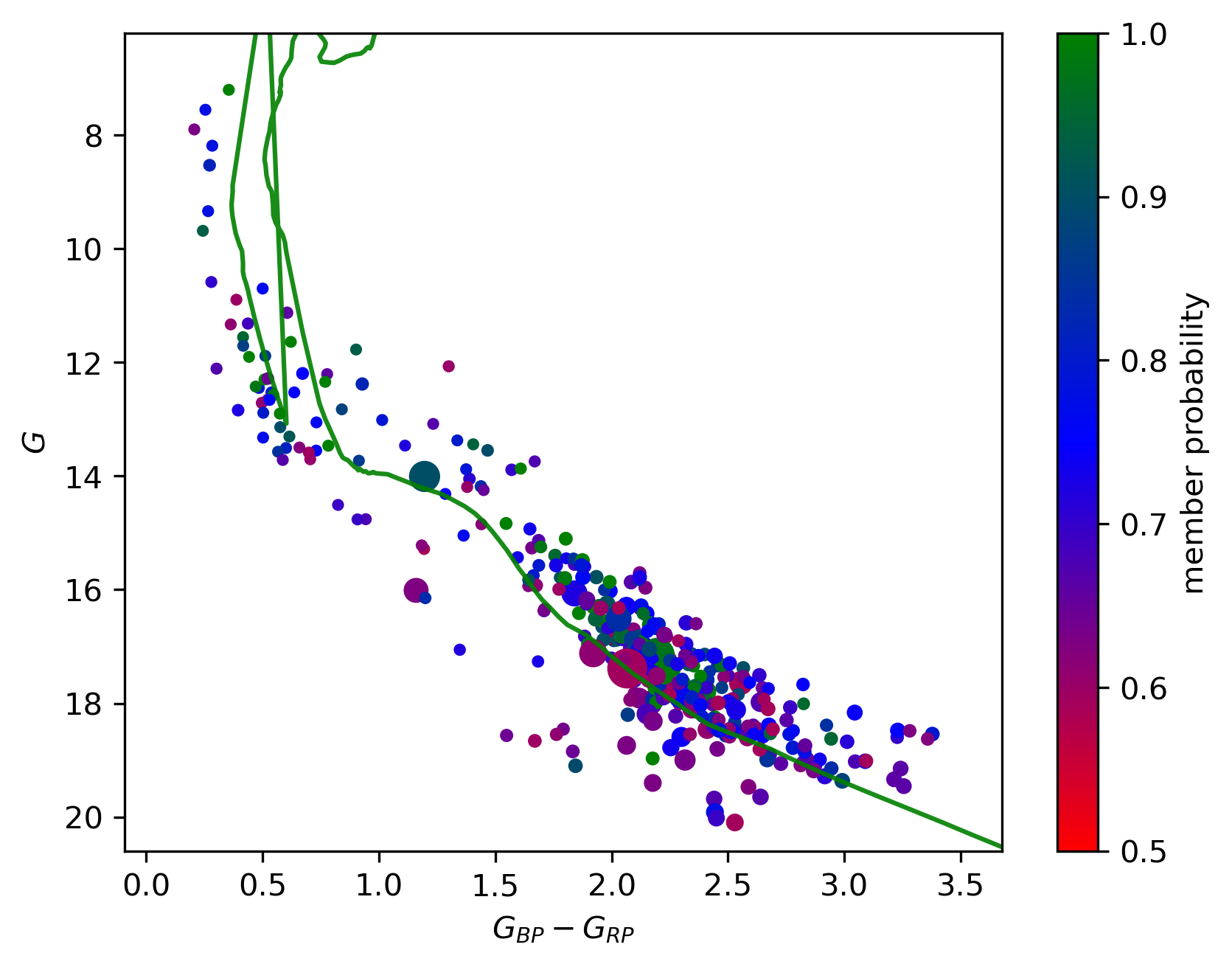}}
%\vspace*{-5mm}
%\hspace*{-1cm}
\centerline{\includegraphics[width=75mm]{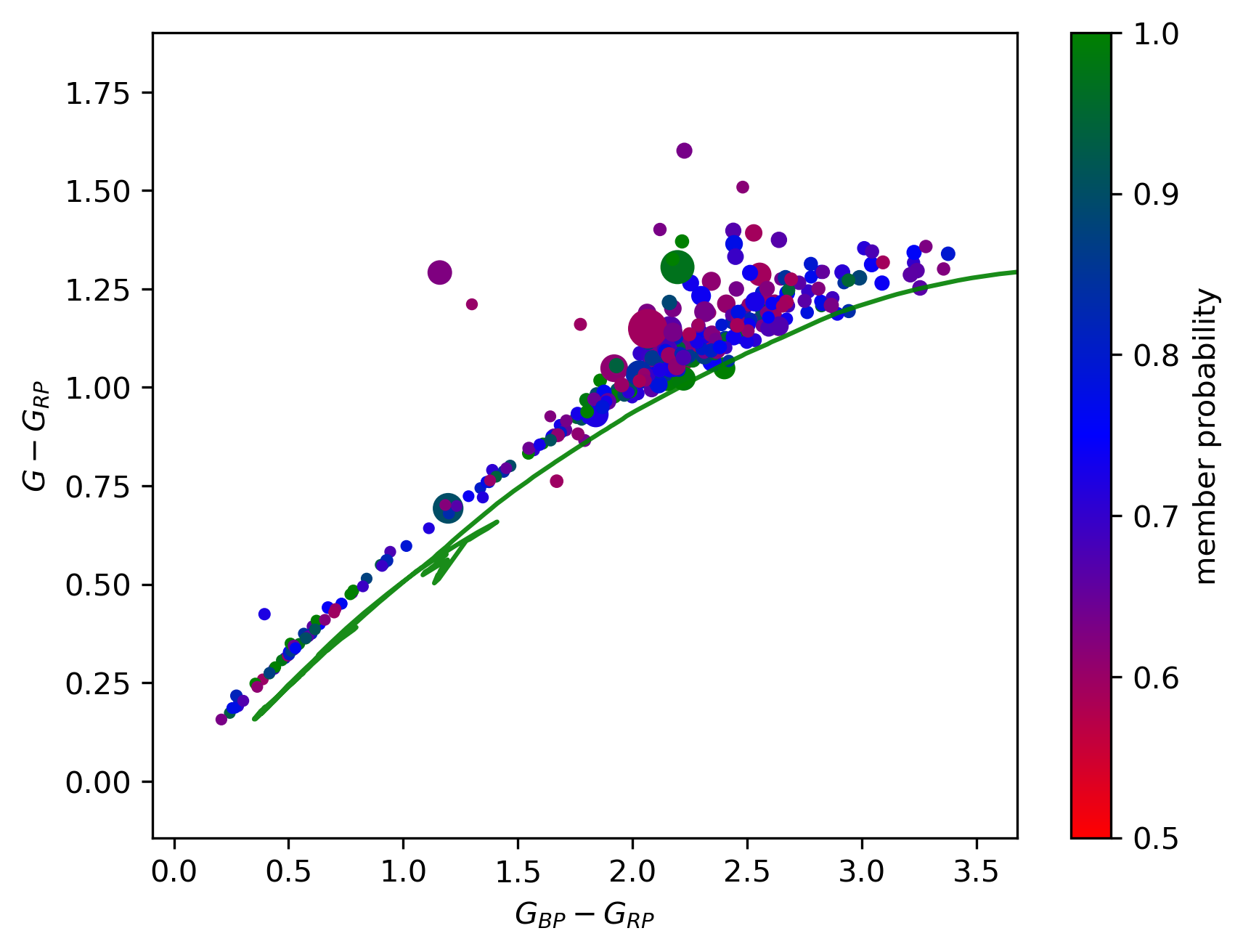} \includegraphics[width=75mm]{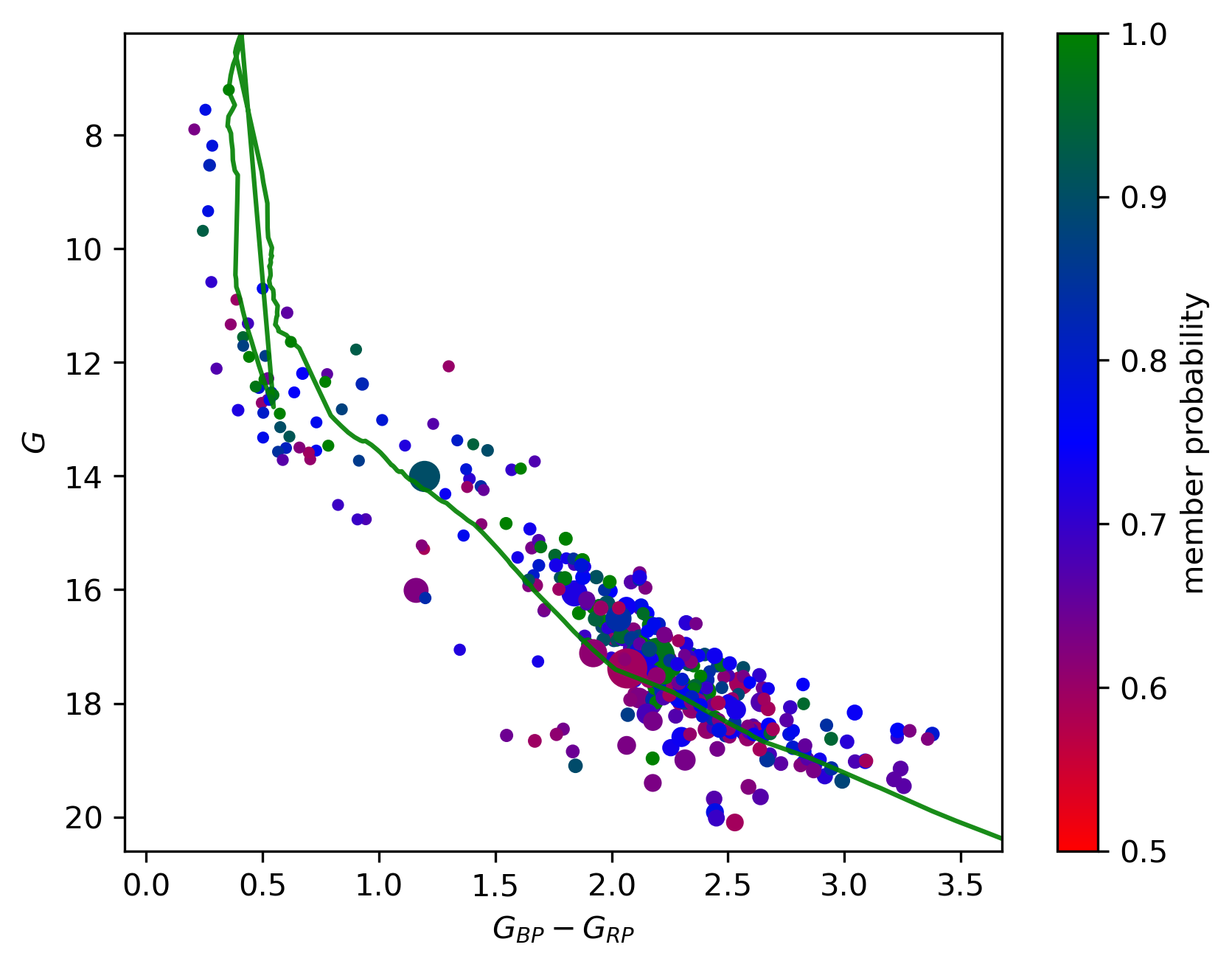}}
\caption{Same as Fig.~\ref{fig:HSC2686_isochrones} but for NGC 2244. }\label{fig:NGC2244_isochrones}
\end{figure}

Fig.~\ref{fig:NGC2244_isochrones} shows good isochrone fits (for both CCD and CMD) for the \citetalias{Hunt2023} OC parameters as well as the parameters fitted with \textit{OCFit}. The comparison of OC parameters by \citetalias{Hunt2023}, \textit{OCFit}, as well as the literature values shows a good agreement within the stated uncertainties. The parameter uncertainties given by \textit{OCFit} are much smaller compared to the uncertainties from fitting HSC~2686 and Lynga~3, giving confidence in the classification of NGC~2244 as a real OC.

\subsection{NGC 6530}
\begin{figure}[h]
%\vspace*{-15mm}%\hspace*{-1cm}
\centerline{\includegraphics[width=75mm]{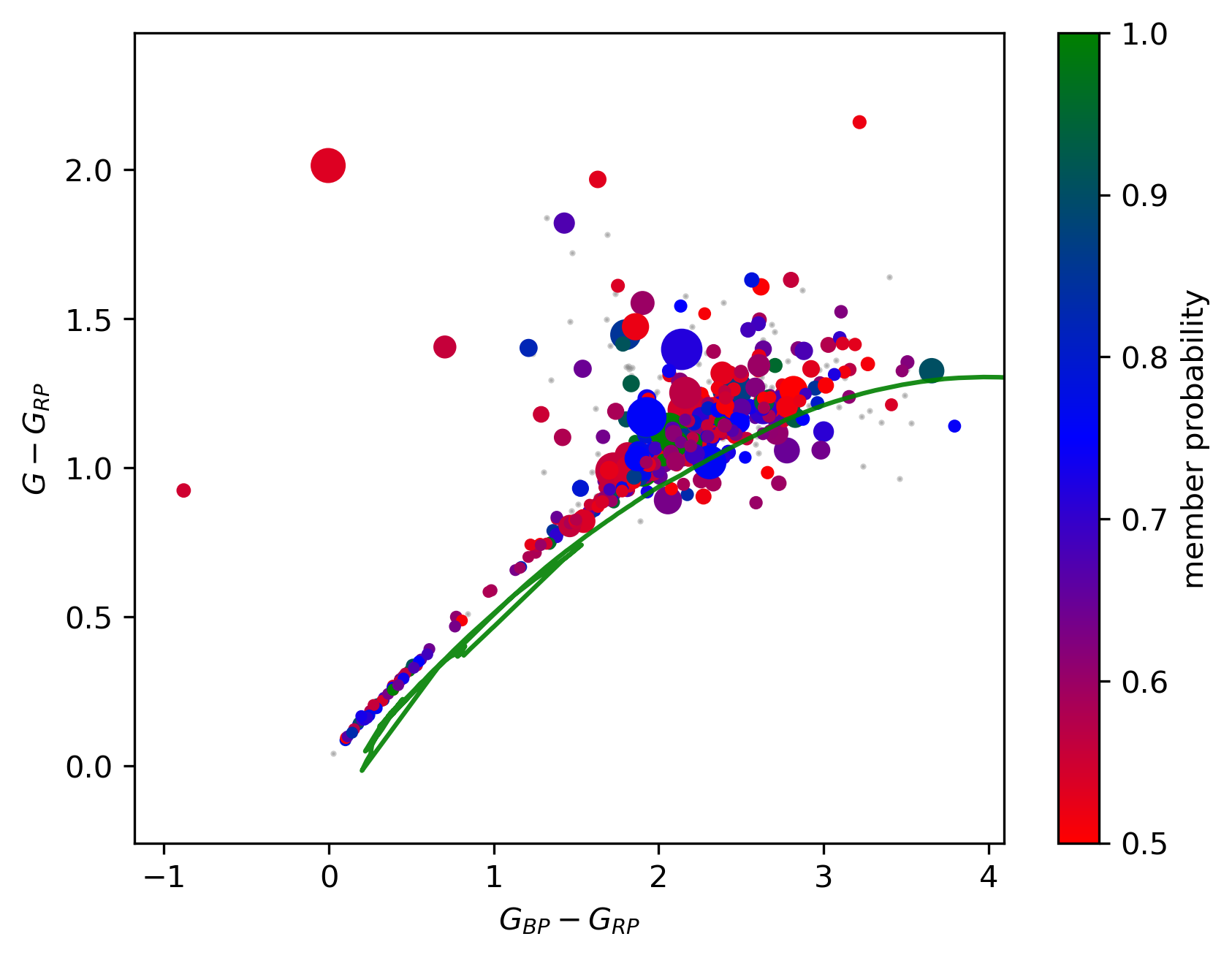} \includegraphics[width=75mm]{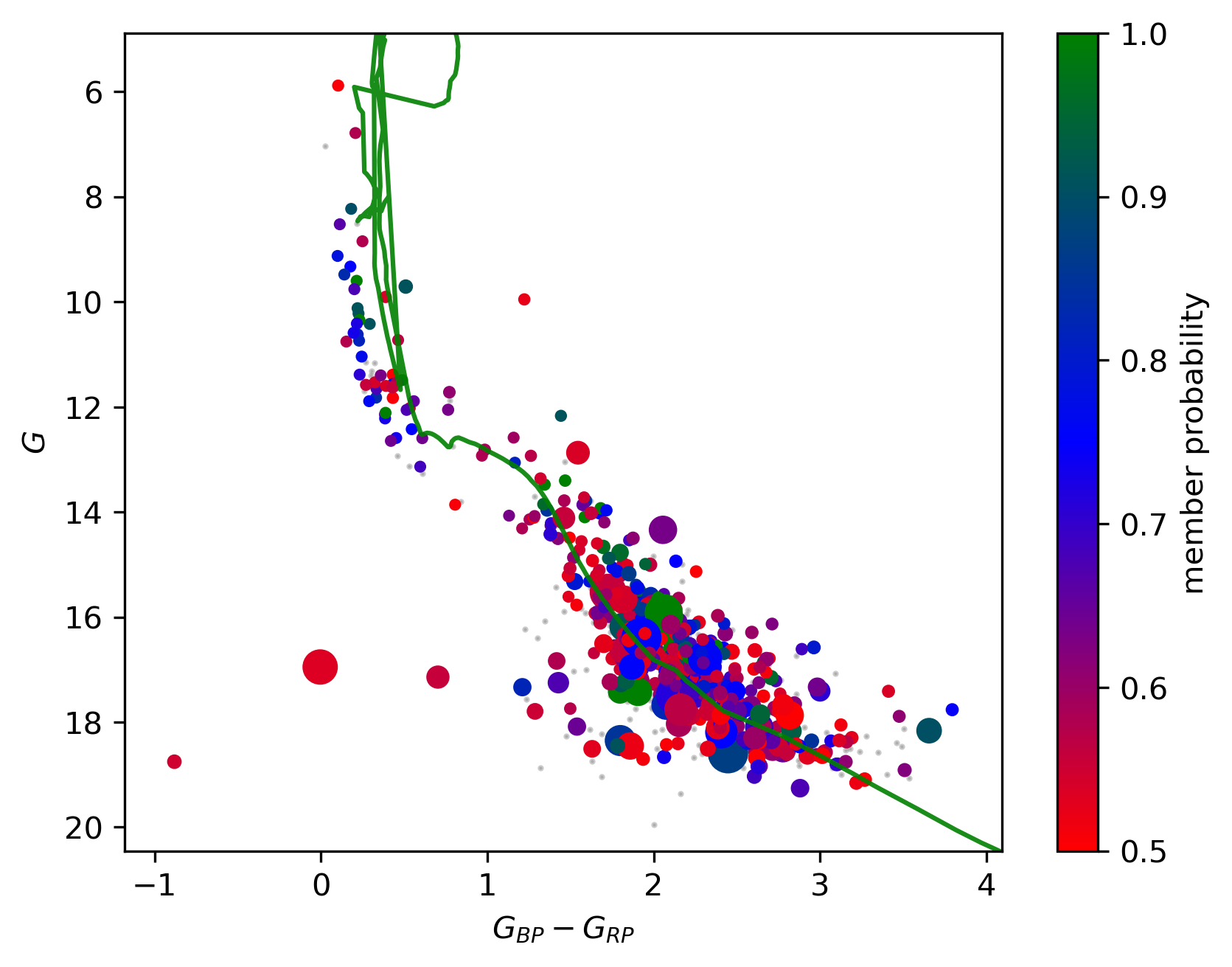}}
\centerline{\includegraphics[width=75mm]{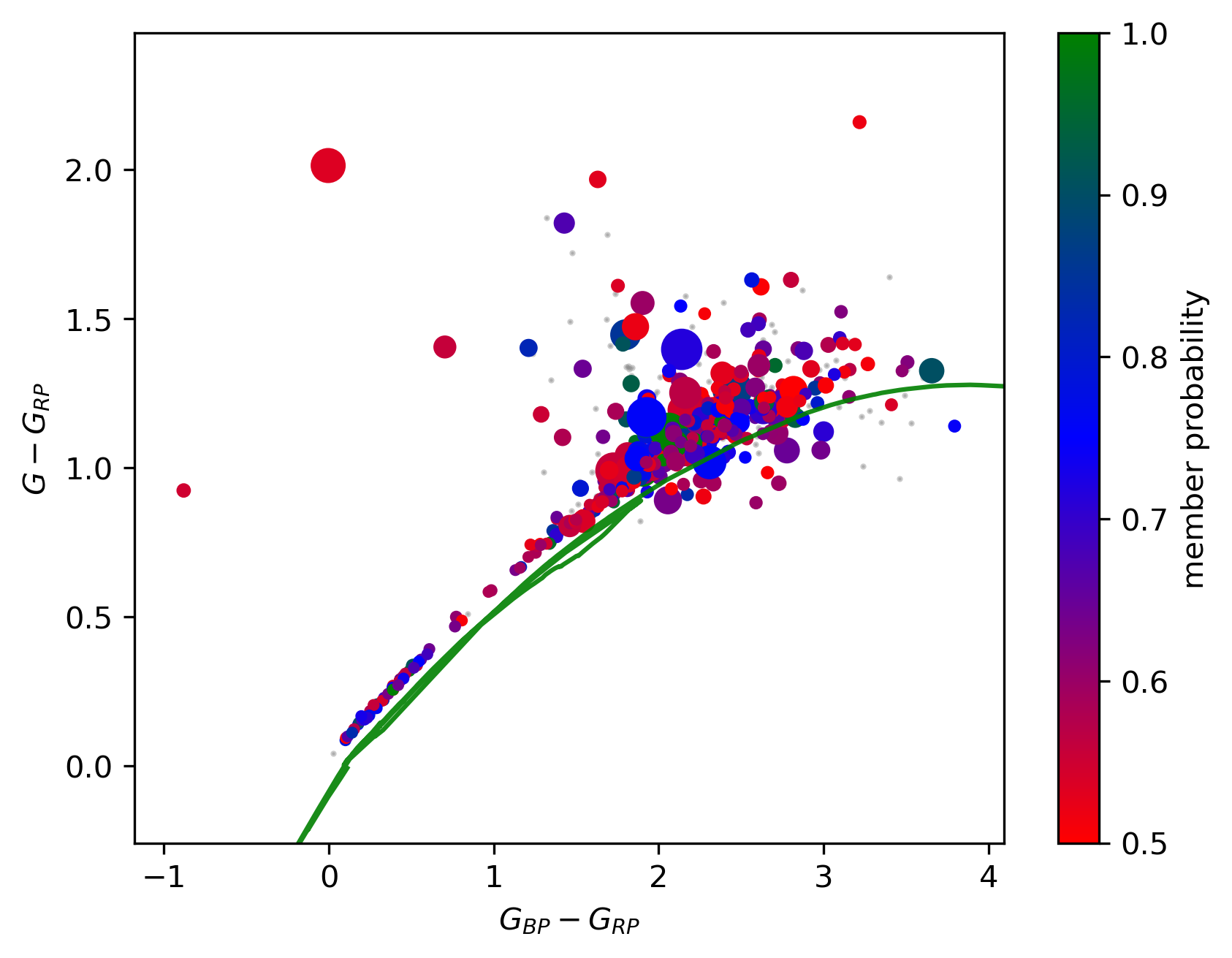} \includegraphics[width=75mm]{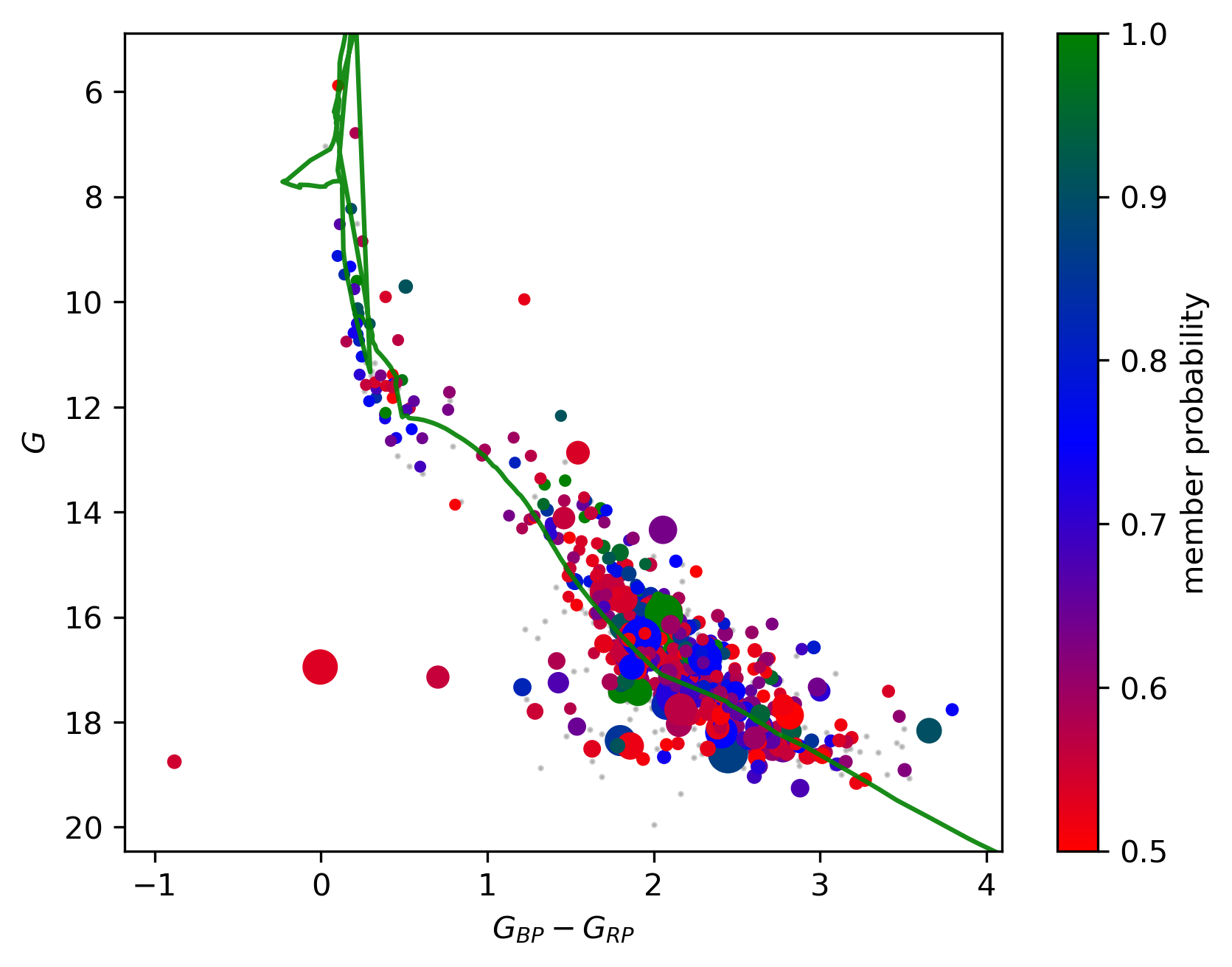}}
\caption{Same as Fig.~\ref{fig:HSC2686_isochrones} but for NGC 6530. \label{fig:NGC6530_isochrones}}
\end{figure}

Similar to NGC 2244, the fitted isochrones and OC parameters for NGC 6530 (Fig.~\ref{fig:NGC6530_isochrones}) match 
the data points very well. While \citetalias{Hunt2023} appear to slightly underestimate the age and overestimate the $A_V$ compared 
to \textit{OCFit}, the \textit{OCFit} parameter values are in good agreement with the literature values within the stated uncertainties.

\section{Discussion and Conclusions}
As part of an investigation into searching for new OC-PN pairs we cross correlated the HASH CSPN list (as a proxy for the PNe themselves) with the large, new catalogue of over 7,000 OC candidates published by \citetalias{Hunt2023}. Contrary to statistical expectations we found only one putative association between HASH likely PN [GKF2010] MN18 and two possible co-located OC candidates HSC~2686 and Lynga~3.\\

On further investigation we find that the PN candidate [GKF2010] MN18 is actually a blue supergiant with a bi-polar nebula and narrow H-alpha and strong [NII] emission lines leading to previous confusion with a PN. It is well described by \citet{Gvaramadze2015} and HASH has been updated to reflect this new identification. Hence, no plausible OC-PN pairs are found in the \citetalias{Hunt2023} catalogue. Furthermore, we find the OC candidates HSC~2686 and Lynga~3 initially thought to possibly harbour a PN are most likely not true OCs given their poor isochrone fits, large differences in OC parameters in the literature, and large radial velocities dispersions shown here that are incompatible with known OC properties. %While the parallaxes of the member stars of both OC candidates are very close to 0 (with uncertainties between 10\% and more than 100\%), leading to large uncertainties in the inferred distances

For comparison, the well known and well studied real OCs NGC~2244 and NGC~6530 
give decent OC parameters as determined by \citetalias{Hunt2023}, though they tend to slightly overestimate the $A_V$ values. 
Uncertainties from \textit{OCFit} are very small and the values agree well with the literature. 

A main indicator from this study, that requires further investigation, is that the OC catalogue by \citetalias{Hunt2023} may have a significant false positive rate, at least for clusters with small numbers ($\leq$100) of cluster members. 

\begin{acknowledgments}
A.R. and Q.A.P. thank the Hong Kong Research Grants Council for GRF research support under grants 17304024, 17326116, and 17300417.
This work made use of the University of Hong Kong/Australian Astronomical Observatory/Strasbourg Observatory H-alpha 
Planetary Nebula (HASH PN) database, hosted by the Laboratory for Space Research at the University of Hong Kong \href{hashpn.space}{http://hashpn.space}, \textit{astropy} \citep{astropy}, the program \textit{OCFit} by (Monteiro \& Dias 2016), and the desktop version of \textit{astrometry.net} \citep{Lang2010}. Data products from observations made with ESO Telescopes at the La Silla Paranal Observatory under public survey programme ID 177.D-3023, as part of the VST Photometric H$\alpha$ Survey of the Southern Galactic Plane and Bulge (VPHAS+, www.vphas.eu), have been used. We would like to thank the anonymous referee for the insightful comments which helped to increase the scientific content of this paper.
\end{acknowledgments}
%\newpage
\begin{contribution}
%%This section gives authors the space to recognize author contributions. The text inside this environment is NOT counted towards the total word quanta. At a minimum, manuscripts are expected to include this text:

%All authors contributed equally to the Terra Mater collaboration.

%% But authors are expected to provide more specific details, e.g. 
%%
AR came up with the initial research concept, supervised the undergraduates, wrote the software, and was responsible for writing and submitting the manuscript. QAP obtained the funding and edited the manuscript. XX and XG searched the cluster catalogue for CSPNe and investigated the cluster parameters and member stars.
%%EBF provided the formal analysis and validation. He also edited the manuscript.
%%GEH  and administers the project github and Zenodo repositories.
%%
%% Authors can use the Contributor Role Taxonomy (CRediT) at
%% for ideas on how write a good statement tailored to their needs.

\end{contribution}

%\facilities{HST(STIS), Swift(XRT and UVOT), AAVSO, CTIO:1.3m, CTIO:1.5m, CXO}

\software{OCFit (Monteiro, H. \& Dias, W. S. (2016) Open Cluster Isochrone Fitting Tool (Version 2.0-beta, available at https://github.com/hektor-monteiro/OCFit/releases, accessed 10 10 2024), 
astrometry.net \citep[][ available at https://astrometry.net, accessed 1 6 2026]{Lang2010}, Source Extractor \citep[][ accessed 1 6 2026]{Bertin1996}, Astropy \citep{astropy:2013, astropy:2018, astropy:2022}.}

%\newpage
\appendix

\begin{figure}[h]
%\vspace*{-15mm}%\hspace*{-1cm}
\centerline{\includegraphics[width=75mm]{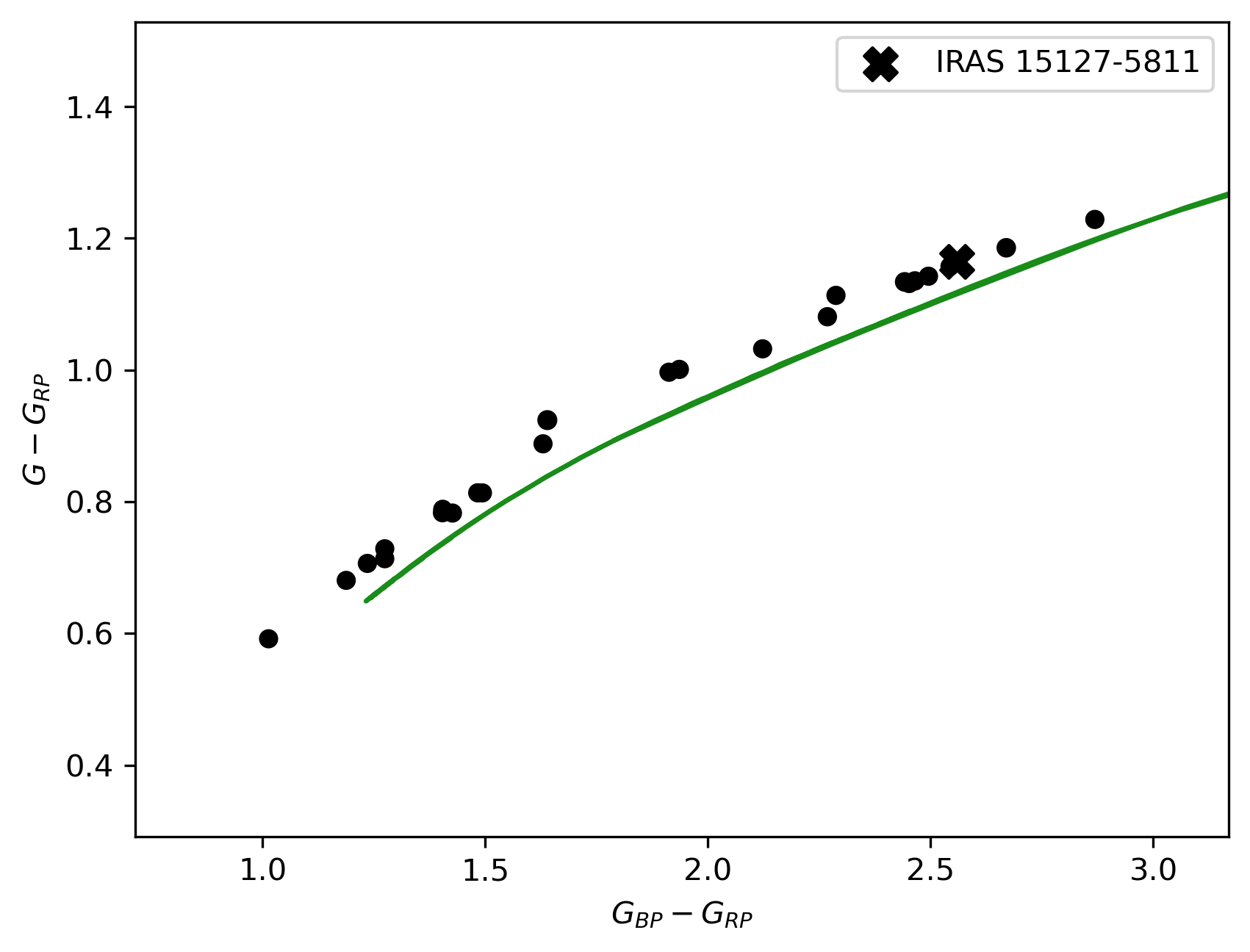} \includegraphics[width=75mm]{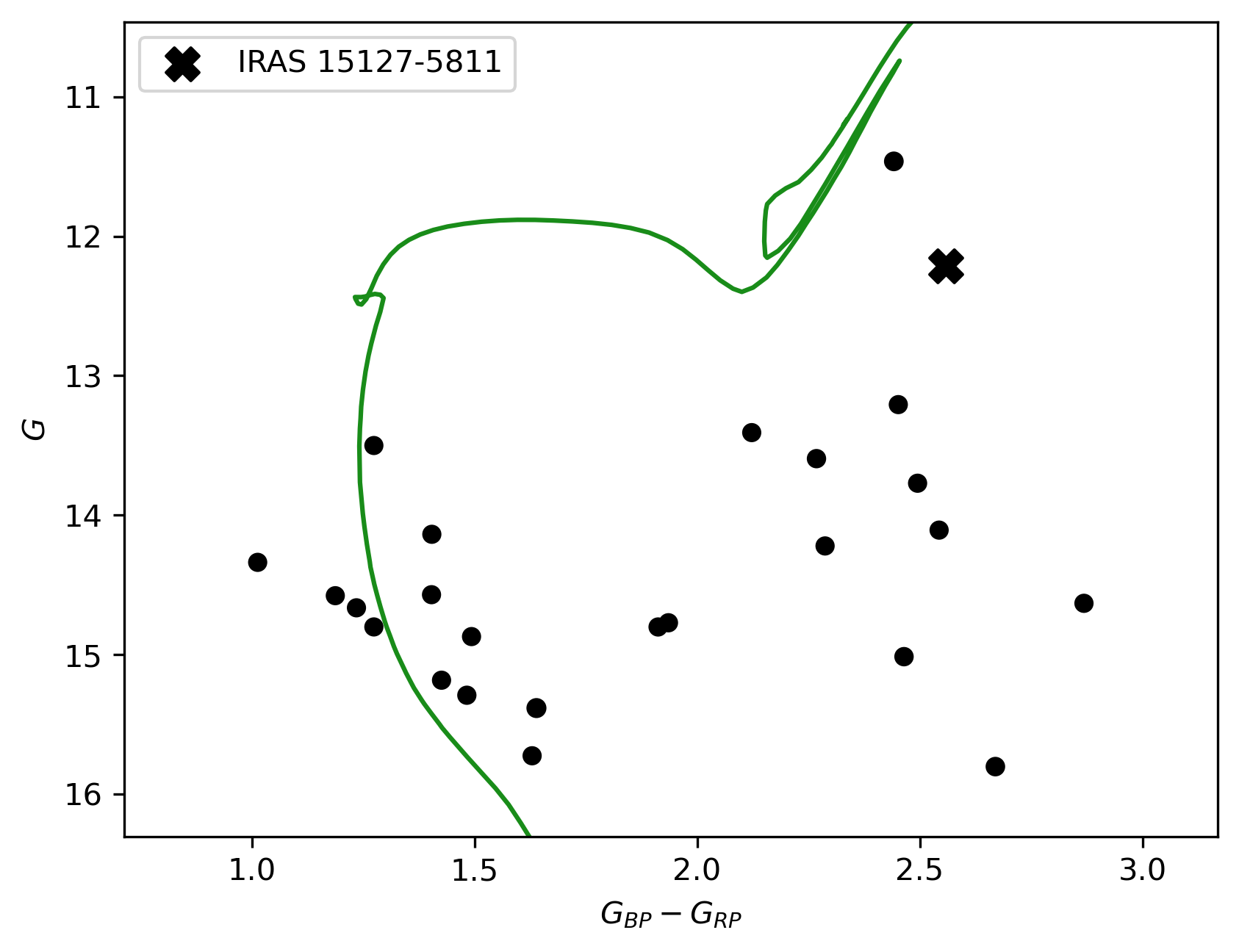}}
\caption{Same as Fig.~\ref{fig:HSC2686_isochrones} but for Lynga~3 member stars from \citetalias{Lynga1964}. Note that the 1st row is missing here as \citetalias{Lynga1964} did not publish OC parameters.}
\label{fig:Lynga3_isochrones_Lynga1964}
\end{figure}   

\begin{figure}[h]
%\vspace*{-15mm}%\hspace*{-1cm}
\centerline{\includegraphics[width=75mm]{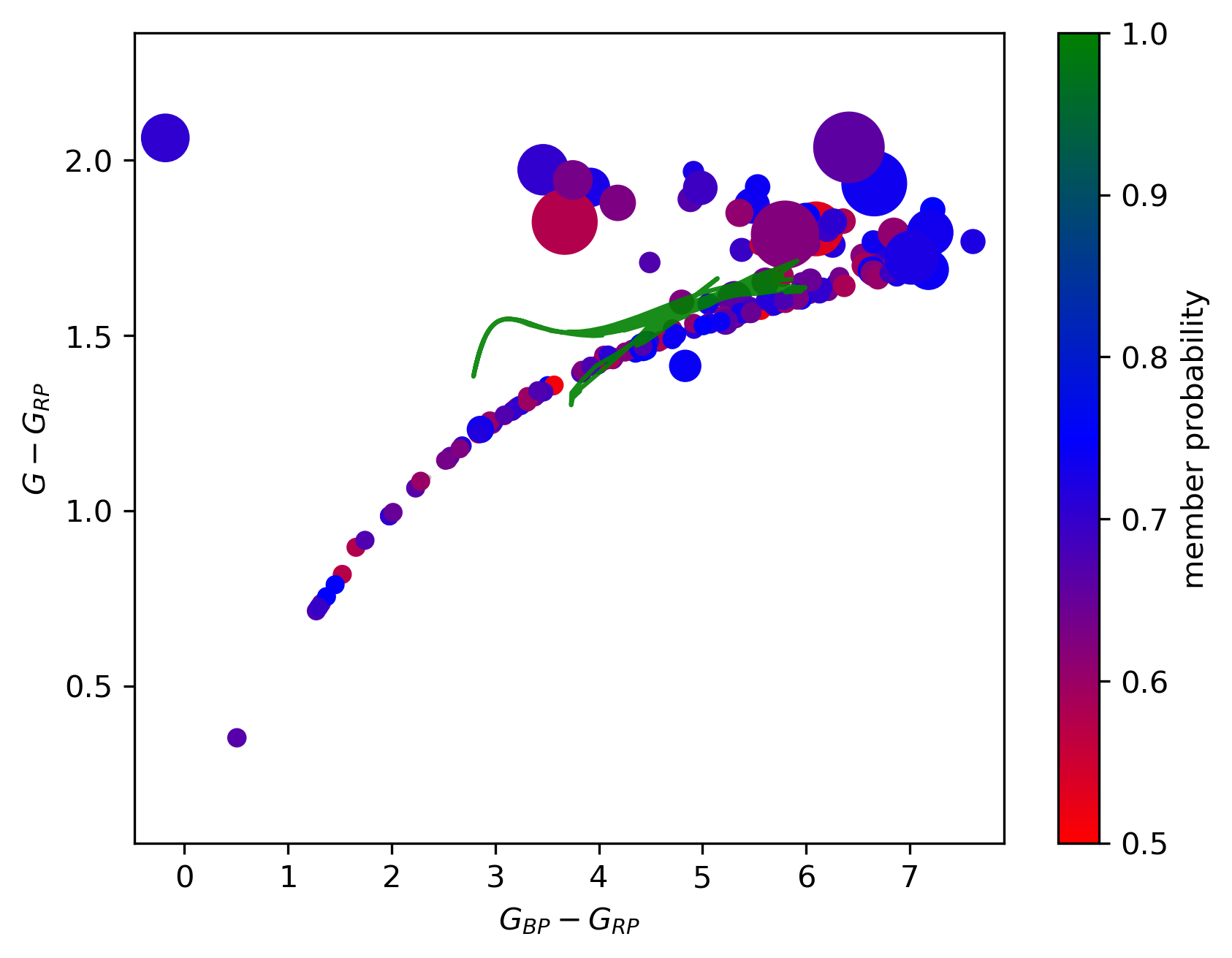} \includegraphics[width=75mm]{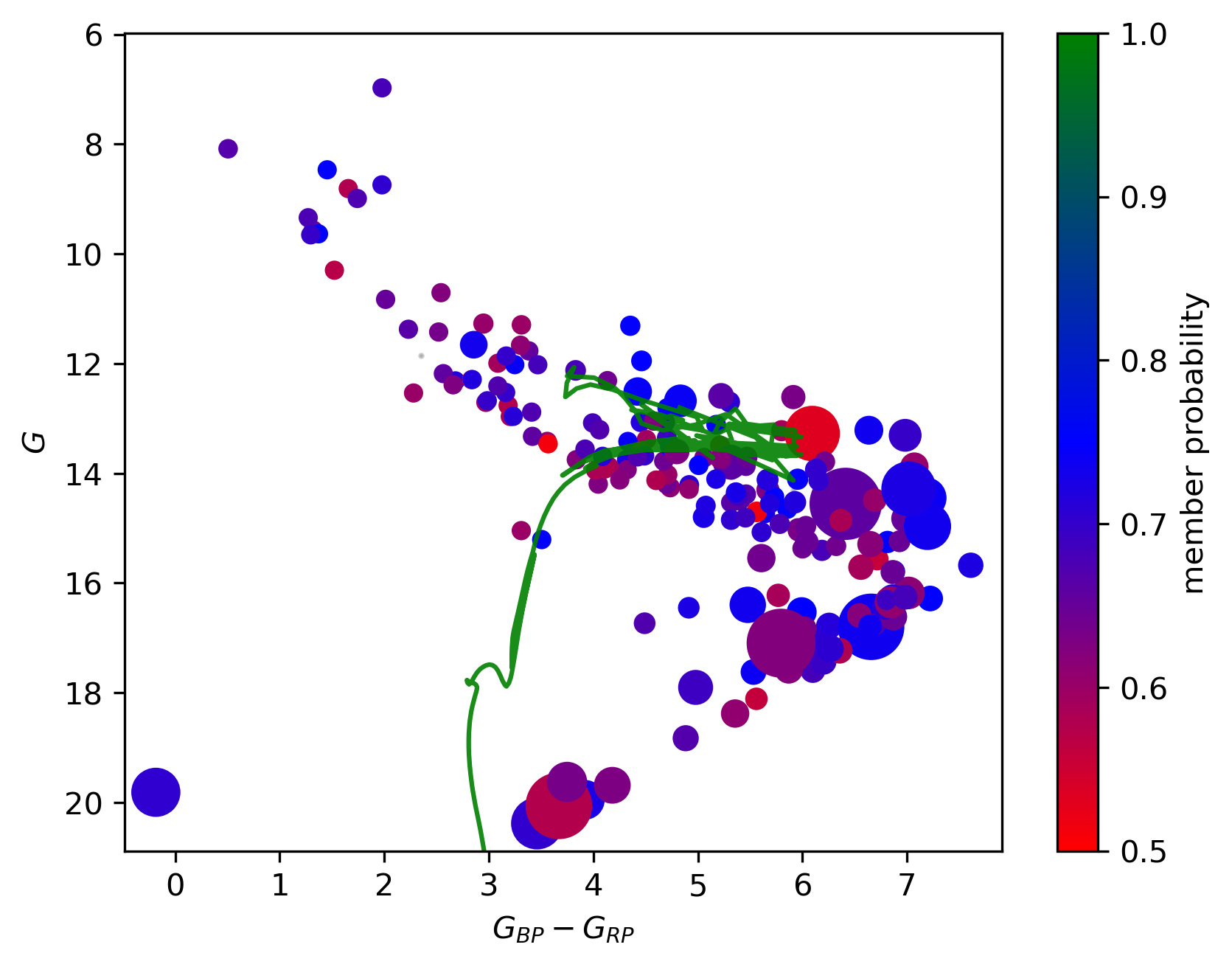}}
\centerline{\includegraphics[width=75mm]{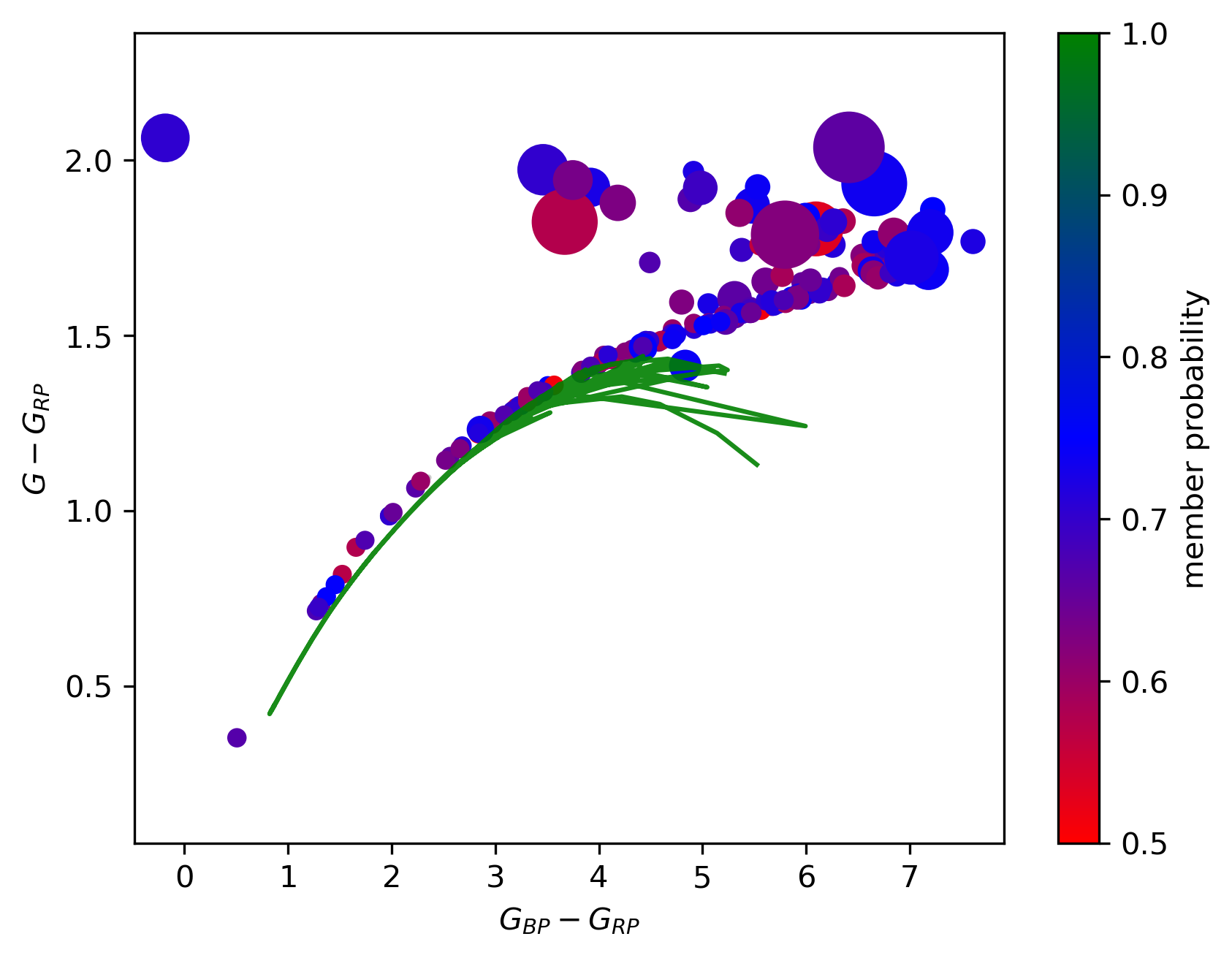} \includegraphics[width=75mm]{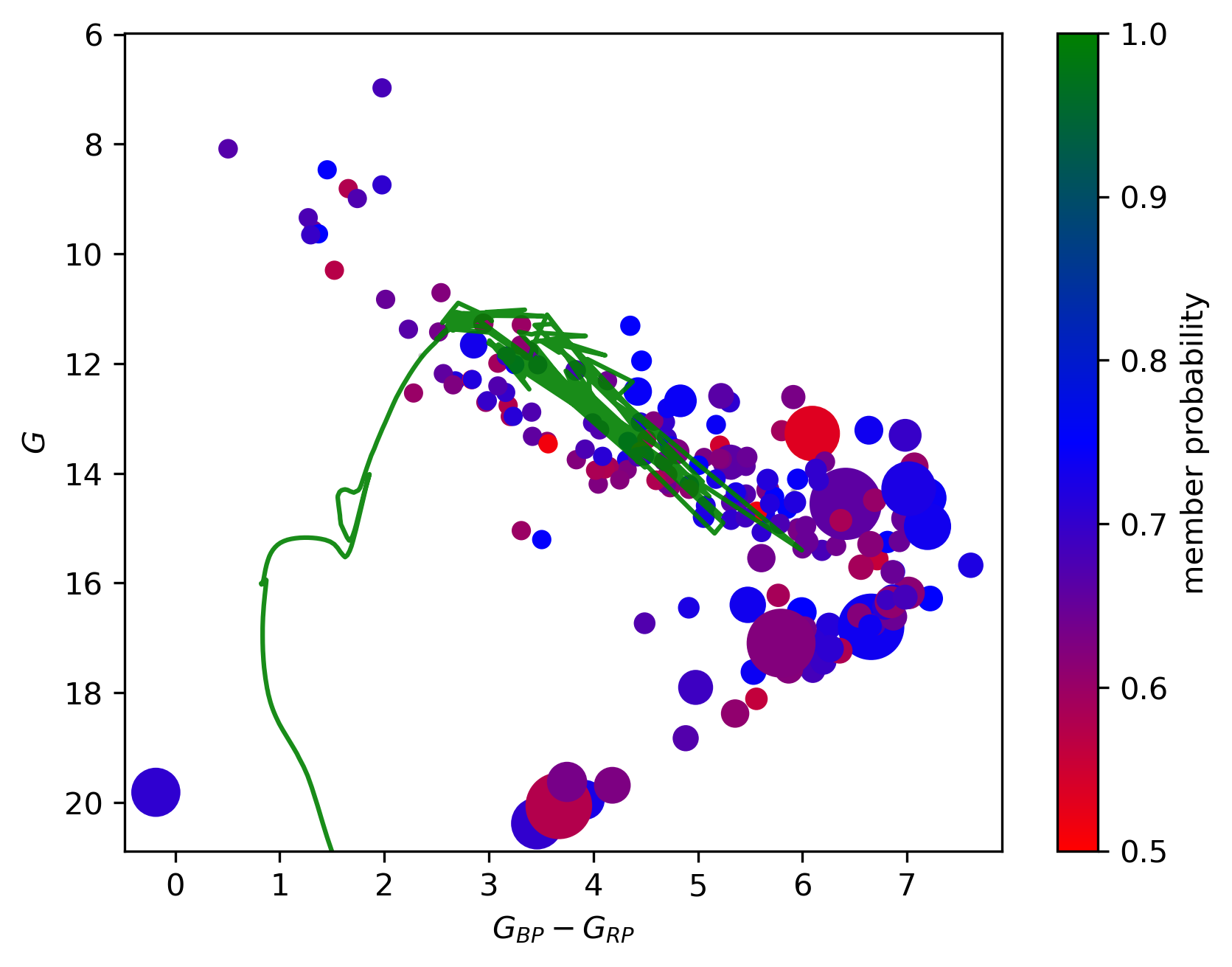}}
\caption{Same as Fig.~\ref{fig:HSC2686_isochrones} but for Lynga~3 member stars from \citetalias{Kharchenko2013}.}\label{fig:Lynga3_isochrones_Kharchenko2013}
\end{figure}

\begin{figure}[h]
%\vspace*{-15mm}%\hspace*{-1cm}
\centerline{\includegraphics[width=75mm]{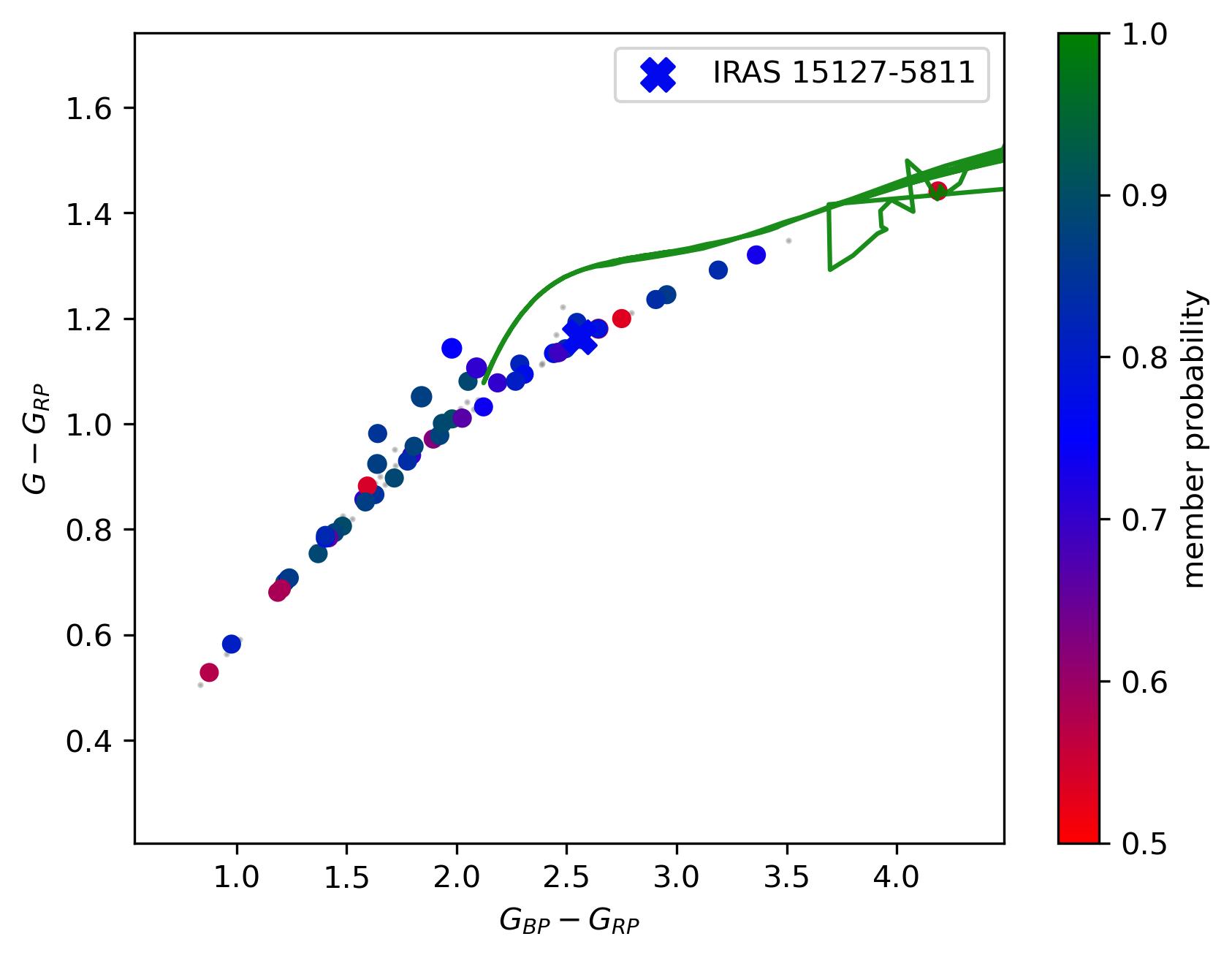} \includegraphics[width=75mm]{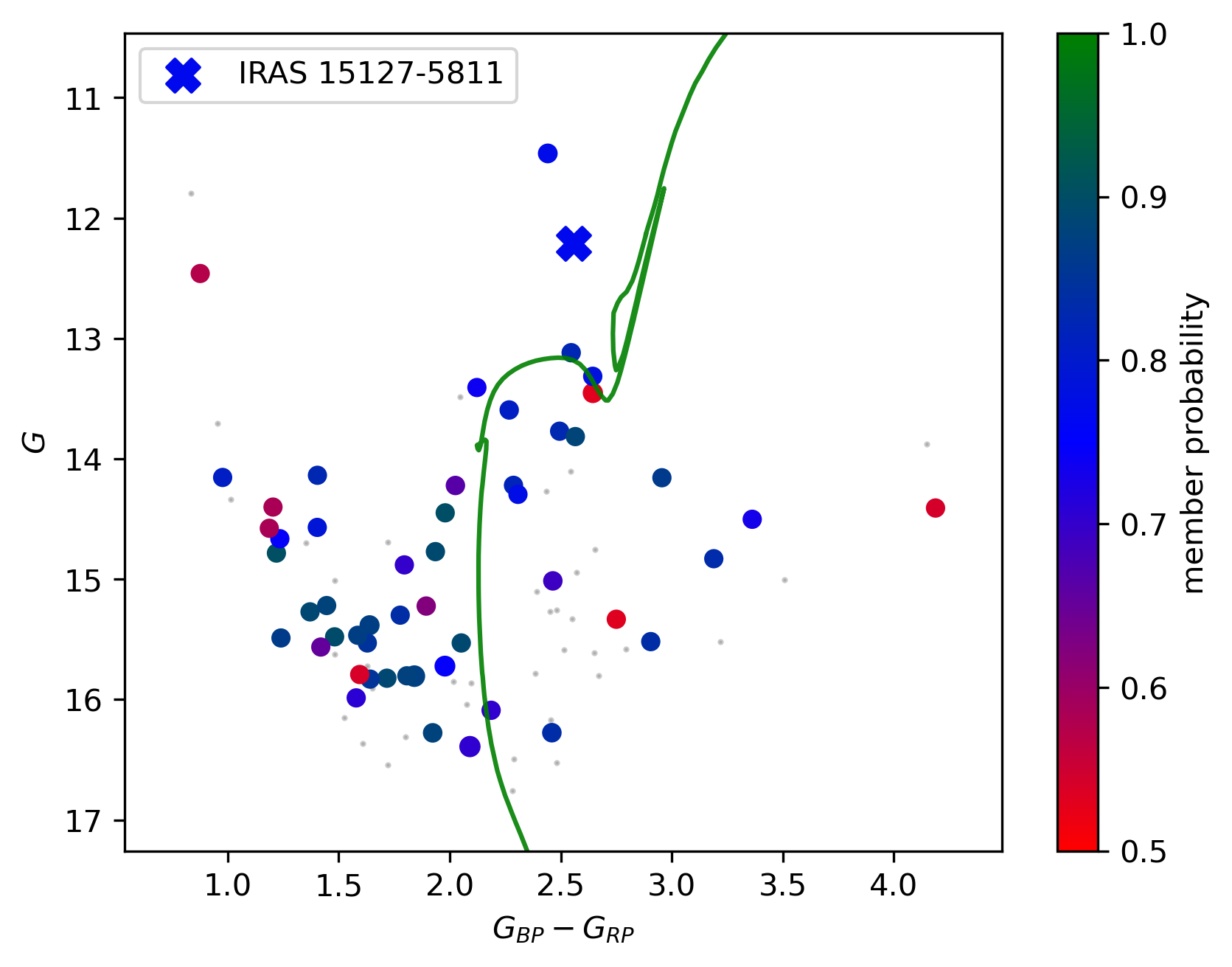}}
\centerline{\includegraphics[width=75mm]{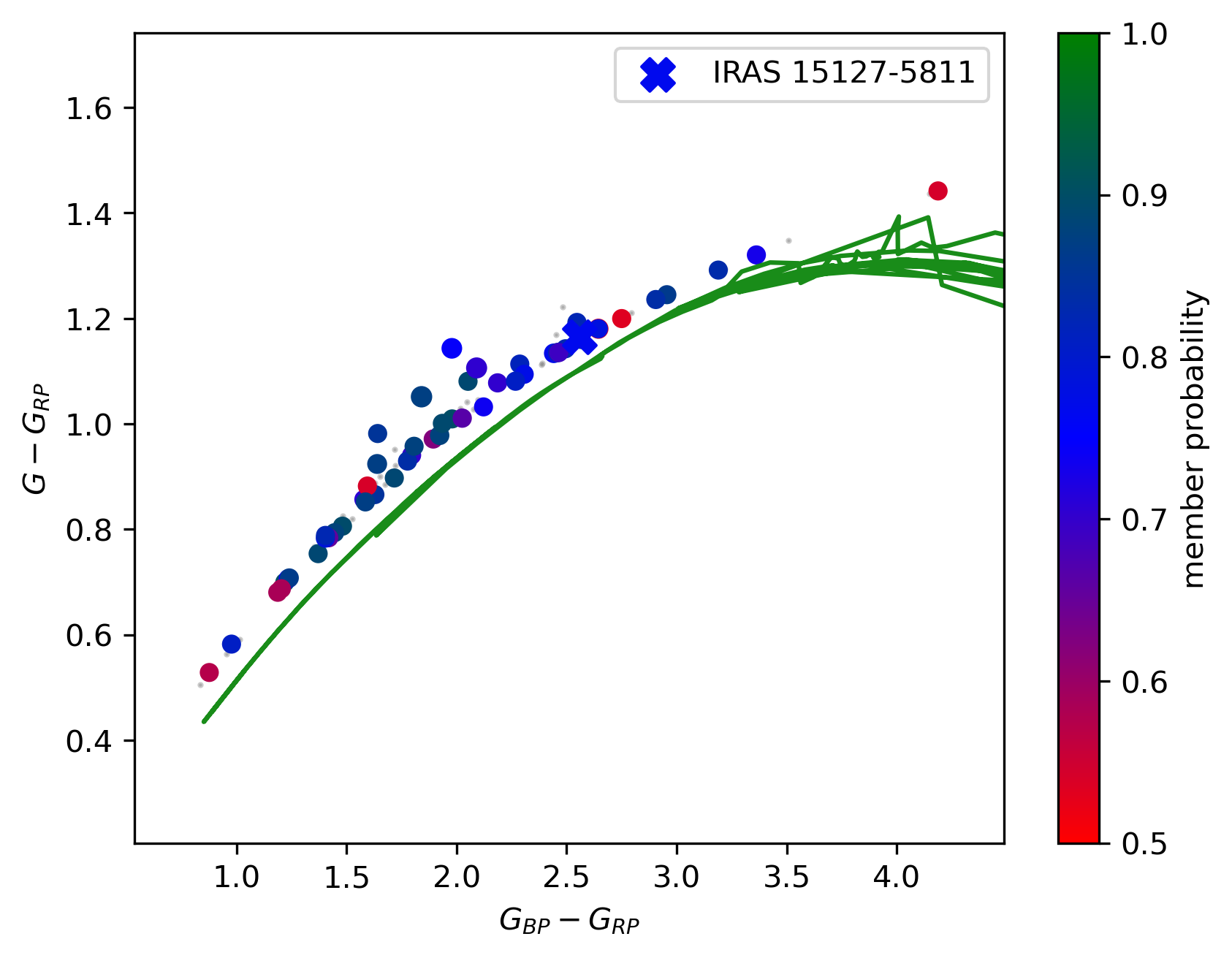} \includegraphics[width=75mm]{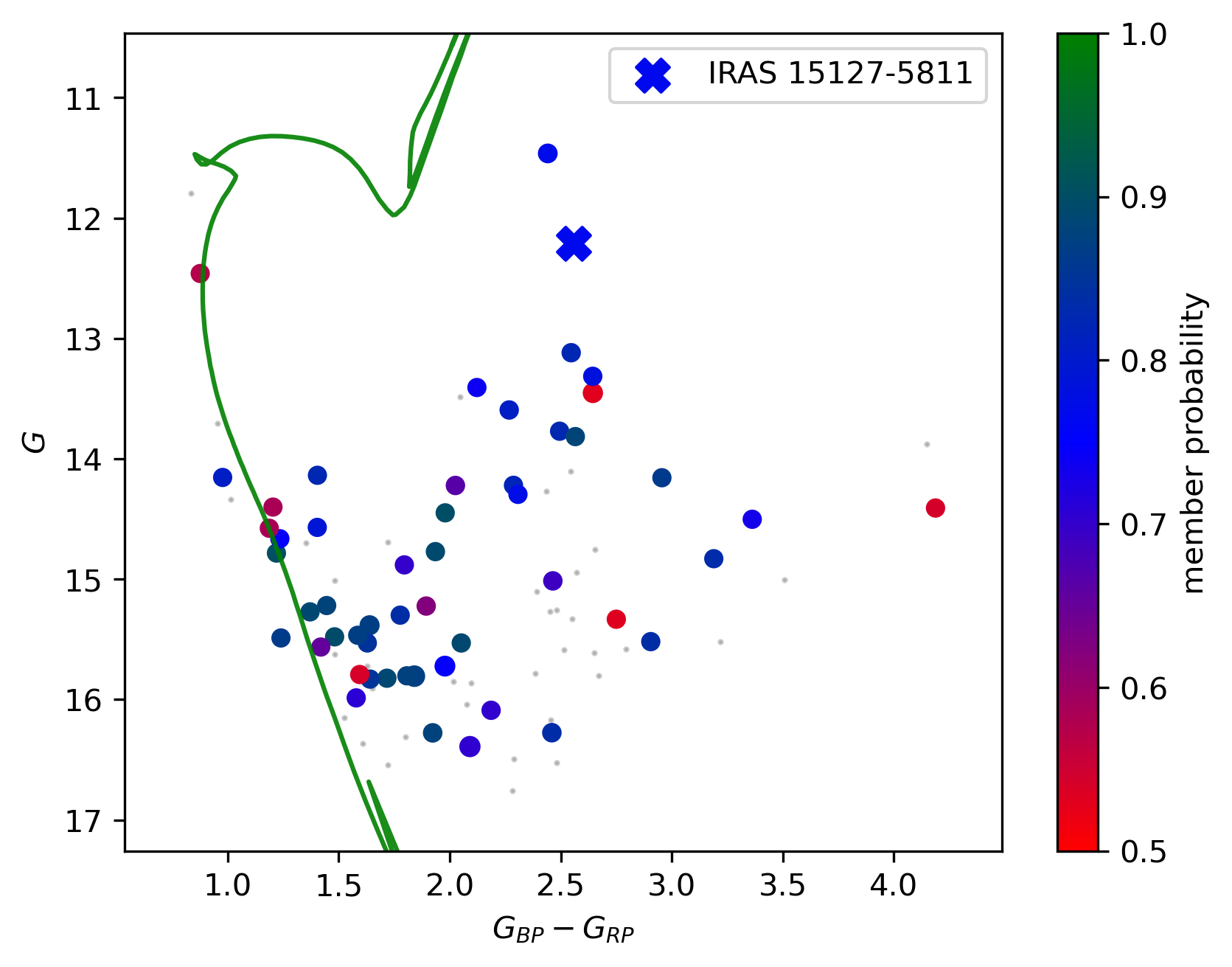}}
\caption{Same as Fig.~\ref{fig:HSC2686_isochrones} but for Lynga~3 member stars from \citetalias{Sampedro2017}. Grey data points are stars with membership probability less than 0.5.}\label{fig:Lynga3_isochrones_Sampedro2017}
\end{figure}

\begin{figure}[h]
\centerline{\includegraphics[width=75mm]{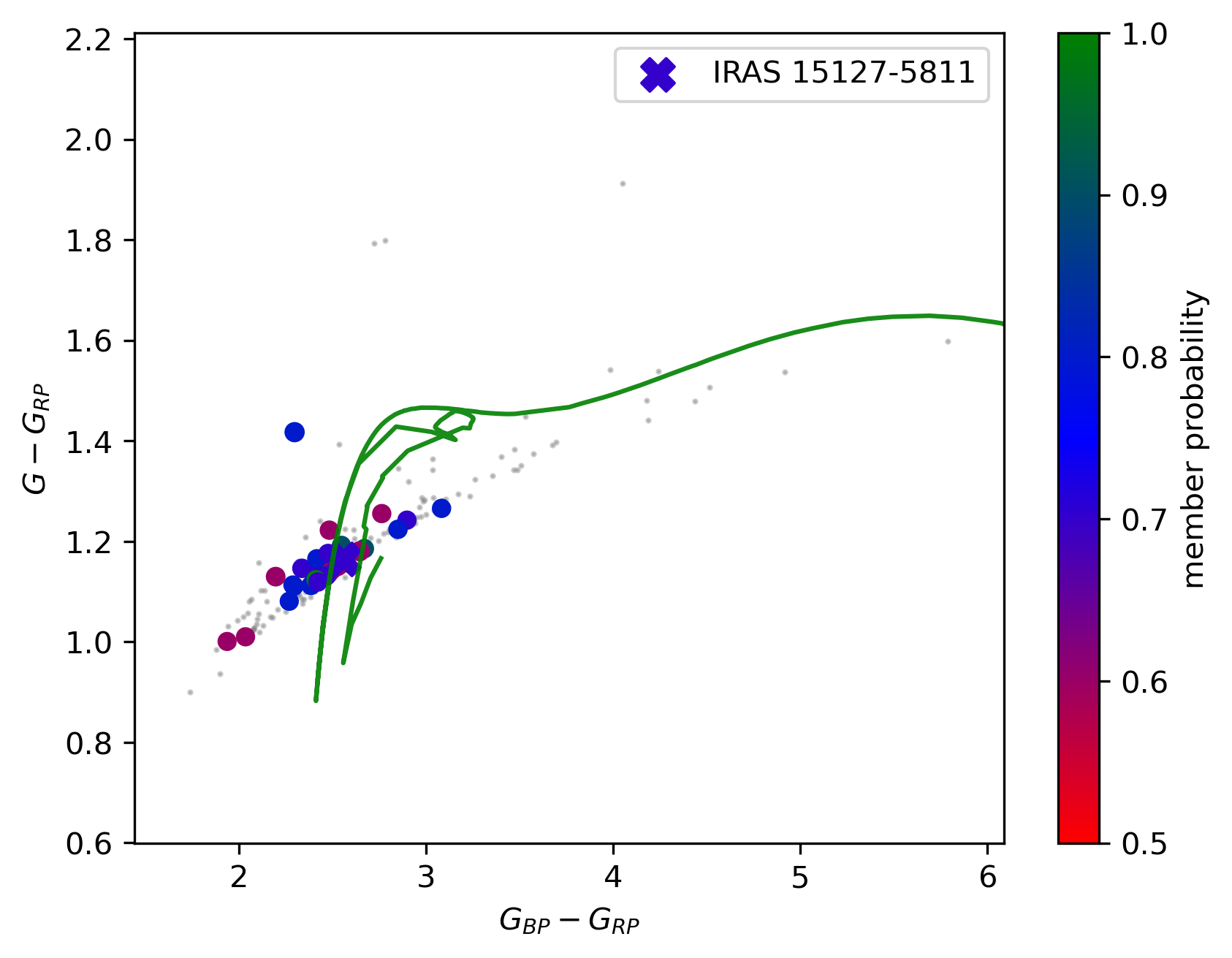} \includegraphics[width=75mm]{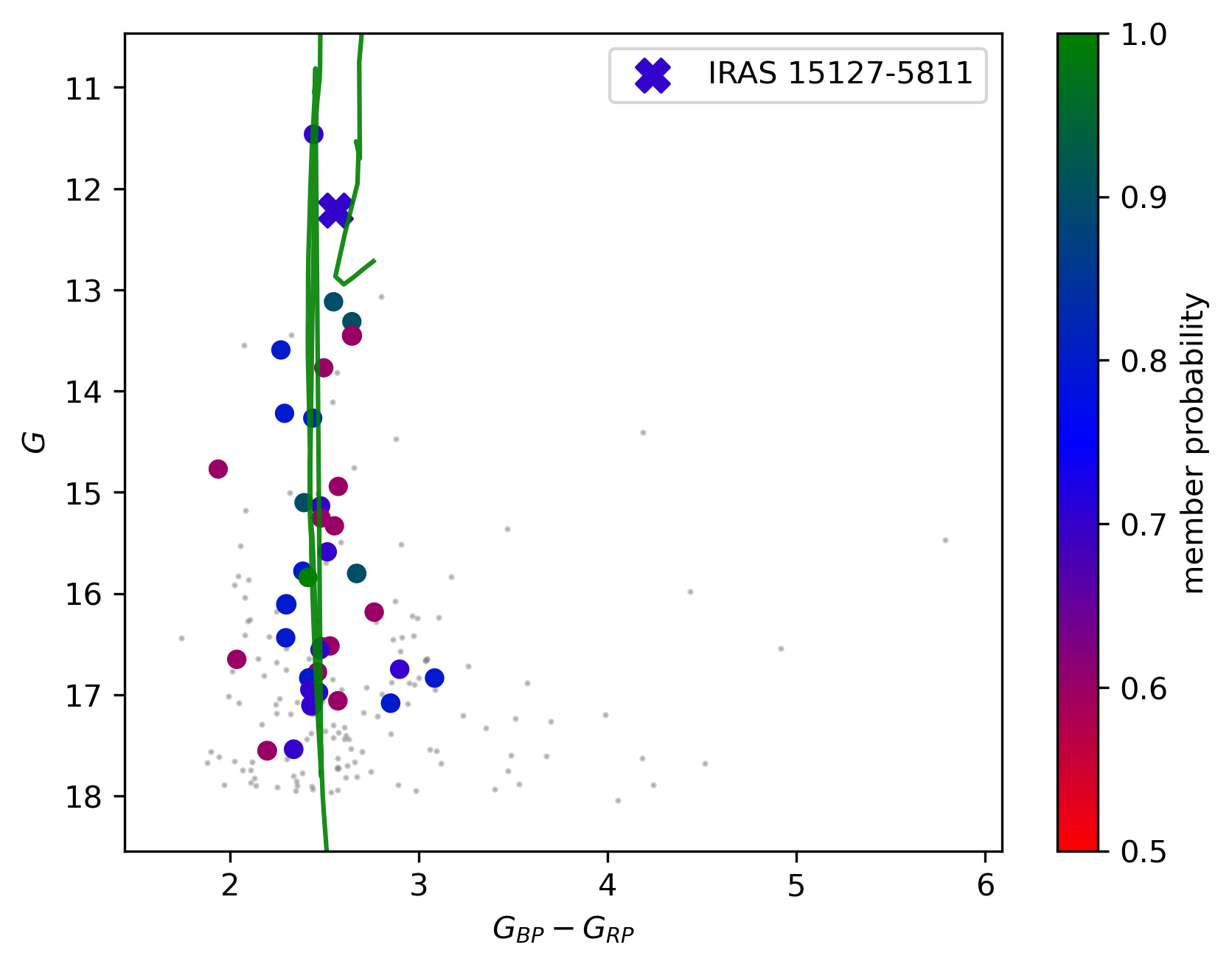}}
\caption{Same as Fig.~\ref{fig:HSC2686_isochrones} but for Lynga~3 member stars from \citetalias{Cantat-Gaudin2018}. Note that the first row is missing here as \citetalias{Cantat-Gaudin2018} only published the distance. Grey data points are stars with membership probability less than 0.5.}
\label{fig:Lynga3_isochrones_Cantat_Gaudin2018}
\end{figure}   

\begin{figure}[h]
\centerline{\includegraphics[width=75mm]{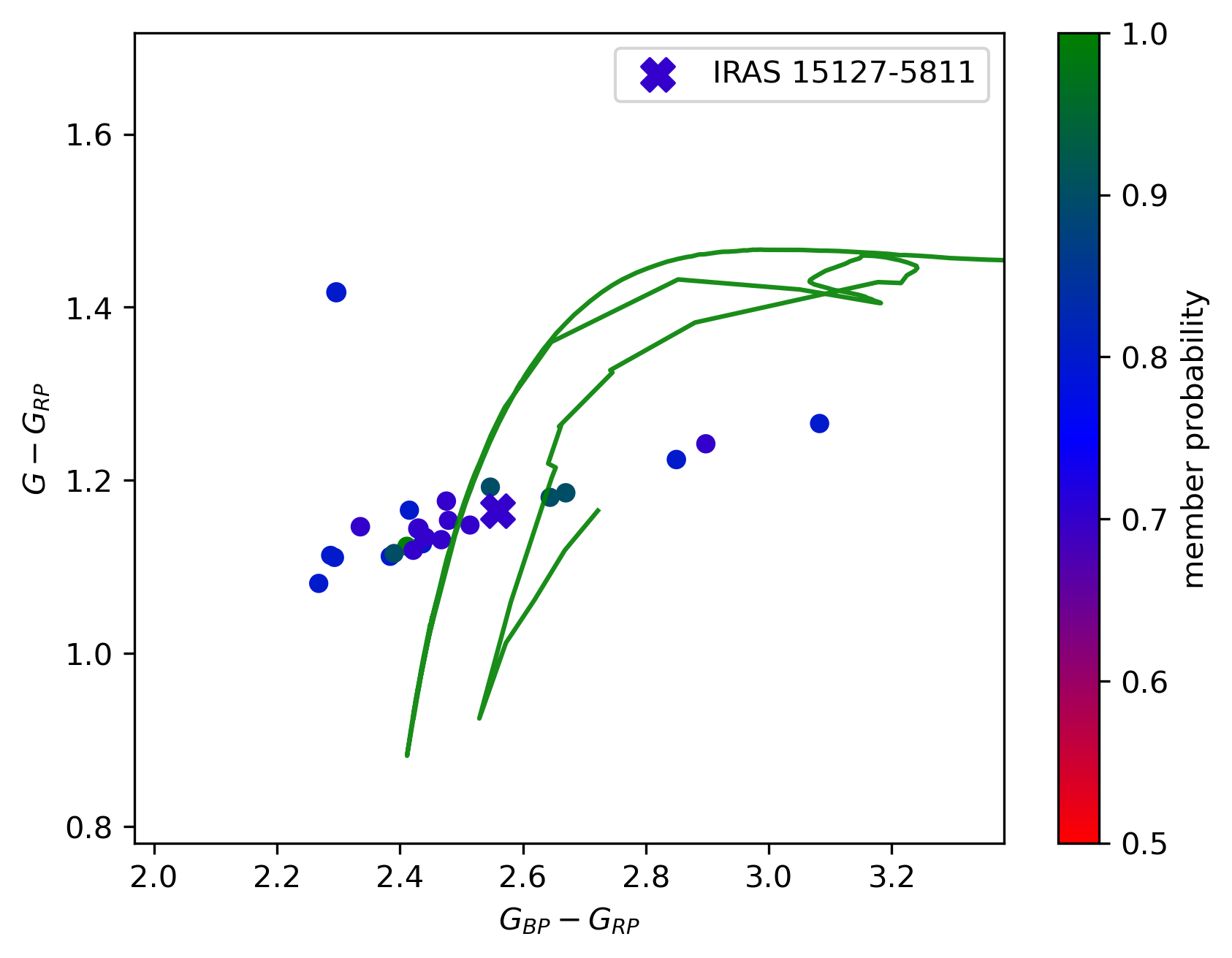} \includegraphics[width=75mm]{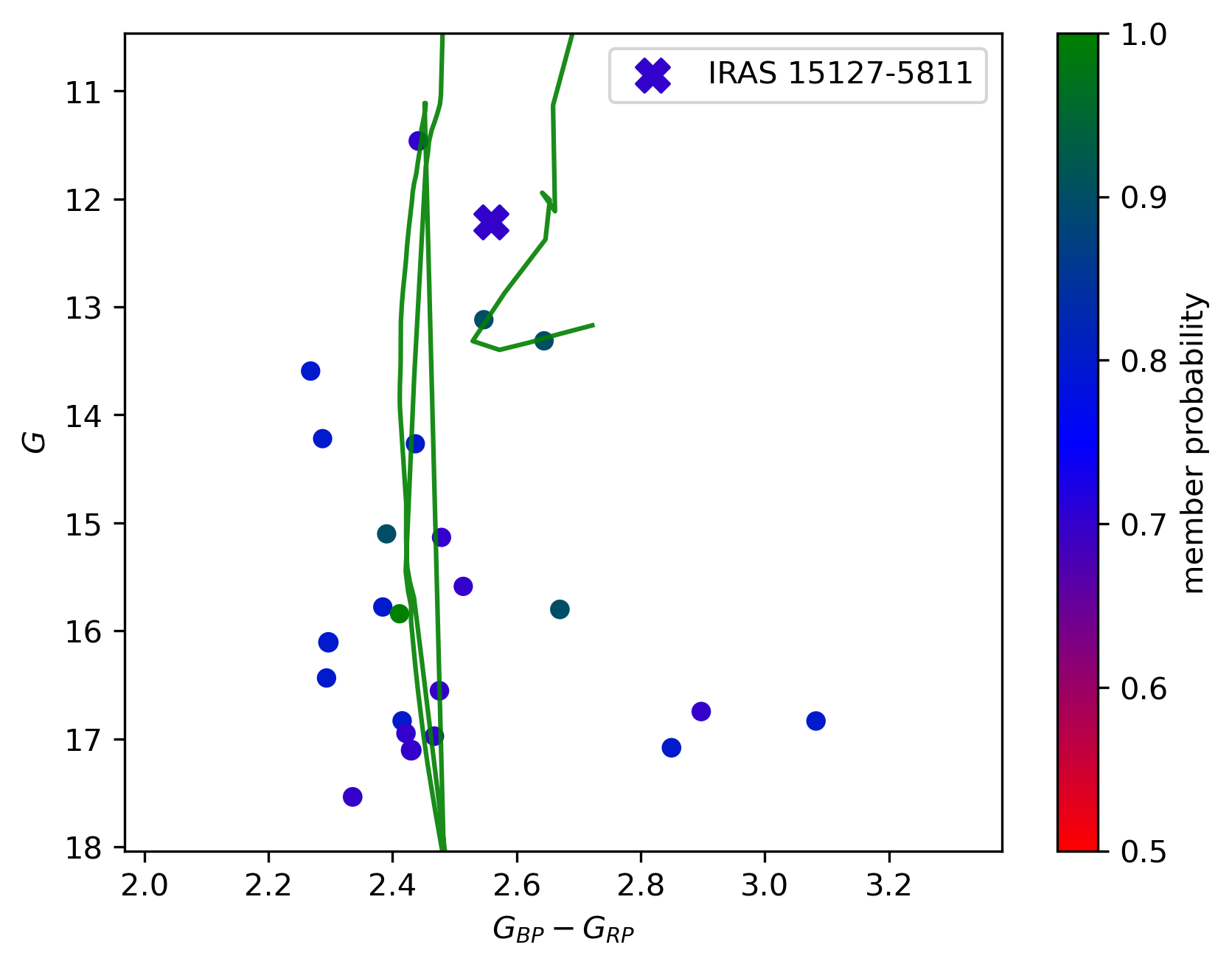}}
\caption{Same as Fig.~\ref{fig:HSC2686_isochrones} but for Lynga~3 member stars from \citetalias{Cantat-Gaudin2020}. Note that the first row is missing here as \citetalias{Cantat-Gaudin2020} did not publish OC parameters.  }\label{fig:Lynga3_isochrones_Cantat_Gaudin2020}
\end{figure}   

\begin{figure}[h]
%\vspace*{-15mm}%\hspace*{-1cm}
\centerline{\includegraphics[width=75mm]{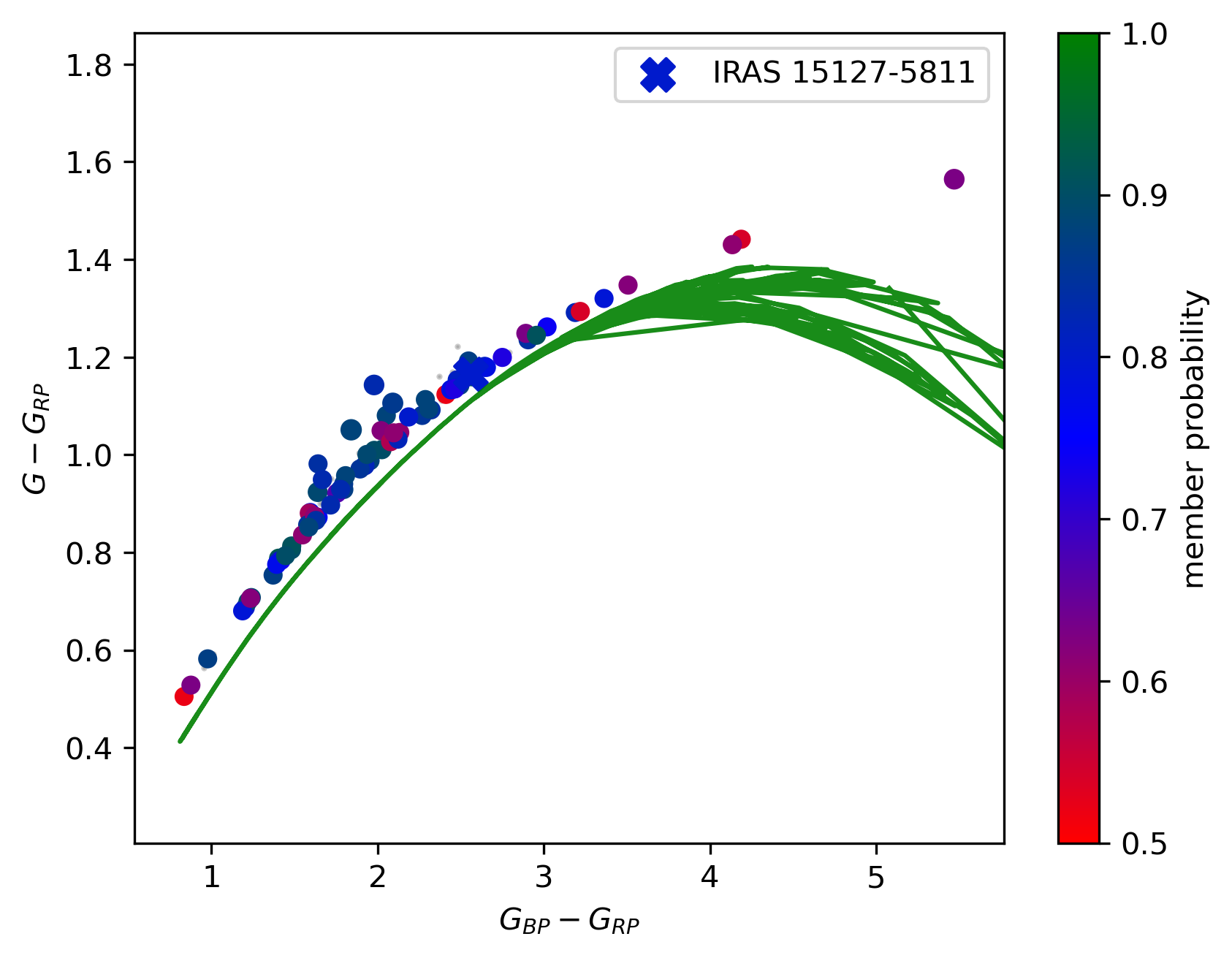} \includegraphics[width=75mm]{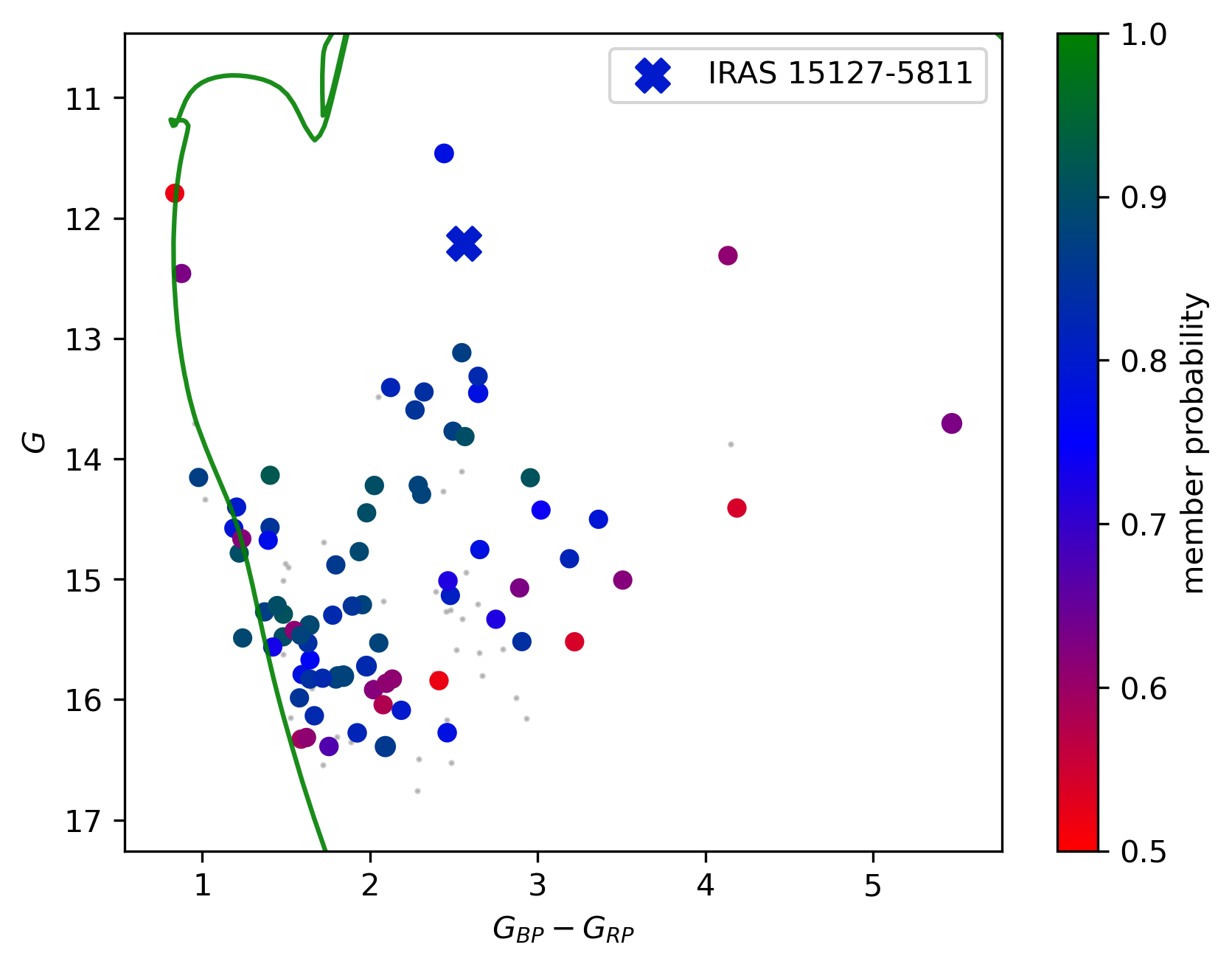}}
\caption{Same as Fig.~\ref{fig:HSC2686_isochrones} but for Lynga~3 member stars from \citetalias{Dias2014}. Note that the first row is missing here as \citetalias{Dias2014} did not publish OC parameters. Grey data points are stars with membership probability less than 0.5.}
\label{fig:Lynga3_isochrones_Dias2014}
\end{figure}   

\begin{figure}[h]
%\vspace*{-15mm}%\hspace*{-1cm}
\centerline{\includegraphics[width=75mm]{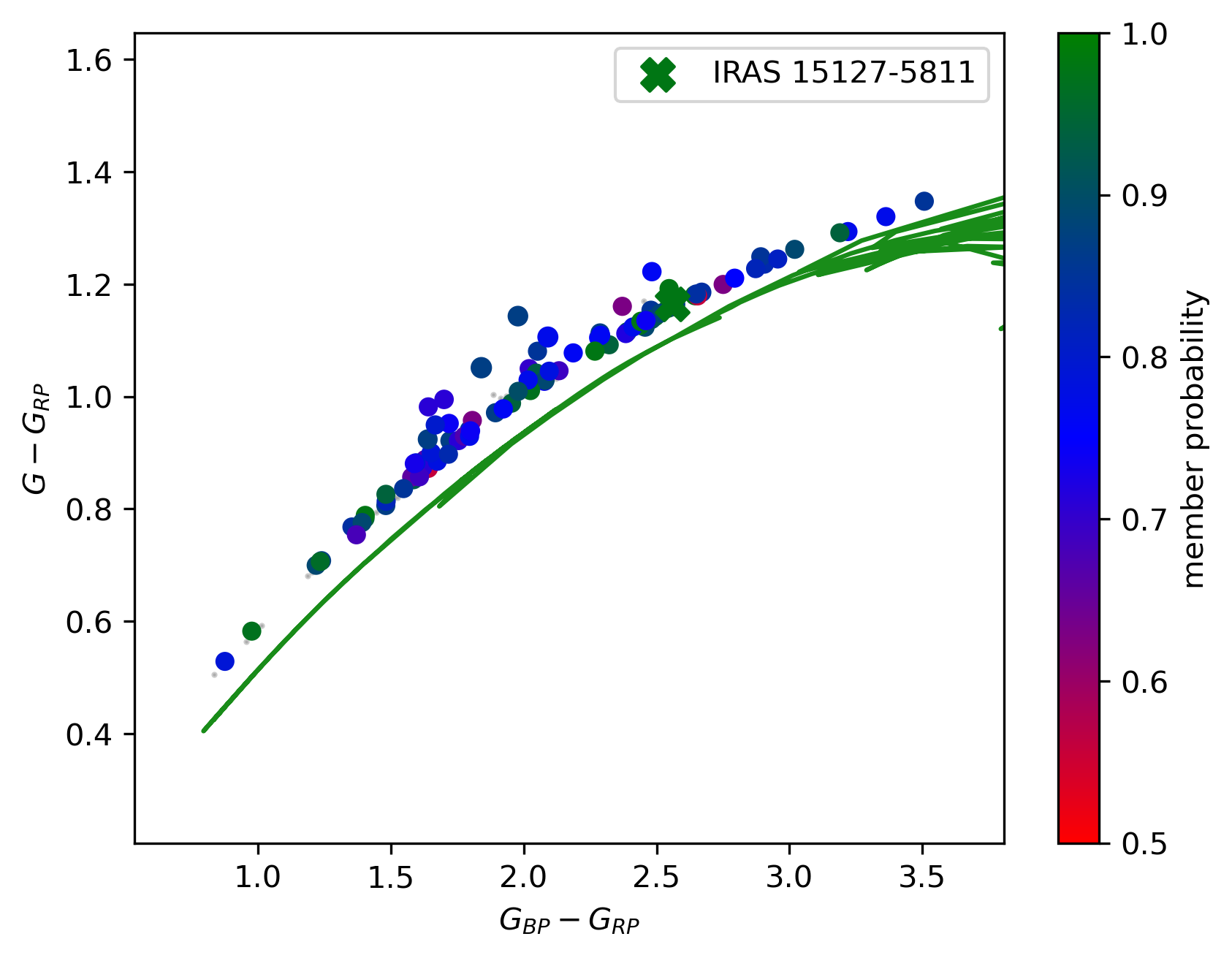} \includegraphics[width=75mm]{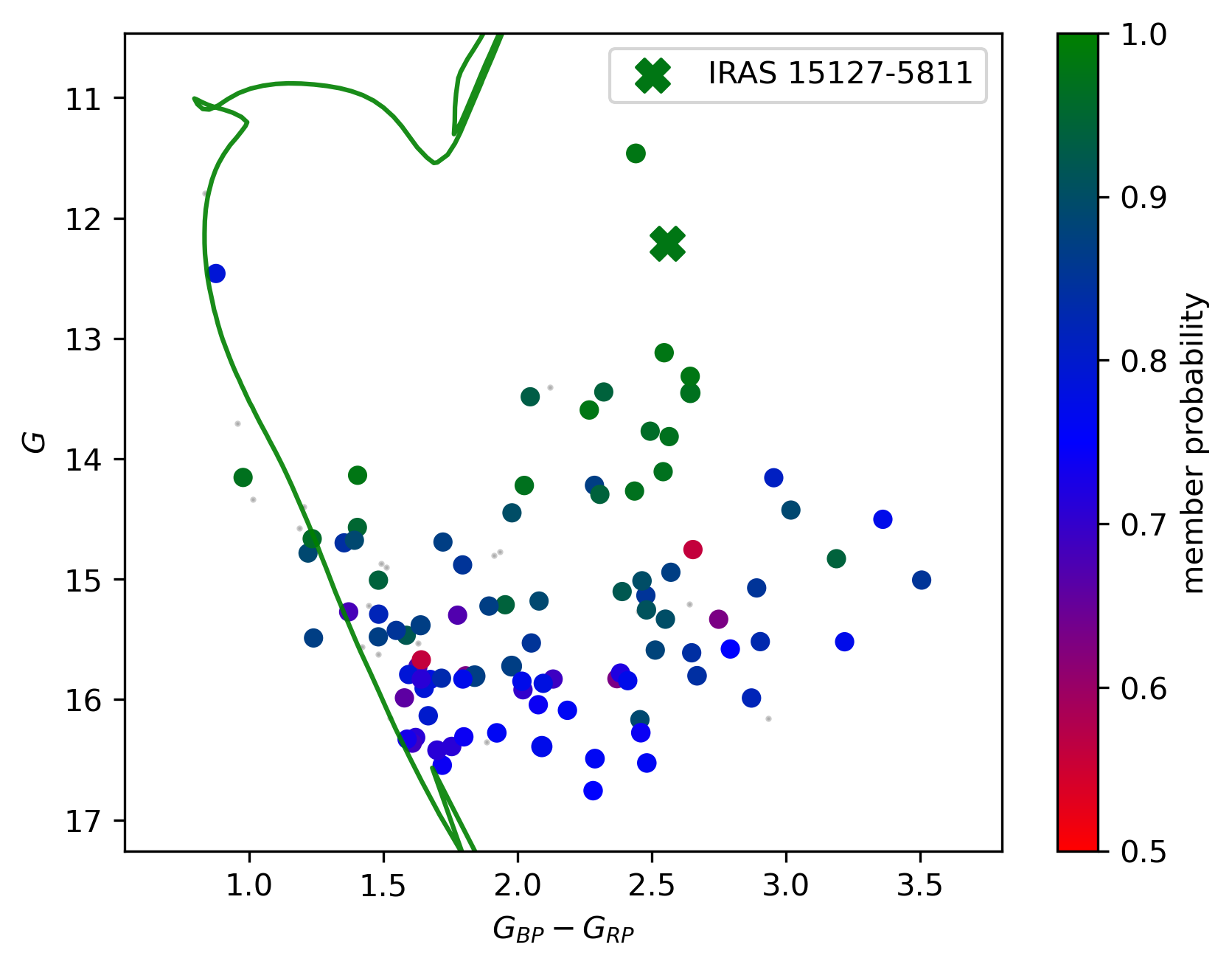}}
\caption{Same as Fig.~\ref{fig:HSC2686_isochrones} but for Lynga~3 member stars from \citetalias{Dias2018}. Note that the first row is missing here as \citetalias{Dias2018} did not publish OC parameters. Grey data points are stars with membership probability less than 0.5.}
\label{fig:Lynga3_isochrones_Dias2018}
\end{figure}

\begin{figure}[h]
%\vspace*{-15mm}%\hspace*{-1cm}
\centerline{\includegraphics[width=75mm]{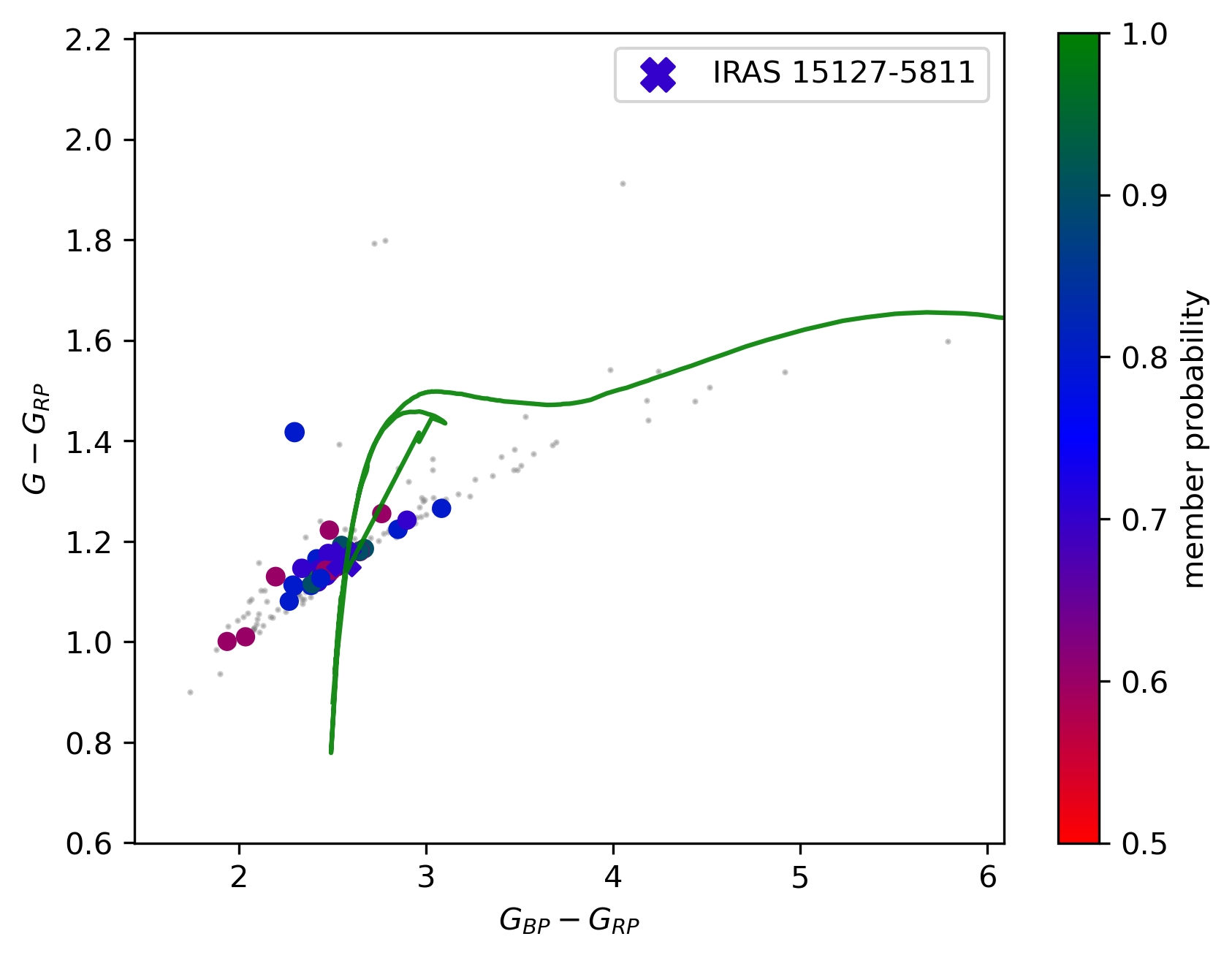} \includegraphics[width=75mm]{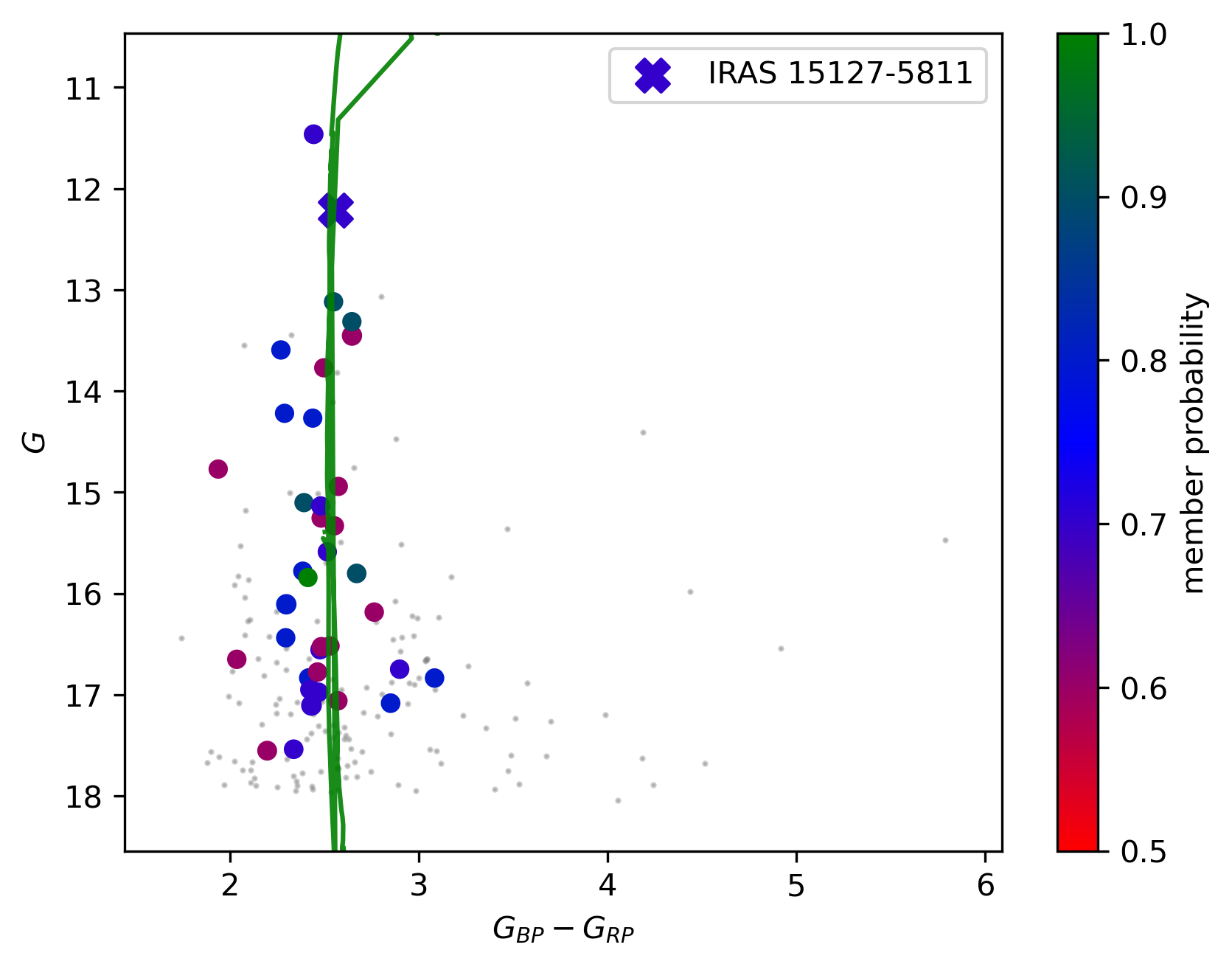}}
\centerline{\includegraphics[width=75mm]{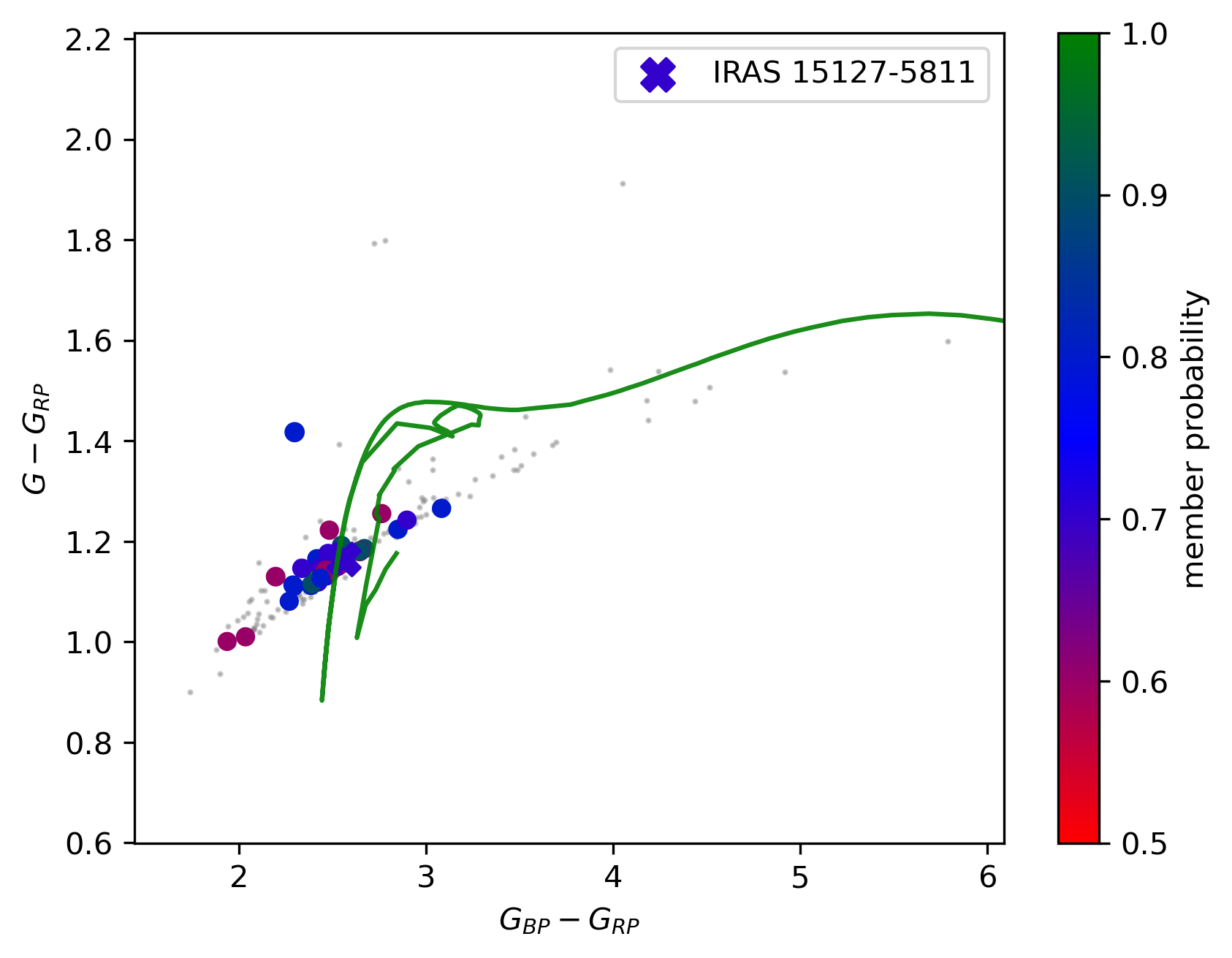} \includegraphics[width=75mm]{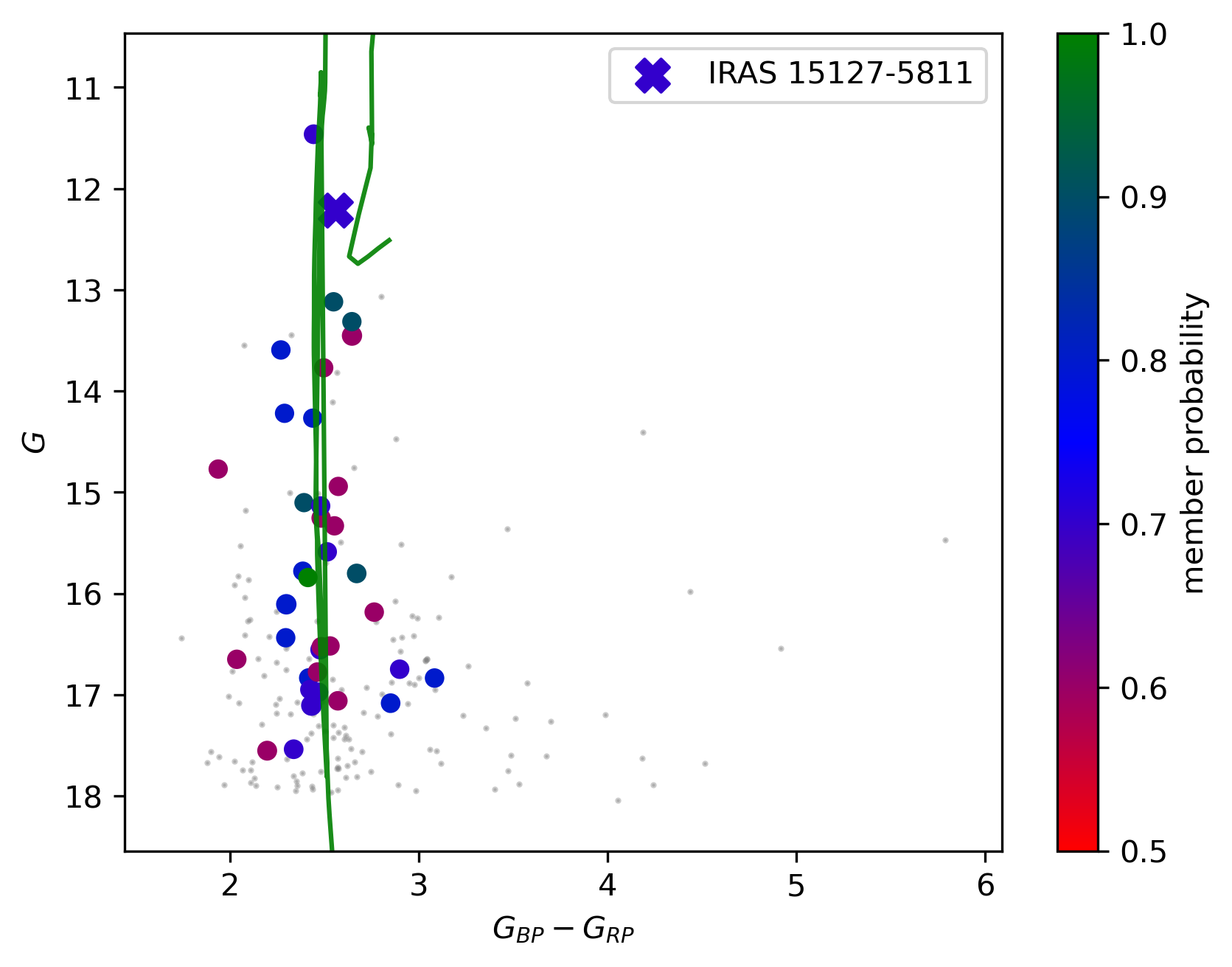}}
\caption{Same as Fig.~\ref{fig:HSC2686_isochrones} but for Lynga 3 member stars from \citetalias{Dias2021}. The grey dots are stars with a membership probability of less than 0.5.} \label{fig:Lynga3_isochrones_Dias2021}
\end{figure}

\begin{figure}[h]
\centerline{\includegraphics[width=75mm]{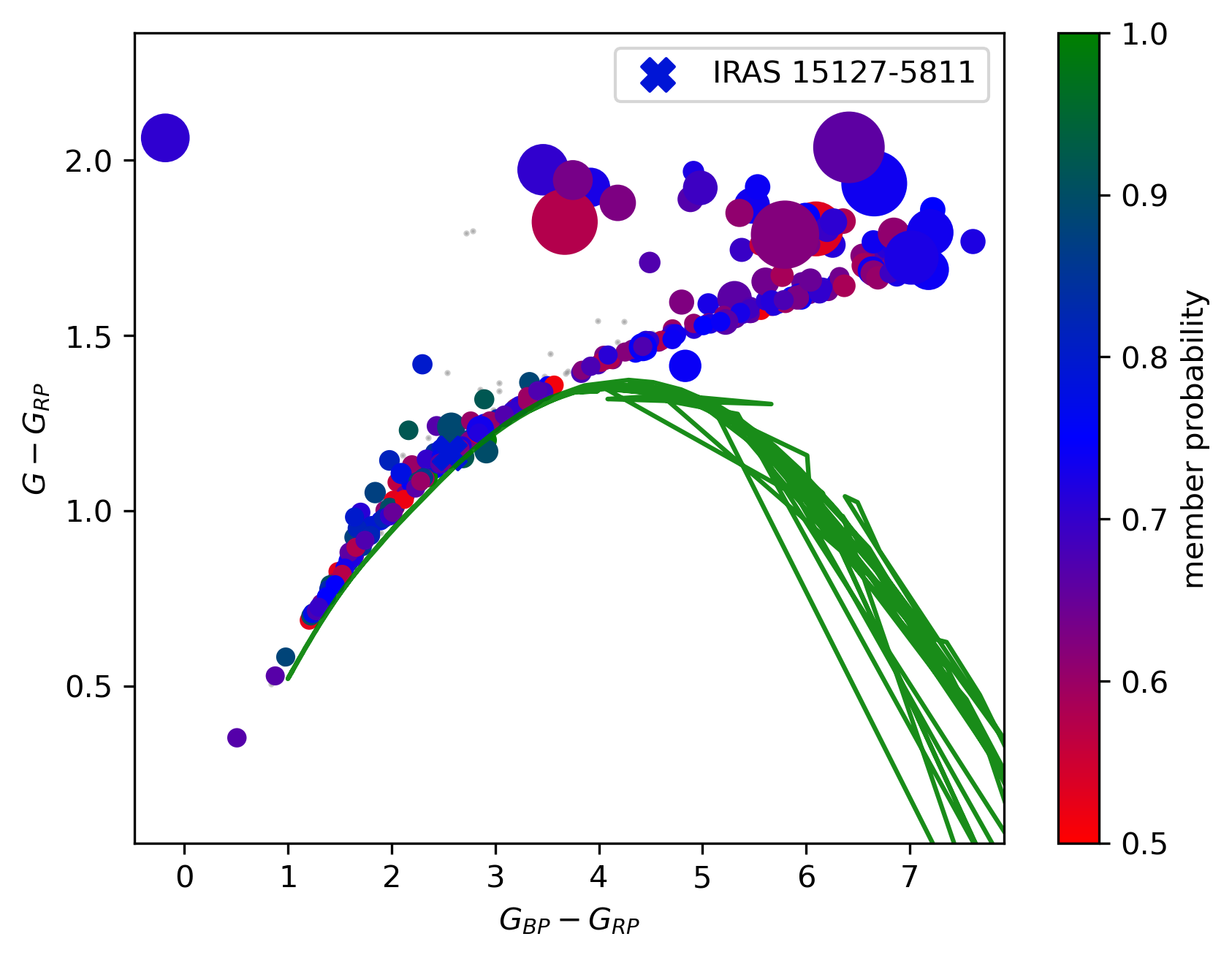} \includegraphics[width=75mm]{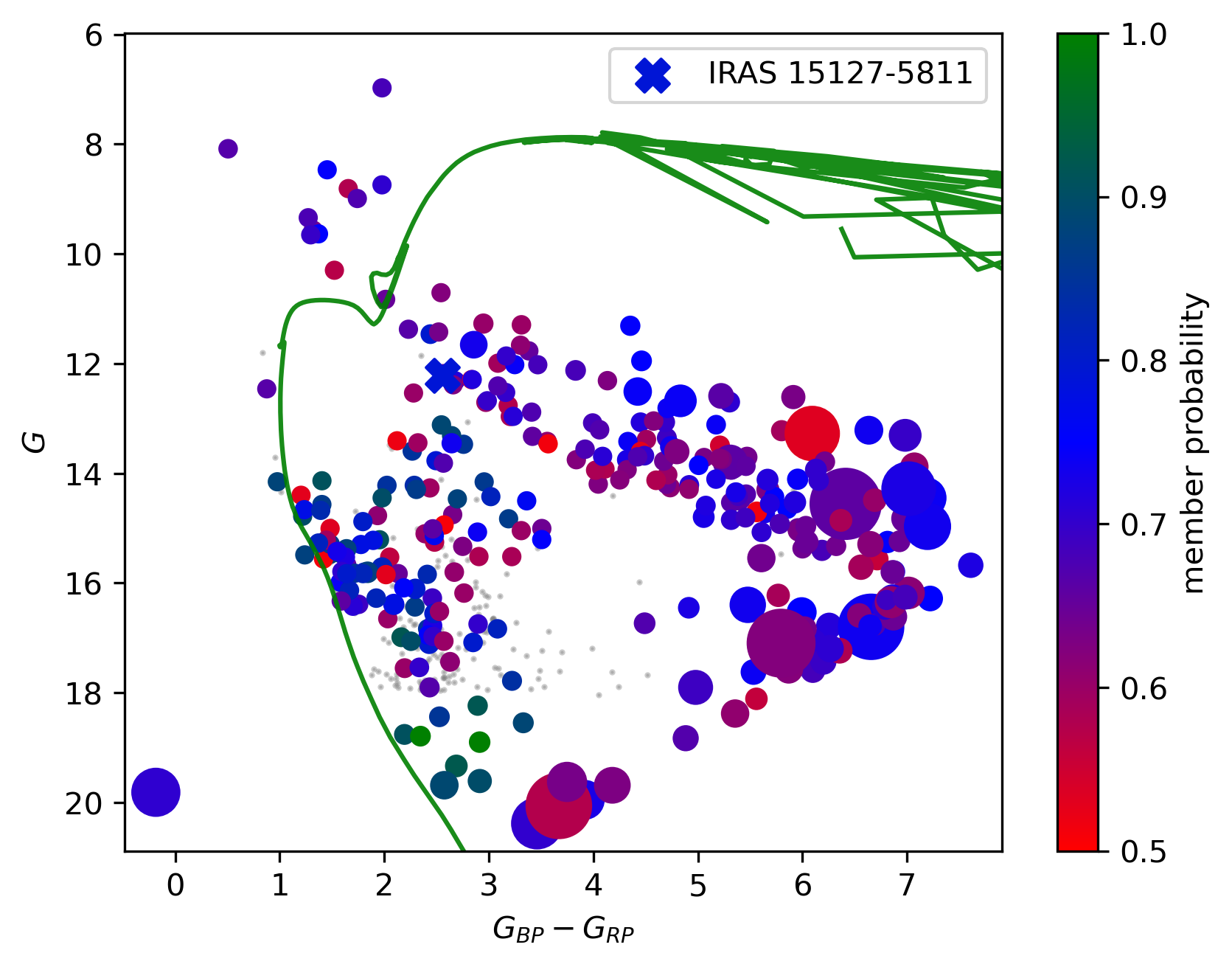}}
\caption{Same as Fig.~\ref{fig:HSC2686_isochrones} but for Lynga 3 member stars combined from all published members lists. The grey dots are stars with a membership probability of less than 0.5.} \label{fig:Lynga3_combined}
\end{figure}   

\clearpage

\bibliography{references}%{}

\begin{thebibliography}{}
\expandafter\ifx\csname natexlab\endcsname\relax\def\natexlab#1{#1}\fi
\providecommand{\url}[1]{\href{#1}{#1}}
\providecommand{\dodoi}[1]{doi:~\href{http://doi.org/#1}{\nolinkurl{#1}}}
\providecommand{\doeprint}[1]{\href{http://ascl.net/#1}{\nolinkurl{http://ascl.net/#1}}}
\providecommand{\doarXiv}[1]{\href{https://arxiv.org/abs/#1}{\nolinkurl{https://arxiv.org/abs/#1}}}

% type= article
\bibitem[{ {Astropy Collaboration} {et~al.}(2013){Astropy Collaboration},
  {Robitaille}, {Tollerud}, {Greenfield}, {Droettboom}, {Bray}, {Aldcroft},
  {Davis}, {Ginsburg}, {Price-Whelan}, {Kerzendorf}, {Conley}, {Crighton},
  {Barbary}, {Muna}, {Ferguson}, {Grollier}, {Parikh}, {Nair}, {Unther},
  {Deil}, {Woillez}, {Conseil}, {Kramer}, {Turner}, {Singer}, {Fox}, {Weaver},
  {Zabalza}, {Edwards}, {Azalee Bostroem}, {Burke}, {Casey}, {Crawford},
  {Dencheva}, {Ely}, {Jenness}, {Labrie}, {Lim}, {Pierfederici}, {Pontzen},
  {Ptak}, {Refsdal}, {Servillat}, \& {Streicher}}]{astropy:2013}
{Astropy Collaboration}, {Robitaille}, T.~P., {Tollerud}, E.~J., {et~al.} 2013,
  \bibinfo{title}{{Astropy: A community Python package for astronomy},} \aap,
  558, A33, \dodoi{10.1051/0004-6361/201322068}

% type= article
\bibitem[{ {Astropy Collaboration} {et~al.}(2018){Astropy Collaboration},
  {Price-Whelan}, {Sip{\H{o}}cz}, {G{\"u}nther}, {Lim}, {Crawford}, {Conseil},
  {Shupe}, {Craig}, {Dencheva}, {Ginsburg}, {Vand erPlas}, {Bradley},
  {P{\'e}rez-Su{\'a}rez}, {de Val-Borro}, {Aldcroft}, {Cruz}, {Robitaille},
  {Tollerud}, {Ardelean}, {Babej}, {Bach}, {Bachetti}, {Bakanov}, {Bamford},
  {Barentsen}, {Barmby}, {Baumbach}, {Berry}, {Biscani}, {Boquien}, {Bostroem},
  {Bouma}, {Brammer}, {Bray}, {Breytenbach}, {Buddelmeijer}, {Burke},
  {Calderone}, {Cano Rodr{\'\i}guez}, {Cara}, {Cardoso}, {Cheedella}, {Copin},
  {Corrales}, {Crichton}, {D'Avella}, {Deil}, {Depagne}, {Dietrich}, {Donath},
  {Droettboom}, {Earl}, {Erben}, {Fabbro}, {Ferreira}, {Finethy}, {Fox},
  {Garrison}, {Gibbons}, {Goldstein}, {Gommers}, {Greco}, {Greenfield},
  {Groener}, {Grollier}, {Hagen}, {Hirst}, {Homeier}, {Horton}, {Hosseinzadeh},
  {Hu}, {Hunkeler}, {Ivezi{\'c}}, {Jain}, {Jenness}, {Kanarek}, {Kendrew},
  {Kern}, {Kerzendorf}, {Khvalko}, {King}, {Kirkby}, {Kulkarni}, {Kumar},
  {Lee}, {Lenz}, {Littlefair}, {Ma}, {Macleod}, {Mastropietro}, {McCully},
  {Montagnac}, {Morris}, {Mueller}, {Mumford}, {Muna}, {Murphy}, {Nelson},
  {Nguyen}, {Ninan}, {N{\"o}the}, {Ogaz}, {Oh}, {Parejko}, {Parley}, {Pascual},
  {Patil}, {Patil}, {Plunkett}, {Prochaska}, {Rastogi}, {Reddy Janga},
  {Sabater}, {Sakurikar}, {Seifert}, {Sherbert}, {Sherwood-Taylor}, {Shih},
  {Sick}, {Silbiger}, {Singanamalla}, {Singer}, {Sladen}, {Sooley},
  {Sornarajah}, {Streicher}, {Teuben}, {Thomas}, {Tremblay}, {Turner},
  {Terr{\'o}n}, {van Kerkwijk}, {de la Vega}, {Watkins}, {Weaver}, {Whitmore},
  {Woillez}, {Zabalza}, \& {Astropy Contributors}}]{astropy:2018}
{Astropy Collaboration}, {Price-Whelan}, A.~M., {Sip{\H{o}}cz}, B.~M., {et~al.}
  2018, \bibinfo{title}{{The Astropy Project: Building an Open-science Project
  and Status of the v2.0 Core Package},} \aj, 156, 123,
  \dodoi{10.3847/1538-3881/aabc4f}

% type= article
\bibitem[{ {Astropy Collaboration} {et~al.}(2022{\natexlab{a}}){Astropy
  Collaboration}, {Price-Whelan}, {Lim}, {Earl}, {Starkman}, {Bradley},
  {Shupe}, {Patil}, {Corrales}, {Brasseur}, {N{\"o}the}, {Donath}, {Tollerud},
  {Morris}, {Ginsburg}, {Vaher}, {Weaver}, {Tocknell}, {Jamieson}, {van
  Kerkwijk}, {Robitaille}, {Merry}, {Bachetti}, {G{\"u}nther}, {Aldcroft},
  {Alvarado-Montes}, {Archibald}, {B{\'o}di}, {Bapat}, {Barentsen},
  {Baz{\'a}n}, {Biswas}, {Boquien}, {Burke}, {Cara}, {Cara}, {Conroy},
  {Conseil}, {Craig}, {Cross}, {Cruz}, {D'Eugenio}, {Dencheva}, {Devillepoix},
  {Dietrich}, {Eigenbrot}, {Erben}, {Ferreira}, {Foreman-Mackey}, {Fox},
  {Freij}, {Garg}, {Geda}, {Glattly}, {Gondhalekar}, {Gordon}, {Grant},
  {Greenfield}, {Groener}, {Guest}, {Gurovich}, {Handberg}, {Hart},
  {Hatfield-Dodds}, {Homeier}, {Hosseinzadeh}, {Jenness}, {Jones}, {Joseph},
  {Kalmbach}, {Karamehmetoglu}, {Ka{\l}uszy{\'n}ski}, {Kelley}, {Kern},
  {Kerzendorf}, {Koch}, {Kulumani}, {Lee}, {Ly}, {Ma}, {MacBride}, {Maljaars},
  {Muna}, {Murphy}, {Norman}, {O'Steen}, {Oman}, {Pacifici}, {Pascual},
  {Pascual-Granado}, {Patil}, {Perren}, {Pickering}, {Rastogi}, {Roulston},
  {Ryan}, {Rykoff}, {Sabater}, {Sakurikar}, {Salgado}, {Sanghi}, {Saunders},
  {Savchenko}, {Schwardt}, {Seifert-Eckert}, {Shih}, {Jain}, {Shukla}, {Sick},
  {Simpson}, {Singanamalla}, {Singer}, {Singhal}, {Sinha}, {Sip{\H{o}}cz},
  {Spitler}, {Stansby}, {Streicher}, {{\v{S}}umak}, {Swinbank}, {Taranu},
  {Tewary}, {Tremblay}, {de Val-Borro}, {Van Kooten}, {Vasovi{\'c}}, {Verma},
  {de Miranda Cardoso}, {Williams}, {Wilson}, {Winkel}, {Wood-Vasey}, {Xue},
  {Yoachim}, {Zhang}, {Zonca}, \& {Astropy Project Contributors}}]{astropy}
{Astropy Collaboration}, {Price-Whelan}, A.~M., {Lim}, P.~L., {et~al.}
  2022{\natexlab{a}}, \bibinfo{title}{{The Astropy Project: Sustaining and
  Growing a Community-oriented Open-source Project and the Latest Major Release
  (v5.0) of the Core Package},} \apj, 935, 167,
  \dodoi{10.3847/1538-4357/ac7c74}

% type= article
\bibitem[{ {Astropy Collaboration} {et~al.}(2022{\natexlab{b}}){Astropy
  Collaboration}, {Price-Whelan}, {Lim}, {Earl}, {Starkman}, {Bradley},
  {Shupe}, {Patil}, {Corrales}, {Brasseur}, {N{"o}the}, {Donath}, {Tollerud},
  {Morris}, {Ginsburg}, {Vaher}, {Weaver}, {Tocknell}, {Jamieson}, {van
  Kerkwijk}, {Robitaille}, {Merry}, {Bachetti}, {G{"u}nther}, {Aldcroft},
  {Alvarado-Montes}, {Archibald}, {B{'o}di}, {Bapat}, {Barentsen}, {Baz{'a}n},
  {Biswas}, {Boquien}, {Burke}, {Cara}, {Cara}, {Conroy}, {Conseil}, {Craig},
  {Cross}, {Cruz}, {D'Eugenio}, {Dencheva}, {Devillepoix}, {Dietrich},
  {Eigenbrot}, {Erben}, {Ferreira}, {Foreman-Mackey}, {Fox}, {Freij}, {Garg},
  {Geda}, {Glattly}, {Gondhalekar}, {Gordon}, {Grant}, {Greenfield}, {Groener},
  {Guest}, {Gurovich}, {Handberg}, {Hart}, {Hatfield-Dodds}, {Homeier},
  {Hosseinzadeh}, {Jenness}, {Jones}, {Joseph}, {Kalmbach}, {Karamehmetoglu},
  {Ka{l}uszy{'n}ski}, {Kelley}, {Kern}, {Kerzendorf}, {Koch}, {Kulumani},
  {Lee}, {Ly}, {Ma}, {MacBride}, {Maljaars}, {Muna}, {Murphy}, {Norman},
  {O'Steen}, {Oman}, {Pacifici}, {Pascual}, {Pascual-Granado}, {Patil},
  {Perren}, {Pickering}, {Rastogi}, {Roulston}, {Ryan}, {Rykoff}, {Sabater},
  {Sakurikar}, {Salgado}, {Sanghi}, {Saunders}, {Savchenko}, {Schwardt},
  {Seifert-Eckert}, {Shih}, {Jain}, {Shukla}, {Sick}, {Simpson},
  {Singanamalla}, {Singer}, {Singhal}, {Sinha}, {Sip{H{o}}cz}, {Spitler},
  {Stansby}, {Streicher}, {{{S}}umak}, {Swinbank}, {Taranu}, {Tewary},
  {Tremblay}, {Val-Borro}, {Van Kooten}, {Vasovi{'c}}, {Verma}, {de Miranda
  Cardoso}, {Williams}, {Wilson}, {Winkel}, {Wood-Vasey}, {Xue}, {Yoachim},
  {Zhang}, {Zonca}, \& {Astropy Project Contributors}}]{astropy:2022}
{Astropy Collaboration}, {Price-Whelan}, A.~M., {Lim}, P.~L., {et~al.}
  2022{\natexlab{b}}, \bibinfo{title}{{The Astropy Project: Sustaining and
  Growing a Community-oriented Open-source Project and the Latest Major Release
  (v5.0) of the Core Package},} \apj, 935, 167,
  \dodoi{10.3847/1538-4357/ac7c74}

% type= article
\bibitem[{C. {Badenes} {et~al.}(2015){Badenes}, {Maoz}, \&
  {Ciardullo}}]{Badenes2015}
{Badenes}, C., {Maoz}, D., \& {Ciardullo}, R. 2015, \bibinfo{title}{{The
  Progenitors and Lifetimes of Planetary Nebulae},} \apjl, 804, L25,
  \dodoi{10.1088/2041-8205/804/1/L25}

% type= article
\bibitem[{E. {Bertin} \& S. {Arnouts}(1996){Bertin} \& {Arnouts}}]{Bertin1996}
{Bertin}, E., \& {Arnouts}, S. 1996, \bibinfo{title}{{SExtractor: Software for
  source extraction.},} \aaps, 117, 393, \dodoi{10.1051/aas:1996164}

% type= article
\bibitem[{C. {Bonatto} \& E. {Bica}(2009){Bonatto} \& {Bica}}]{Bonatto2009}
{Bonatto}, C., \& {Bica}, E. 2009, \bibinfo{title}{{Probing the age and
  structure of the nearby very young open clusters NGC2244 and 2239},} \mnras,
  394, 2127, \dodoi{10.1111/j.1365-2966.2009.14474.x}

% type= article
\bibitem[{A. {Bressan} {et~al.}(2012{\natexlab{a}}){Bressan}, {Marigo},
  {Girardi}, {Salasnich}, {Dal Cero}, {Rubele}, \&
  {Nanni}}]{2012MNRAS.427..127B}
{Bressan}, A., {Marigo}, P., {Girardi}, L., {et~al.} 2012{\natexlab{a}},
  \bibinfo{title}{{PARSEC: stellar tracks and isochrones with the PAdova and
  TRieste Stellar Evolution Code},} \mnras, 427, 127,
  \dodoi{10.1111/j.1365-2966.2012.21948.x}

% type= article
\bibitem[{A. {Bressan} {et~al.}(2012{\natexlab{b}}){Bressan}, {Marigo},
  {Girardi}, {Salasnich}, {Dal Cero}, {Rubele}, \& {Nanni}}]{Bressan2012}
{Bressan}, A., {Marigo}, P., {Girardi}, L., {et~al.} 2012{\natexlab{b}},
  \bibinfo{title}{{PARSEC: stellar tracks and isochrones with the PAdova and
  TRieste Stellar Evolution Code},} \mnras, 427, 127,
  \dodoi{10.1111/j.1365-2966.2012.21948.x}

% type= article
\bibitem[{T. {Cantat-Gaudin} \& F. {Anders}(2020){Cantat-Gaudin} \&
  {Anders}}]{Cantat-Gaudin2020}
{Cantat-Gaudin}, T., \& {Anders}, F. 2020, \bibinfo{title}{{Clusters and
  mirages: cataloguing stellar aggregates in the Milky Way},} \aap, 633, A99,
  \dodoi{10.1051/0004-6361/201936691}

% type= article
\bibitem[{T. {Cantat-Gaudin} {et~al.}(2018){Cantat-Gaudin}, {Jordi},
  {Vallenari}, {Bragaglia}, {Balaguer-N{\'u}{\~n}ez}, {Soubiran}, {Bossini},
  {Moitinho}, {Castro-Ginard}, {Krone-Martins}, {Casamiquela}, {Sordo}, \&
  {Carrera}}]{Cantat-Gaudin2018}
{Cantat-Gaudin}, T., {Jordi}, C., {Vallenari}, A., {et~al.} 2018,
  \bibinfo{title}{{A Gaia DR2 view of the open cluster population in the Milky
  Way},} \aap, 618, A93, \dodoi{10.1051/0004-6361/201833476}

% type= article
\bibitem[{T. {Cantat-Gaudin} {et~al.}(2019){Cantat-Gaudin}, {Krone-Martins},
  {Sedaghat}, {Farahi}, {de Souza}, {Skalidis}, {Malz}, {Mac{\^e}do}, {Moews},
  {Jordi}, {Moitinho}, {Castro-Ginard}, {Ishida}, {Heneka}, {Boucaud}, \&
  {Trindade}}]{Cantat-Gaudin2019}
{Cantat-Gaudin}, T., {Krone-Martins}, A., {Sedaghat}, N., {et~al.} 2019,
  \bibinfo{title}{{Gaia DR2 unravels incompleteness of nearby cluster
  population: new open clusters in the direction of Perseus},} \aap, 624, A126,
  \dodoi{10.1051/0004-6361/201834453}

% type= article
\bibitem[{G. {Carraro} {et~al.}(2006){Carraro}, {Janes}, {Costa}, \&
  {M{\'e}ndez}}]{Carraro2006}
{Carraro}, G., {Janes}, K.~A., {Costa}, E., \& {M{\'e}ndez}, R.~A. 2006,
  \bibinfo{title}{{Photometry of seven overlooked open clusters in the first
  and fourth Galactic quadrants},} \mnras, 368, 1078,
  \dodoi{10.1111/j.1365-2966.2006.10187.x}

% type= article
\bibitem[{R. {Carrera} {et~al.}(2019){Carrera}, {Bragaglia}, {Cantat-Gaudin},
  {Vallenari}, {Balaguer-N{\'u}{\~n}ez}, {Bossini}, {Casamiquela}, {Jordi},
  {Sordo}, \& {Soubiran}}]{Carrera2019}
{Carrera}, R., {Bragaglia}, A., {Cantat-Gaudin}, T., {et~al.} 2019,
  \bibinfo{title}{{Open clusters in APOGEE and GALAH. Combining Gaia and
  ground-based spectroscopic surveys},} \aap, 623, A80,
  \dodoi{10.1051/0004-6361/201834546}

% type= article
\bibitem[{L. {Cavallo} {et~al.}(2024){Cavallo}, {Spina}, {Carraro}, {Magrini},
  {Poggio}, {Cantat-Gaudin}, {Pasquato}, {Lucatello}, {Ortolani}, \&
  {Schiappacasse-Ulloa}}]{Cavallo2024}
{Cavallo}, L., {Spina}, L., {Carraro}, G., {et~al.} 2024,
  \bibinfo{title}{{Parameter Estimation for Open Clusters using an Artificial
  Neural Network with a QuadTree-based Feature Extractor},} \aj, 167, 12,
  \dodoi{10.3847/1538-3881/ad07e5}

% type= article
\bibitem[{H. {Chi} {et~al.}(2023){Chi}, {Wang}, {Wang}, {Deng}, \&
  {Li}}]{Chi2023}
{Chi}, H., {Wang}, F., {Wang}, W., {Deng}, H., \& {Li}, Z. 2023,
  \bibinfo{title}{{Blind Search of the Solar Neighborhood Galactic Disk within
  5 kpc: 1179 New Star Clusters Found in Gaia DR3},} \apjs, 266, 36,
  \dodoi{10.3847/1538-4365/accb50}

% type= article
\bibitem[{N. {Chornay} \& N.~A. {Walton}(2021){Chornay} \&
  {Walton}}]{Chornay2021}
{Chornay}, N., \& {Walton}, N.~A. 2021, \bibinfo{title}{{One star, two star,
  red star, blue star: an updated planetary nebula central star distance
  catalogue from Gaia EDR3},} \aap, 656, A110,
  \dodoi{10.1051/0004-6361/202142008}

% type= article
\bibitem[{W.~S. {Dias} {et~al.}(2002){Dias}, {Alessi}, {Moitinho}, \&
  {L{\'e}pine}}]{Dias2002}
{Dias}, W.~S., {Alessi}, B.~S., {Moitinho}, A., \& {L{\'e}pine}, J.~R.~D. 2002,
  \bibinfo{title}{{New catalogue of optically visible open clusters and
  candidates},} \aap, 389, 871, \dodoi{10.1051/0004-6361:20020668}

% type= article
\bibitem[{W.~S. {Dias} {et~al.}(2018){Dias}, {Monteiro}, \&
  {Assafin}}]{Dias2018}
{Dias}, W.~S., {Monteiro}, H., \& {Assafin}, M. 2018, \bibinfo{title}{{Update
  of membership and mean proper motion of open clusters from UCAC5 catalogue},}
  \mnras, 478, 5184, \dodoi{10.1093/mnras/sty1456}

% type= article
\bibitem[{W.~S. {Dias} {et~al.}(2014){Dias}, {Monteiro}, {Caetano},
  {L{\'e}pine}, {Assafin}, \& {Oliveira}}]{Dias2014}
{Dias}, W.~S., {Monteiro}, H., {Caetano}, T.~C., {et~al.} 2014,
  \bibinfo{title}{{Proper motions of the optically visible open clusters based
  on the UCAC4 catalog},} \aap, 564, A79, \dodoi{10.1051/0004-6361/201323226}

% type= article
\bibitem[{W.~S. {Dias} {et~al.}(2021){Dias}, {Monteiro}, {Moitinho},
  {L{\'e}pine}, {Carraro}, {Paunzen}, {Alessi}, \& {Villela}}]{Dias2021}
{Dias}, W.~S., {Monteiro}, H., {Moitinho}, A., {et~al.} 2021,
  \bibinfo{title}{{Updated parameters of 1743 open clusters based on Gaia
  DR2},} \mnras, 504, 356, \dodoi{10.1093/mnras/stab770}

% type= article
\bibitem[{J.~E. {Drew} {et~al.}(2014){Drew}, {Gonzalez-Solares}, {Greimel},
  {Irwin}, {K{\"u}pc{\"u} Yoldas}, {Lewis}, {Barentsen}, {Eisl{\"o}ffel},
  {Farnhill}, {Martin}, {Walsh}, {Walton}, {Mohr-Smith}, {Raddi}, {Sale},
  {Wright}, {Groot}, {Barlow}, {Corradi}, {Drake}, {Fabregat}, {Frew},
  {G{\"a}nsicke}, {Knigge}, {Mampaso}, {Morris}, {Naylor}, {Parker},
  {Phillipps}, {Ruhland}, {Steeghs}, {Unruh}, {Vink}, {Wesson}, \&
  {Zijlstra}}]{Drew2014}
{Drew}, J.~E., {Gonzalez-Solares}, E., {Greimel}, R., {et~al.} 2014,
  \bibinfo{title}{{The VST Photometric H{\ensuremath{\alpha}} Survey of the
  Southern Galactic Plane and Bulge (VPHAS+)},} \mnras, 440, 2036,
  \dodoi{10.1093/mnras/stu394}

% type= article
\bibitem[{G.~G. {Fazio} {et~al.}(2004){Fazio}, {Hora}, {Allen}, {Ashby},
  {Barmby}, {Deutsch}, {Huang}, {Kleiner}, {Marengo}, {Megeath}, {Melnick},
  {Pahre}, {Patten}, {Polizotti}, {Smith}, {Taylor}, {Wang}, {Willner},
  {Hoffmann}, {Pipher}, {Forrest}, {McMurty}, {McCreight}, {McKelvey},
  {McMurray}, {Koch}, {Moseley}, {Arendt}, {Mentzell}, {Marx}, {Losch},
  {Mayman}, {Eichhorn}, {Krebs}, {Jhabvala}, {Gezari}, {Fixsen}, {Flores},
  {Shakoorzadeh}, {Jungo}, {Hakun}, {Workman}, {Karpati}, {Kichak}, {Whitley},
  {Mann}, {Tollestrup}, {Eisenhardt}, {Stern}, {Gorjian}, {Bhattacharya},
  {Carey}, {Nelson}, {Glaccum}, {Lacy}, {Lowrance}, {Laine}, {Reach},
  {Stauffer}, {Surace}, {Wilson}, {Wright}, {Hoffman}, {Domingo}, \&
  {Cohen}}]{Fazio2004}
{Fazio}, G.~G., {Hora}, J.~L., {Allen}, L.~E., {et~al.} 2004,
  \bibinfo{title}{{The Infrared Array Camera (IRAC) for the Spitzer Space
  Telescope},} \apjs, 154, 10, \dodoi{10.1086/422843}

% type= article
\bibitem[{V. {Fragkou} {et~al.}(2026){Fragkou}, {Parker}, \&
  {Gon{\c{c}}alves}}]{2026ApJ...996...90F}
{Fragkou}, V., {Parker}, Q.~A., \& {Gon{\c{c}}alves}, D.~R. 2026,
  \bibinfo{title}{{PHR J1724-3859: A Bipolar Planetary Nebula in Open Cluster
  Trumpler 25},} \apj, 996, 90, \dodoi{10.3847/1538-4357/ae1f0c}

% type= article
\bibitem[{V. {Fragkou} {et~al.}(2022{\natexlab{a}}){Fragkou}, {Parker},
  {Zijlstra}, {Crause}, {Sabin}, \& {V{\'a}zquez}}]{Fragkou2022a}
{Fragkou}, V., {Parker}, Q.~A., {Zijlstra}, A.~A., {et~al.} 2022{\natexlab{a}},
  \bibinfo{title}{{Further Studies of the Association of Planetary Nebula BMP
  J16135406 with Galactic Open Cluster NGC 6067},} Galaxies, 10, 44,
  \dodoi{10.3390/galaxies10020044}

% type= article
\bibitem[{V. {Fragkou} {et~al.}(2022{\natexlab{b}}){Fragkou}, {Parker},
  {Zijlstra}, {V{\'a}zquez}, {Sabin}, \& {Rechy-Garcia}}]{Fragkou2022}
{Fragkou}, V., {Parker}, Q.~A., {Zijlstra}, A.~A., {et~al.} 2022{\natexlab{b}},
  \bibinfo{title}{{The Planetary Nebula in the 500 Myr Old Open Cluster M37},}
  \apjl, 935, L35, \dodoi{10.3847/2041-8213/ac88c1}

% type= article
\bibitem[{V. {Fragkou} {et~al.}(2025){Fragkou}, {V{\'a}zquez}, {Parker},
  {Gon{\c{c}}alves}, \& {Lomel{\'\i}-N{\'u}{\~n}ez}}]{Fragkou2025}
{Fragkou}, V., {V{\'a}zquez}, R., {Parker}, Q.~A., {Gon{\c{c}}alves}, D.~R., \&
  {Lomel{\'\i}-N{\'u}{\~n}ez}, L. 2025, \bibinfo{title}{{The physical
  association of planetary nebula NGC 2818 with open cluster NGC 2818A},} \aap,
  696, A146, \dodoi{10.1051/0004-6361/202453031}

% type= article
\bibitem[{X. {Fu} {et~al.}(2022){Fu}, {Bragaglia}, {Liu}, {Zhang}, {Xu},
  {Wang}, {Zhang}, {Zhong}, {Chang}, {Li}, {Chen}, {Chen}, {Wang}, {Gjergo},
  {Wang}, {Yue}, \& {Zhang}}]{Fu2022}
{Fu}, X., {Bragaglia}, A., {Liu}, C., {et~al.} 2022, \bibinfo{title}{{LAMOST
  meets Gaia: The Galactic open clusters},} \aap, 668, A4,
  \dodoi{10.1051/0004-6361/202243590}

% type= article
\bibitem[{I. {Gonz{\'a}lez-Santamar{\'\i}a}
  {et~al.}(2021){Gonz{\'a}lez-Santamar{\'\i}a}, {Manteiga}, {Manchado}, {Ulla},
  {Dafonte}, \& {L{\'o}pez Varela}}]{Gonzalez-Santamaria2021}
{Gonz{\'a}lez-Santamar{\'\i}a}, I., {Manteiga}, M., {Manchado}, A., {et~al.}
  2021, \bibinfo{title}{{Planetary nebulae in Gaia EDR3: Central star
  identification, properties, and binarity},} \aap, 656, A51,
  \dodoi{10.1051/0004-6361/202141916}

% type= article
\bibitem[{V.~V. {Gvaramadze} {et~al.}(2015){Gvaramadze}, {Kniazev},
  {Bestenlehner}, {Bodensteiner}, {Langer}, {Greiner}, {Grebel}, {Berdnikov},
  \& {Beletsky}}]{Gvaramadze2015}
{Gvaramadze}, V.~V., {Kniazev}, A.~Y., {Bestenlehner}, J.~M., {et~al.} 2015,
  \bibinfo{title}{{The blue supergiant MN18 and its bipolar circumstellar
  nebula},} \mnras, 454, 219, \dodoi{10.1093/mnras/stv1995}

% type= article
\bibitem[{E.~L. {Hunt} \& S. {Reffert}(2023){Hunt} \& {Reffert}}]{Hunt2023}
{Hunt}, E.~L., \& {Reffert}, S. 2023, \bibinfo{title}{{Improving the open
  cluster census. II. An all-sky cluster catalogue with Gaia DR3},} \aap, 673,
  A114, \dodoi{10.1051/0004-6361/202346285}

% type= article
\bibitem[{E.~L. {Hunt} \& S. {Reffert}(2024){Hunt} \& {Reffert}}]{Hunt2024}
{Hunt}, E.~L., \& {Reffert}, S. 2024, \bibinfo{title}{{Improving the open
  cluster census. III. Using cluster masses, radii, and dynamics to create a
  cleaned open cluster catalogue},} \aap, 686, A42,
  \dodoi{10.1051/0004-6361/202348662}

% type= article
\bibitem[{A. {Just} {et~al.}(2023){Just}, {Piskunov}, {Klos}, {Kovaleva}, \&
  {Polyachenko}}]{Just2023}
{Just}, A., {Piskunov}, A.~E., {Klos}, J.~H., {Kovaleva}, D.~A., \&
  {Polyachenko}, E.~V. 2023, \bibinfo{title}{{Global survey of star clusters in
  the Milky Way. VII. Tidal parameters and mass function},} \aap, 672, A187,
  \dodoi{10.1051/0004-6361/202244723}

% type= article
\bibitem[{N.~V. {Kharchenko} {et~al.}(2013){Kharchenko}, {Piskunov},
  {Schilbach}, {R{\"o}ser}, \& {Scholz}}]{Kharchenko2013}
{Kharchenko}, N.~V., {Piskunov}, A.~E., {Schilbach}, E., {R{\"o}ser}, S., \&
  {Scholz}, R.~D. 2013, \bibinfo{title}{{Global survey of star clusters in the
  Milky Way. II. The catalogue of basic parameters},} \aap, 558, A53,
  \dodoi{10.1051/0004-6361/201322302}

% type= article
\bibitem[{M. {Kounkel} {et~al.}(2020){Kounkel}, {Covey}, \&
  {Stassun}}]{Kounkel2020}
{Kounkel}, M., {Covey}, K., \& {Stassun}, K.~G. 2020,
  \bibinfo{title}{{Untangling the Galaxy. II. Structure within 3 kpc},} \aj,
  160, 279, \dodoi{10.3847/1538-3881/abc0e6}

% type= article
\bibitem[{K.~B. {Kwitter} \& R.~B.~C. {Henry}(2022){Kwitter} \&
  {Henry}}]{Kwitter2022}
{Kwitter}, K.~B., \& {Henry}, R.~B.~C. 2022, \bibinfo{title}{{Planetary
  Nebulae: Sources of Enlightenment},} \pasp, 134, 022001,
  \dodoi{10.1088/1538-3873/ac32b1}

% type= article
\bibitem[{D. {Lang} {et~al.}(2010){Lang}, {Hogg}, {Mierle}, {Blanton}, \&
  {Roweis}}]{Lang2010}
{Lang}, D., {Hogg}, D.~W., {Mierle}, K., {Blanton}, M., \& {Roweis}, S. 2010,
  \bibinfo{title}{{Astrometry.net: Blind Astrometric Calibration of Arbitrary
  Astronomical Images},} \aj, 139, 1782, \dodoi{10.1088/0004-6256/139/5/1782}

% type= article
\bibitem[{G. {Lynga}(1964){Lynga}}]{Lynga1964}
{Lynga}, G. 1964, \bibinfo{title}{{Studies of the Milky Way from Centaurus to
  Norma. I. UBV photometry.},} Meddelanden fran Lunds Astronomiska
  Observatorium Serie II, 139, 1

% type= article
\bibitem[{G. Lynga(1987)Lynga}]{Lynga1987}
Lynga, G. 1987, \bibinfo{title}{Catalogue of Open Cluster Data (5th Ed.),}
  VII/92A.
\newblock \url{https://cdsarc.cds.unistra.fr/viz-bin/cat/VII/92A#/article}

% type= article
\bibitem[{H.~B. Mann \& D.~R. Whitney(1947)Mann \& Whitney}]{Mann1947}
Mann, H.~B., \& Whitney, D.~R. 1947, \bibinfo{title}{{On a Test of Whether one
  of Two Random Variables is Stochastically Larger than the Other},} The Annals
  of Mathematical Statistics, 18, 50 , \dodoi{10.1214/aoms/1177730491}

% type= article
\bibitem[{J.-C. {Mermilliod} {et~al.}(2009){Mermilliod}, {Mayor}, \&
  {Udry}}]{Mermilliod2009}
{Mermilliod}, J.-C., {Mayor}, M., \& {Udry}, S. 2009,
  \bibinfo{title}{{Catalogues of radial and rotational velocities of 1253 F-K
  dwarfs in 13 nearby open clusters},} \aap, 498, 949,
  \dodoi{10.1051/0004-6361/200810244}

% type= article
\bibitem[{M.~M. {Miller Bertolami}(2016){Miller
  Bertolami}}]{2016A&A...588A..25M}
{Miller Bertolami}, M.~M. 2016, \bibinfo{title}{{New models for the evolution
  of post-asymptotic giant branch stars and central stars of planetary
  nebulae},} \aap, 588, A25, \dodoi{10.1051/0004-6361/201526577}

% type= article
\bibitem[{Q.~A. {Parker}(2022){Parker}}]{2022FrASS...9.5287P}
{Parker}, Q.~A. 2022, \bibinfo{title}{{Planetary nebulae and how to find them:
  A concise review},} Frontiers in Astronomy and Space Sciences, 9, 895287,
  \dodoi{10.3389/fspas.2022.895287}

% type= inproceedings
\bibitem[{Q.~A. {Parker} {et~al.}(2016){Parker}, {Boji{\v{c}}i{\'c}}, \&
  {Frew}}]{Parker2016}
{Parker}, Q.~A., {Boji{\v{c}}i{\'c}}, I.~S., \& {Frew}, D.~J. 2016,
  \bibinfo{title}{{HASH: the Hong Kong/AAO/Strasbourg H{\ensuremath{\alpha}}
  planetary nebula database},} in Journal of Physics Conference Series, Vol.
  728, Journal of Physics Conference Series (IOP), 032008,
  \dodoi{10.1088/1742-6596/728/3/032008}

% type= article
\bibitem[{Q.~A. {Parker} {et~al.}(2011){Parker}, {Frew}, {Miszalski},
  {Kovacevic}, {Frinchaboy}, {Dobbie}, \& {K{\"o}ppen}}]{Parker2011}
{Parker}, Q.~A., {Frew}, D.~J., {Miszalski}, B., {et~al.} 2011,
  \bibinfo{title}{{PHR 1315-6555: a bipolar planetary nebula in the compact
  Hyades-age open cluster ESO 96-SC04},} \mnras, 413, 1835,
  \dodoi{10.1111/j.1365-2966.2011.18259.x}

% type= article
\bibitem[{Q.~A. {Parker} {et~al.}(2022){Parker}, {Xiang}, \&
  {Ritter}}]{Parker2022}
{Parker}, Q.~A., {Xiang}, Z., \& {Ritter}, A. 2022, \bibinfo{title}{{A
  Preliminary Investigation of CSPN in the HASH Database},} Galaxies, 10, 32,
  \dodoi{10.3390/galaxies10010032}

% type= article
\bibitem[{G.~I. {Perren} {et~al.}(2023){Perren}, {Pera}, {Navone}, \&
  {V{\'a}zquez}}]{Perren2023}
{Perren}, G.~I., {Pera}, M.~S., {Navone}, H.~D., \& {V{\'a}zquez}, R.~A. 2023,
  \bibinfo{title}{{The Unified Cluster Catalogue: towards a comprehensive and
  homogeneous data base of stellar clusters},} \mnras, 526, 4107,
  \dodoi{10.1093/mnras/stad2826}

% type= article
\bibitem[{O. {Plevne} \& F. {Akbaba}(2026){Plevne} \& {Akbaba}}]{Plevne2026}
{Plevne}, O., \& {Akbaba}, F. 2026, \bibinfo{title}{{Exploring open cluster
  properties with Gaia and LAMOST},} \doarXiv{2605.23802}

% type= article
\bibitem[{L. {Prisinzano} {et~al.}(2019){Prisinzano}, {Damiani}, {Kalari},
  {Jeffries}, {Bonito}, {Micela}, {Wright}, {Jackson}, {Tognelli}, {Guarcello},
  {Vink}, {Klutsch}, {Jim{\'e}nez-Esteban}, {Roccatagliata},
  {Tautvai{\v{s}}ien{\.{e}}}, {Gilmore}, {Randich}, {Alfaro}, {Flaccomio},
  {Koposov}, {Lanzafame}, {Pancino}, {Bergemann}, {Carraro}, {Franciosini},
  {Frasca}, {Gonneau}, {Hourihane}, {Jofr{\'e}}, {Lewis}, {Magrini}, {Monaco},
  {Morbidelli}, {Sacco}, {Worley}, \& {Zaggia}}]{Prisinzano2019}
{Prisinzano}, L., {Damiani}, F., {Kalari}, V., {et~al.} 2019,
  \bibinfo{title}{{The Gaia-ESO Survey: Age spread in the star forming region
  NGC 6530 from the HR diagram and gravity indicators},} \aap, 623, A159,
  \dodoi{10.1051/0004-6361/201834870}

% type= article
\bibitem[{S. {Randich} {et~al.}(2022){Randich}, {Gilmore}, {Magrini}, {Sacco},
  {Jackson}, {Jeffries}, {Worley}, {Hourihane}, {Gonneau}, {Viscasillas
  Vazquez}, {Franciosini}, {Lewis}, {Alfaro}, {Allende Prieto}, {Bensby},
  {Blomme}, {Bragaglia}, {Flaccomio}, {Fran{\c{c}}ois}, {Irwin}, {Koposov},
  {Korn}, {Lanzafame}, {Pancino}, {Recio-Blanco}, {Smiljanic}, {Van Eck},
  {Zwitter}, {Asplund}, {Bonifacio}, {Feltzing}, {Binney}, {Drew}, {Ferguson},
  {Micela}, {Negueruela}, {Prusti}, {Rix}, {Vallenari}, {Bayo}, {Bergemann},
  {Biazzo}, {Carraro}, {Casey}, {Damiani}, {Frasca}, {Heiter}, {Hill},
  {Jofr{\'e}}, {de Laverny}, {Lind}, {Marconi}, {Martayan}, {Masseron},
  {Monaco}, {Morbidelli}, {Prisinzano}, {Sbordone}, {Sousa}, {Zaggia},
  {Adibekyan}, {Bonito}, {Caffau}, {Daflon}, {Feuillet}, {Gebran}, {Gonzalez
  Hernandez}, {Guiglion}, {Herrero}, {Lobel}, {Maiz Apellaniz}, {Merle},
  {Mikolaitis}, {Montes}, {Morel}, {Soubiran}, {Spina}, {Tabernero},
  {Tautvai{\v{s}}iene}, {Traven}, {Valentini}, {Van der Swaelmen}, {Villanova},
  {Wright}, {Abbas}, {Aguirre B{\o}rsen-Koch}, {Alves}, {Balaguer-Nunez},
  {Barklem}, {Barrado}, {Berlanas}, {Binks}, {Bressan}, {Capuzzo-Dolcetta},
  {Casagrande}, {Casamiquela}, {Collins}, {D'Orazi}, {Dantas}, {Debattista},
  {Delgado-Mena}, {Di Marcantonio}, {Drazdauskas}, {Evans}, {Famaey},
  {Franchini}, {Fr{\'e}mat}, {Friel}, {Fu}, {Geisler}, {Gerhard}, {Gonzalez
  Solares}, {Grebel}, {Gutierrez Albarran}, {Hatzidimitriou}, {Held},
  {Jim{\'e}nez-Esteban}, {J{\"o}nsson}, {Jordi}, {Khachaturyants},
  {Kordopatis}, {Kos}, {Lagarde}, {Mahy}, {Mapelli}, {Marfil}, {Martell},
  {Messina}, {Miglio}, {Minchev}, {Moitinho}, {Montalban}, {Monteiro},
  {Morossi}, {Mowlavi}, {Mucciarelli}, {Murphy}, {Nardetto}, {Ortolani},
  {Paletou}, {Palou{\v{s}}}, {Paunzen}, {Pickering}, {Quirrenbach}, {Re
  Fiorentin}, {Read}, {Romano}, {Ryde}, {Sanna}, {Santos}, {Seabroke},
  {Spagna}, {Steinmetz}, {Stonkut{\'e}}, {Sutorius}, {Th{\'e}venin}, {Tosi},
  {Tsantaki}, {Vink}, {Wright}, {Wyse}, {Zoccali}, {Zorec}, {Zucker}, \&
  {Walton}}]{Randich2022}
{Randich}, S., {Gilmore}, G., {Magrini}, L., {et~al.} 2022,
  \bibinfo{title}{{The Gaia-ESO Public Spectroscopic Survey: Implementation,
  data products, open cluster survey, science, and legacy},} \aap, 666, A121,
  \dodoi{10.1051/0004-6361/202243141}

% type= article
\bibitem[{L. {Sampedro} {et~al.}(2017){Sampedro}, {Dias}, {Alfaro}, {Monteiro},
  \& {Molino}}]{Sampedro2017}
{Sampedro}, L., {Dias}, W.~S., {Alfaro}, E.~J., {Monteiro}, H., \& {Molino}, A.
  2017, \bibinfo{title}{{A multimembership catalogue for 1876 open clusters
  using UCAC4 data},} \mnras, 470, 3937, \dodoi{10.1093/mnras/stx1485}

% type= article
\bibitem[{M.~W. {Werner} {et~al.}(2004){Werner}, {Roellig}, {Low}, {Rieke},
  {Rieke}, {Hoffmann}, {Young}, {Houck}, {Brandl}, {Fazio}, {Hora}, {Gehrz},
  {Helou}, {Soifer}, {Stauffer}, {Keene}, {Eisenhardt}, {Gallagher}, {Gautier},
  {Irace}, {Lawrence}, {Simmons}, {Van Cleve}, {Jura}, {Wright}, \&
  {Cruikshank}}]{Werner2004}
{Werner}, M.~W., {Roellig}, T.~L., {Low}, F.~J., {et~al.} 2004,
  \bibinfo{title}{{The Spitzer Space Telescope Mission},} \apjs, 154, 1,
  \dodoi{10.1086/422992}

% type= article
\bibitem[{J. {Zhong} {et~al.}(2020){Zhong}, {Chen}, {Wu}, {Li}, {Bai}, \&
  {Hou}}]{Zhong2020}
{Zhong}, J., {Chen}, L., {Wu}, D., {et~al.} 2020, \bibinfo{title}{{Exploring
  open cluster properties with Gaia and LAMOST},} \aap, 640, A127,
  \dodoi{10.1051/0004-6361/201937131}

\end{thebibliography}
\bibliographystyle{aasjournalv7}
\clearpage


\end{document}